\documentclass[a4paper,onecolumn,11pt,unpublished]{quantumarticle}
\pdfoutput=1
\usepackage[utf8]{inputenc}
\usepackage[english]{babel}
\usepackage[T1]{fontenc}
\usepackage{amsmath,amssymb,amsfonts,amsthm,mathrsfs}
\usepackage{graphicx}
\usepackage{booktabs,array,multirow}
\usepackage{xcolor}
\usepackage{hyperref}
\usepackage[numbers,sort&compress]{natbib}

\theoremstyle{plain}
\newtheorem{theorem}{Theorem}
\newtheorem{lemma}[theorem]{Lemma}
\newtheorem{proposition}[theorem]{Proposition}
\newtheorem{corollary}[theorem]{Corollary}
\theoremstyle{definition}
\newtheorem{definition}[theorem]{Definition}

\renewcommand{\figurename}{Fig.}

\newcommand{\ket}[1]{|#1\rangle}
\newcommand{\bra}[1]{\langle#1|}
\newcommand{\braket}[2]{\langle#1|#2\rangle}
\newcommand{\Ot}{\mathcal{O}_t}
\newcommand{\hw}{\mathrm{hw}}
\newcommand{\rank}{\mathrm{rank}}
\newcommand{\supp}{\mathrm{supp}}
\newcommand{\Gam}{\Gamma}
\newcommand{\IP}{\mathrm{IP}}
\newcommand{\idx}{\mathtt{idx}}
\newcommand{\inp}{\mathtt{inp}}
\newcommand{\rem}{\mathtt{rem}}
\newcommand{\est}{\mathtt{est}}
\newcommand{\Otil}{\tilde{O}}

\usepackage{adjustbox}
\usepackage{placeins}
\usepackage{multibib}
\newcites{supp}{Supplementary References}
\robustify{\citesupp}
\graphicspath{{./}{figures/}}
\theoremstyle{plain}
\newtheorem{conjecture}[theorem]{Conjecture}
\begin{document}

\title{Linear-depth quantum oracles for clique problems from edge colorings and graph states, with linear non-Clifford cost and a provably bounded-error $k$-clique search}

\author{Payman Kazemikhah}
\affiliation{School of Electrical and Computer Engineering, College of Engineering, University of Tehran, Tehran, Iran}
\author{Ali Hadizadeh Moghadam}
\affiliation{School of Electrical and Computer Engineering, College of Engineering, University of Tehran, Tehran, Iran}
\author{Hossein Aghababa}
\affiliation{School of Electrical and Computer Engineering, College of Engineering, University of Tehran, Tehran, Iran}
\maketitle

\begin{abstract}
Quantum algorithms for clique problems are usually analyzed in the query model, where the graph is accessed through an oracle of unit cost. When the oracle must itself be compiled into gates, published $k$-clique circuits have depth quadratic in the number $n$ of vertices, and every oracle that counts edges spends at least one non-Clifford gate per edge. We first show that scheduling the commuting edge gates by a proper edge coloring yields adjacency-type oracles, including an exact $k$-clique oracle, whose depth and width are linear in $n$, within a logarithmic factor of a depth--width lower bound. Our central result is a $k$-clique oracle in which the graph enters only through the Clifford gate that prepares a graph state, so that its non-Clifford cost is linear in $n$ for every graph. We analyze this oracle exactly: on a candidate vertex set it acts as a reflection whose overlap is $-1$ for cliques and, as we prove, at most $5/8$ in modulus otherwise, a tight bound that tends to $1/2$ as the candidate set grows. A two-qubit phase-estimation filter turns the reflection into a one-sided bounded-error phase oracle, and amplitude amplification then finds a $k$-clique with $O(\sqrt{\binom nk})$ oracle calls, each of depth $O(\Delta+\log n)$ for maximum degree $\Delta$. The guarantee holds for two or more estimation registers; exact simulations on $312$ induced subgraphs of two brain networks confirm it and show that the cheaper single-register variant performs nearly as well.
\end{abstract}

\section{Introduction}\label{sec:intro}

Finding cliques (complete subgraphs) is a basic primitive of graph analysis with applications from computational biochemistry to network science \cite{Butenko2006,Bomze1999}. $k$-\textsc{clique} is one of Karp's original NP-complete problems \cite{Karp1972}, and \textsc{triangle finding} ($k=3$) is tied to Boolean matrix multiplication \cite{Itai1978,Williams2018}. Classically, triangles are found in $O(n^{\omega})$ time, $\omega<2.373$ \cite{Itai1978,Alman2021}, $k$-cliques in $O(n^{\omega k/3})$ \cite{Nesetril1985,Eisenbrand2004}, and branch-and-bound solvers answer instances with thousands of vertices in milliseconds \cite{Bron1973,Tomita2006,SanSegundo2011}.

Quantum algorithms for these problems come in two kinds. In the \emph{query model} the adjacency matrix is an oracle of unit cost: Grover search \cite{Grover1996,Boyer1998,Brassard2002} over the $\binom nk$ vertex subsets needs $\Otil(n^{k/2})$ queries, and quantum-walk and learning-graph algorithms have brought triangle finding down to $\Otil(n^{5/4})$ queries \cite{Magniez2007,Belovs2012,Lee2013,LeGall2014,LeGall2017}. In the \emph{hard-wired circuit model}, the white-box counterpart of the query model, the graph must be compiled into gates and the oracle is no longer free: the explicit $k$-clique circuits of Chang et al.~\cite{Chang2018}, of Metwalli et al.~\cite{Metwalli2020} and of Sanyal Bhaduri et al.~\cite{Sanyal2022} have depth $\Theta(m)=\Theta(n^{2})$ per query, because the $m$ gates that encode the edges all act on a common counter or flag register, a property of those circuits rather than of the problem, as we show below. The recent QAA and Johnson-walk circuits of Perriello~\cite{Perriello2025} lower the per-query depth to $O(n\log n)$ at width $O(n\log n)$ by binary-encoded row sums, reporting up to $2^{15}$ in depth$\times$width over Metwalli et al.~\cite{Metwalli2020}; the graph still enters through one controlled adder per edge, so the non-Clifford count per query remains $\Theta(m\log n)$. Between the two lies the QRAM/QRAG model, in which random access to the $n^{2}$ adjacency bits costs $O(1)$ \cite{Ambainis2007,Buhrman2022,Allcock2023,Jaques2025}. We make the model explicit throughout and place all of our statements in the hard-wired circuit model, without random-access hardware.

Our starting point is an elementary observation: the gates of an adjacency oracle correspond to edges of $G$, gates on disjoint edges commute and can run in parallel, and a proper edge coloring therefore partitions them into $\chi'(G)\le\Delta(G)+1\le n$ layers (Vizing's theorem; \cite{Vizing1964,Misra1992}), with the 1-factorization of the complete graph \cite{Wallis1997} as the worst case. This is how graph states are prepared in depth $O(\Delta)$ \cite{Hein2004,Hein2006,Raussendorf2001,Cabello2011}, how commuting terms are scheduled in linear-depth fermionic simulation and in general-purpose circuit compilation \cite{Kivlichan2018,Guerreschi2018}, and the known depth bound for the phase separator of the quantum approximate optimization algorithm \cite{Herrman2021,Majumdar2021}, and the general reduction of depth minimization for commuting gates to graph coloring is treated in Lee and others~\cite{Lee2025}. The principle is not ours; we use it only as a scheduling tool for oracle circuits. On its own it settles the depth question: applied to the edge-counting oracle of Metwalli et al.~\cite{Metwalli2020} with one small degree counter per vertex, it gives an \emph{exact} $k$-clique oracle of depth $\Theta(n)$ and width $\Theta(n)$ for fixed $k$, and of depth $O(\Delta+\log n)$ on sparse graphs (Proposition~\ref{prop:count}), so depth and width alone do not distinguish oracle constructions. What does distinguish them is where the graph enters the circuit.

In every counting oracle each edge is tested by a Toffoli-class gate, so the number of non-Clifford gates (the quantity that sets the cost of fault-tolerant execution, where Clifford gates are cheap and every $T$ gate consumes a distilled magic state; \cite{Bravyi2005,Fowler2012,Babbush2018,Gidney2018,Beverland2022}) is $\Theta(m)$, i.e.\ $\Theta(n^{2})$ on dense graphs. Our main contribution is a clique test in which the graph enters only through the graph-state unitary $\prod_{uv\in E}\mathrm{CZ}_{uv}$, a Clifford operation, so that the non-Clifford gates are confined to graph-independent blocks of size $\Theta(n)$. Applied to a suitable symmetric test state supported on a vertex set $H$, the graph-state unitary returns the state with overlap $a(H)=\pm1$ exactly when $H$ is complete or empty, and with $|a(H)|$ bounded away from~1 otherwise. We prove $|a(H)|\le5/8$ for every non-clique and every size of the queried set, by reducing $a(H)$ to Gauss sums of quadratic forms over $\mathrm{GF}(2)$ and $\mathbb{Z}_{4}$ \cite{Lidl1997,Brown1972,Aaronson2004}, the machinery by which stabilizer-state amplitudes are evaluated \cite{Dehaene2003,VandenNest2010,BuKoh2022}, and bounding them by the ranks of two matrices attached to the complement of $H$. Because the composite operator $\Gam S_{0}$ is then an exact rotation whose angle is zero for cliques and at least $102.6^{\circ}$ otherwise, a two-qubit phase-estimation filter \cite{Kitaev1995,Cleve1998} converts the oracle into a one-sided bounded-error phase oracle with error at most $2/27$, and the search algorithm of H{\o}yer et al.~\cite{Hoyer2003} for bounded-error inputs gives $O(\sqrt{\binom{n}{k}})$ oracle calls with no logarithmic overhead. Over the whole search the non-Clifford cost is $O(n\sqrt{\binom nk})$, against $O(n^{2}\sqrt{\binom nk})$ with any counting oracle. Two further ingredients belong to the hard-wired circuit model: a two-level edge-detection circuit that removes the common-target bottleneck of the hypergraph-state form \cite{Rossi2013,Qu2013,Guhne2014} of the naive adjacency oracle, and a counting argument showing that at width $\Theta(n)$ no hard-wired oracle family can have depth $o(n/\log n)$, which explains our one-hot encoding of vertex sets.

\begin{figure}[tp]
\centering
\includegraphics[width=\textwidth]{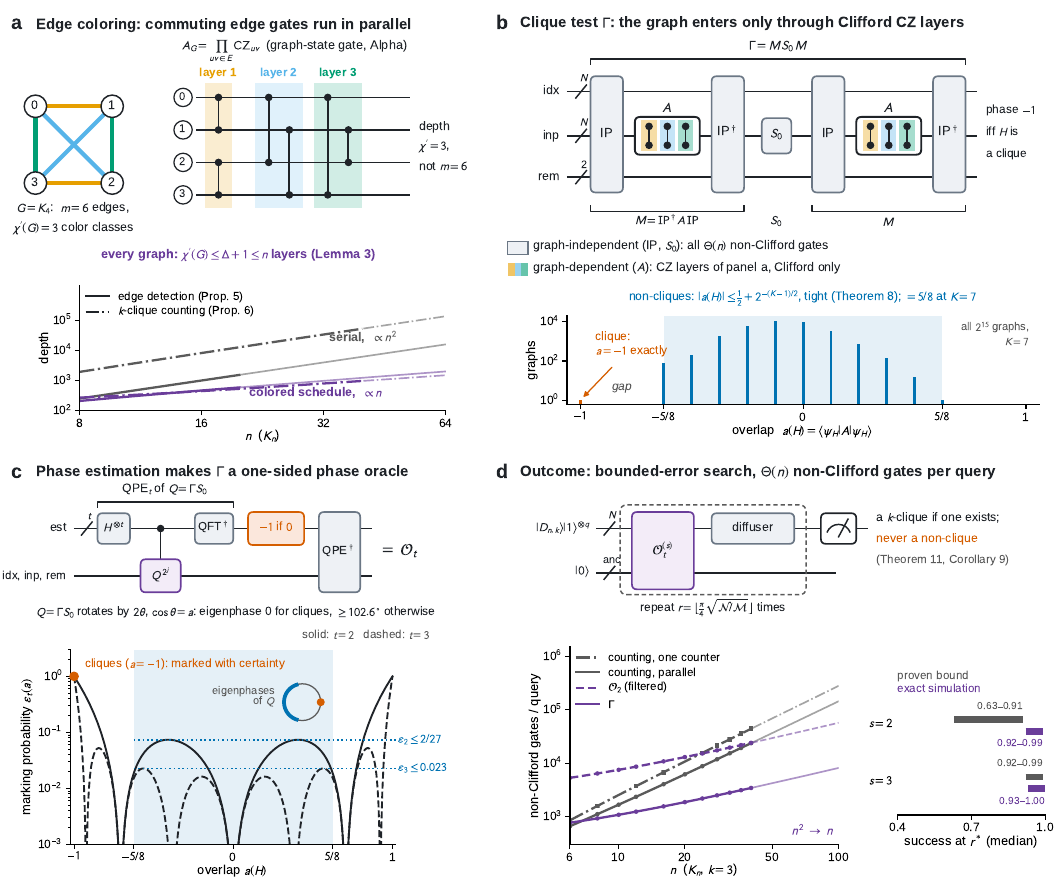}
\caption{\textbf{The idea and the results at a glance.} \textbf{a}, The gates of an adjacency oracle are indexed by edges and commute, so a proper edge coloring schedules them in $\chi'(G)\le\Delta+1\le n$ layers (Definition~\ref{def:schedule}, Lemma~\ref{lem:colouring}); shown for the graph-state gate Alpha of $K_{4}$, whose three color classes are its three layers. Inset: transpiled depth on $K_{n}$ of edge detection and of exact $k$-clique counting with serial and with colored schedules (Propositions~\ref{prop:edge} and~\ref{prop:count}); solid lines are fits over the compiled range ($n\le20$ for edge detection, $n\le40$ for counting), faded lines extrapolate. \textbf{b}, The clique oracle $\Gam=MS_{0}M$ with $M=\IP^{\dagger}A\,\IP$: the graph enters only through the Clifford gate $A$, built from the colored CZ layers of \textbf{a}, while all $\Theta(n)$ non-Clifford gates sit in the graph-independent blocks $\IP$ and $S_{0}$ (Proposition~\ref{prop:tcount}). The overlap $a(H)$ equals $-1$ exactly when $H$ is a clique and satisfies $|a(H)|\le\frac12+2^{-(K-1)/2}$ otherwise (Theorem~\ref{thm:gap}); the histogram shows the exact overlaps of all $2^{15}$ graphs on $K=7$ vertices with one universal vertex, where the bound $5/8$ is attained. \textbf{c}, Phase estimation of $Q=\Gam S_{0}$, a phase flip on the zero estimate and un-computation give the filtered oracle $\Ot$ (Theorem~\ref{thm:filter}); a controlled $Q$ needs only a controlled $S_{0}$ because $M^{2}=I$. Cliques have eigenphase $0$ and are marked with certainty; a non-clique is marked with probability $\varepsilon_{t}(a)=[U_{L-1}(a)/L]^{2}$, at most $2/27$ for $t=2$ (shaded: the range of $a$ for non-cliques; inset: the eigenphases of $Q$, the arc covering all non-cliques). \textbf{d}, Amplitude amplification with $\Ot^{(s)}$ returns a $k$-clique after $O(\sqrt{\mathcal{N}/\mathcal{M}})$ oracle calls and never a non-clique (Theorem~\ref{thm:search}, Corollary~\ref{cor:nofp}). Left: non-Clifford gates per query on $K_{n}$, $k=3$; the counting oracles grow as $n^{2}$ (the sequential counter as its $T$-count after Toffoli-only resynthesis), $\Gam$ and $\mathcal{O}_{2}$ as $n$ (points: compiled circuits; lines: fits; Table~\ref{tab:tradeoff}). Right: median success probability at the Grover count $r^{*}$ in the guaranteed regime $s\ge2$, proven lower bound against exact simulation on brain-network subgraphs (Table~\ref{tab:bench}).}
\label{fig:overview}
\end{figure}

Figure~\ref{fig:overview} summarizes the idea and the results. Section~\ref{sec:prelim} fixes notation and the cost model. Section~\ref{sec:partI} presents Part~I of the paper, the linear-depth oracles: the edge-detection oracle and the exact parallel counting oracle. Part~II introduces the graph-state clique test and proves its spectral gap (Sect.~\ref{sec:partII}), constructs the phase-estimation filter (Sect.~\ref{sec:filter}) and the search algorithm (Sect.~\ref{sec:search}), and reports compiled resources (Sect.~\ref{sec:resources}) and exact simulations on induced subgraphs of two brain networks \cite{Rossi2015} (Sect.~\ref{sec:sims}); Sect.~\ref{sec:discussion} discusses the results. Complete proofs, extended data and the computational protocols are in the Supplementary Material appended after the references, which contains Supplementary Notes~1--16, Figs.~S1--S17 and Tables~S1--S22; the data that produce every number are available upon reasonable request.

\section{Preliminaries and cost model}\label{sec:prelim}

Throughout, $G=(V,E)$ is a simple undirected graph with $n$ vertices, $m$ edges and maximum degree $\Delta$; $\chi'(G)$ is its chromatic index, the minimum number of matchings that partition $E$. Vertex sets are encoded one-hot on an $n$-qubit register $\idx$, so that a basis state $\ket{x}$ represents $\supp(x)=\{u: x_{u}=1\}$, and $\hw(x)$ denotes the Hamming weight; henceforth, ``weight'' always means Hamming weight. Dicke states \cite{Dicke1954} $\ket{D_{n,k}}$, the uniform superposition of all weight-$k$ strings, are prepared with the circuits of B{\"a}rtschi and Eidenbenz~\cite{Bartschi2019} and B{\"a}rtschi and Eidenbenz~\cite{Bartschi2022} and Aktar et al.~\cite{Aktar2022}. Circuit costs are counted after transpilation to the basis $\{R_{z},\sqrt{X},X,\mathrm{CX}\}$ under all-to-all connectivity unless stated otherwise; multi-controlled gates are synthesized with ancillas borrowed from qubits of the circuit that are idle at the time of the gate \cite{Barenco1995,Iten2016}, with depth linear in the number of controls (Sect.~\ref{sec:protocol}; Supplementary Fig.~S11).

Three cost models are in use for graph problems, and we fix terminology. In the \emph{query model} (also called the black-box model), the graph is accessed only through an oracle treated as a black box, and cost counts oracle calls; in the QRAG model, an adjacency bit can be read coherently in $O(1)$ time. We work in the \emph{hard-wired circuit model}, the white-box counterpart of the query model: an oracle for $G$ is an explicit circuit $C_{G}$ over a fixed finite gate set with the graph compiled into its gates, and cost is counted in gates, depth and width of the full circuit, oracle included. Our constructions are invisible in the query model, where oracle calls are free, and unnecessary in the QRAG model; the hard-wired circuit model is the one in which they matter.

\begin{definition}[Edge-detection oracle]\label{def:edge}
An \emph{edge-detection oracle} for $G$ is a unitary on the registers $\idx$ ($n$ qubits), $\mathtt{out}$ (one qubit) and possibly ancillas that, for every two-hot input $x=e_{a}+e_{b}$, maps $\ket{x}\ket{o}\mapsto\ket{x}\ket{o\oplus[ab\in E]}$, with all ancillas starting and ending in $\ket{0}$.
\end{definition}

\begin{definition}[Schedules and the graph-state gate]\label{def:schedule}
A \emph{schedule} of $G$ is an ordered partition $\mathcal{M}=(M_{1},\dots,M_{\ell})$ of $E$ into $\ell$ matchings. A circuit that applies one gate $g_{uv}$ per edge $uv\in E$, where the gates mutually commute and those of edges in a common matching act on pairwise disjoint qubits, is \emph{scheduled by} $\mathcal{M}$ if it applies the gates of $M_{1}$ as one layer, then those of $M_{2}$, and so on; its depth is then $\ell$ layers. The \emph{graph-state gate} of $G$, which we call Alpha, is the circuit $A_{G}=\prod_{uv\in E}\mathrm{CZ}_{uv}$ with one qubit per vertex; it is a diagonal, self-inverse Clifford unitary that prepares the graph state $\ket{G}=A_{G}\ket{+}^{\otimes n}$ \cite{Hein2004}.
\end{definition}

Every circuit in this paper that depends on $G$ is built from such scheduled per-edge layers (the edge detector and the counting oracle of Part~I and Alpha; Fig.~\ref{fig:overview}a), so each construction is parametric in the schedule. The next lemma supplies the schedule used throughout.

\begin{lemma}[Edge partition]\label{lem:colouring}
Every $G$ has a schedule with $\chi'(G)\le\Delta(G)+1\le n$ matchings, computable classically in $O(mn)$ time; $\chi'(K_{n})=n-1$ for even $n$ and $n$ for odd $n$. Consequently, every circuit scheduled by it, in particular $A_{G}$, has depth $\chi'(G)$ layers, the minimum over all schedules of $G$.
\end{lemma}

The first statement is Vizing's theorem with the constructive proof of Misra and Gries \cite{Vizing1964,Misra1992}, and $\chi'(K_{n})$ is attained by the round-robin 1-factorization \cite{Wallis1997}. Neither the partition nor its use as a depth schedule is new \cite{Hein2004,Kivlichan2018,Guerreschi2018}; the lemma is stated for general graphs because every cost in this paper is charged for it (Supplementary Note~1).

The following counting argument fixes what can be hoped for in the hard-wired circuit model.

\begin{proposition}[Depth--width trade-off]\label{prop:lower}
Fix a finite gate set. Let $\{C_{G}\}$ be a family of circuits, one for every simple graph $G$ on $n$ vertices, implementing the edge-detection oracle of Definition~\ref{def:edge} with width at most $w\le n^{O(1)}$. Then some $C_{G}$ has depth $\Omega(n^{2}/(w\log n))$. In particular, at width $\Theta(n)$ the depth is $\Omega(n/\log n)$, and a binary (log-width) encoding of the vertices forces depth $\tilde{\Omega}(n^{2})$.
\end{proposition}

Proposition~\ref{prop:lower} is a Shannon-type counting argument \cite{Shannon1949,Basu2023}; adding one universal vertex extends it verbatim to $k$-clique oracles (Supplementary Note~14).

\section{Linear-depth oracles from edge colorings}\label{sec:partI}

\subsection{Edge detection}\label{sec:edge}

The naive circuit places one Toffoli gate per edge on the common target $\mathtt{out}$, so its depth is $m$. Our circuit is scheduled by the matchings of Lemma~\ref{lem:colouring}: the Toffolis of one matching write into $\lfloor n/2\rfloor$ distinct ancillas, an OR collapses the ancillas onto $\mathtt{out}$, and the gate-by-gate mirror image of the first block un-computes them (both circuits are drawn in Supplementary Fig.~S16a,b).

\begin{proposition}[Linear-depth edge detection]\label{prop:edge}
For every $G$ and every schedule of $G$ with $\ell$ matchings, the partitioned circuit of Supplementary Fig.~S16b scheduled by it is an edge-detection oracle for $G$ (Definition~\ref{def:edge}) with $2m$ Toffoli gates in $2\ell$ layers plus one $\lfloor n/2\rfloor$-controlled X. With the schedule of Lemma~\ref{lem:colouring}, $\ell=\chi'(G)$, its depth is $2\chi'(G)+O(n)\subseteq O(n)$, with $\Theta(m)$ gates and $n+\lfloor n/2\rfloor+O(n)$ qubits. Correctness uses only that at most one Toffoli fires under the two-hot promise, so that the ancilla parities equal the OR.
\end{proposition}

Transpiled, the partitioned circuit is exactly $32n-49$ deep on $K_{n}$ for $n\ge8$, against $\approx3.9n^{2}$ for the naive circuit, and by Proposition~\ref{prop:lower} it is optimal to within a logarithmic factor at its width; the compiled details and their qualifications, and the use of the oracle inside Le Gall's triangle-finding algorithm \cite{LeGall2014} as a bound on oracle depth only, are in Supplementary Notes~7, 8 and~13.

\subsection{Clique detection by parallel counting}\label{sec:count}

The same schedule applies to the edge-counting $k$-clique oracles of Chang et al.~\cite{Chang2018} and Metwalli et al.~\cite{Metwalli2020}, whose depth $\Theta(m)$ comes from a single counter that every edge increments. Giving every vertex $v$ a $b=\lceil\log_{2}k\rceil$-bit counter $\mathtt{cnt}_{v}$ removes the bottleneck: for each edge $uv$ of a color class, $\idx_{u}$ is added to $\mathtt{cnt}_{v}$ and $\idx_{v}$ to $\mathtt{cnt}_{u}$, and since the two updates act on disjoint qubits and two edges of a class share no vertex, a whole class is one layer. After the last class, every vertex of $H=\supp(y)$ whose counter differs from $k-1$ is flagged, an AND tree over the flags applies the phase $-1$ exactly when no flag is set, and everything is un-computed (Supplementary Note~4 and Supplementary Fig.~S15a).

\begin{proposition}[Exact parallel counting oracle]\label{prop:count}
For every $G$ and $k$, the circuit of Supplementary Fig.~S15a implements the exact phase oracle $\ket{y}\mapsto(-1)^{[\supp(y)\ \text{is a }k\text{-clique}]}\ket{y}$ on weight-$k$ inputs, with $2m$ controlled increments in $\chi'(G)\le\Delta+1$ layers, $O(n)$ further Toffolis, $n(4+\lceil\log_{2}k\rceil)-1$ qubits, and depth $O(\chi'(G)\,\mathrm{poly}\log k+\log n)$, i.e.\ $O(\Delta+\log n)$ for fixed $k$ and $O(n)$ on every graph. Without the tree ancillas the final test is an $n$-controlled $Z$ on dirty ancillas, at $n(3+\lceil\log_{2}k\rceil)$ qubits and depth $O(\chi'(G)\,\mathrm{poly}\log k+n)$.
\end{proposition}

The proof is given in Supplementary Notes~4 and~14 and the compiled cost in Supplementary Note~13 (Fig.~S15b,c): on $K_{n}$ the depth is $\approx23n+76$ for $k=3,4$, the shallowest construction in this paper up to $n=36$, by a factor of up to $2.3$ relative to the log-depth compilation of $\Gam$ (Proposition~\ref{prop:logdepth}), and on random graphs it follows $\chi'(G)$; both variants were verified by statevector simulation. Proposition~\ref{prop:count} settles the depth question for counting oracles: for fixed $k$ nothing is gained in depth or width by giving up exactness, and the regime left open by Proposition~\ref{prop:lower} is only $k=\omega(1)$.

What the schedule does not change is the non-Clifford cost. Each of the $2m$ controlled increments contains at least one Toffoli-class gate, so the oracle uses $\Theta(m\log k)$ non-Clifford gates, $\Theta(n^{2})$ on dense graphs ($14n^{2}+28n-14$ after transpilation for $k=3$), exactly like the sequential counter of Metwalli et al.~\cite{Metwalli2020} and the edge detectors of Sect.~\ref{sec:edge}: in all of them the graph is written into the circuit one Toffoli per edge. The graph-state test of Sect.~\ref{sec:partII} removes this factor~$n$.

\section{A graph-state clique test}\label{sec:partII}

Let $k\ge3$ be the requested clique size, let $q=4-((k-3)\bmod4)\in\{1,2,3,4\}$, and let $\hat G$ be $G$ augmented by $q$ \emph{dummy} vertices adjacent to every other vertex. A queried vertex set is $H=R\cup D$ with $R\subseteq V(G)$, $|R|=k$, and $D$ the set of dummies, so that $K:=|H|=k+q\equiv3\pmod4$ and $H$ is complete in $\hat G$ iff $R$ is a $k$-clique of $G$; the dummies both exclude empty subgraphs and enforce the congruence. The registers are $\idx$ and $\inp$ ($N=n+q$ qubits each) and $\rem$ (2 qubits); $2N+2$ qubits in total.

\subsection{Construction}\label{sec:construction}

\paragraph{Alpha.} The oracle applies the graph-state gate of $\hat G$ (Definition~\ref{def:schedule}), $A:=A_{\hat G}$, with $A_{\hat G}=\prod_{uv\in E(\hat G)}\mathrm{CZ}_{uv}$, to $\inp$. It is diagonal, $A\ket{x}=(-1)^{e(x)}\ket{x}$, where $e(x)$ is the number of edges of $\hat G$ inside $\supp(x)$, and $A^{2}=I$. Scheduled by Lemma~\ref{lem:colouring}, its depth is $\chi'(\hat G)\le N$ (Fig.~\ref{fig:overview}a; Supplementary Fig.~S16c); the dummies make $\chi'(\hat G)\ge N-1$ whatever $G$ is, but their gates depend on $x$ only through the parity of its real part, so Proposition~\ref{prop:logdepth} replaces them by a Clifford parity tree of depth $O(\log n)$ and only the real edges cost $\chi'(G)\le\Delta+1$ layers. $A$ is a Clifford unitary: it carries the entire dependence of the oracle on $G$ and none of its non-Clifford gates, and on platforms where graph states are the native resource \cite{Raussendorf2001,Bartolucci2023} it is a state preparation rather than a circuit.

\paragraph{Input Preparator.} $\IP$ (Supplementary Fig.~S16d) maps $\ket{y}_{\idx}\ket{0}_{\inp}\ket{00}_{\rem}$ to $\ket{y}\ket{\psi_{H}}$, $H=\supp(y)$, where on $\inp$
\begin{equation}
\ket{\psi_{H}}\propto\sum_{x\subseteq H,\ \hw(x)\equiv2,3\ (\mathrm{mod}\ 4)}\ket{x}\otimes\ket{s_{x}}_{\rem},
\label{eq:psiH}
\end{equation}
each of the $2^{K-1}$ admissible subsets carrying equal weight and a two-dimensional attachment $\ket{s_{x}}$ on $\rem$. Step 3 of the circuit maps every subset of weight $\equiv0,1\pmod4$ onto its complement within $H$, of weight $\equiv3,2$, which is where $K\equiv3\pmod4$ is used. $\IP$ contains no information about $G$; it uses $N$ controlled-Hadamards, $2N$ Toffolis and $N$ CNOTs, and has depth $\Theta(N)$ from the counter chain \cite{Omer2005} when compiled without ancillas, or $O(\log N)$ with the adder tree of Proposition~\ref{prop:logdepth}. The state \eqref{eq:psiH} is derived in Supplementary Note~16.

\paragraph{The overlap and the oracle.} Write $M=\IP^{\dagger}A\,\IP$, let $\ket{0}$ denote the ancilla reference state ($\inp=0^{N}$, $\rem=00$), let $\Pi_{0}$ project onto $\inp=0^{N}$ and $S_{0}=I-2\Pi_{0}$ (an $(N-1)$-controlled Z with inverted controls: $O(N)$ Toffolis with dirty ancillas \cite{Barenco1995,Iten2016}, or an AND tree at depth $\lceil\log_{2}N\rceil$; Proposition~\ref{prop:logdepth}). The clique oracle is
\begin{equation}
\Gam=M\,S_{0}\,M,\qquad a(H):=\bra{0}M\ket{0}=\bra{\psi_{H}}A\ket{\psi_{H}}=2^{-(K-1)}\!\!\sum_{\substack{x\subseteq H\\ \hw(x)\equiv2,3\,(4)}}(-1)^{e(x)}
\label{eq:aH}
\end{equation}
(Fig.~\ref{fig:overview}b; Supplementary Fig.~S17a). Since $(-1)^{\binom{h}{2}}=-1$ exactly when $h\equiv2,3\pmod4$, $a(H)=-1$ when $H$ is complete and $a(H)=+1$ when $H$ is empty.

\subsection{Exact action}\label{sec:exact}

The following proposition makes the action on every candidate set exact.

\begin{proposition}[Exact action of $\Gam$]\label{prop:exact}
For every basis state $y$ of $\idx$ with $H=\supp(y)$: $M^{2}=I$; the plane $P_{H}=\mathrm{span}\{\ket{0},M\ket{0}\}$ of the ancilla space is invariant under $S_{0}$, $M$ and $\Gam$; writing $a=a(H)$, $b=\sqrt{1-a^{2}}$ and $M\ket{0}=a\ket{0}+b\ket{\perp}$,
\begin{equation}
\Gam\ket{0}=(1-2a^{2})\ket{0}-2ab\ket{\perp},\qquad Q:=\Gam S_{0}\big|_{P_{H}}=\begin{pmatrix}\cos2\theta&-\sin2\theta\\ \sin2\theta&\cos2\theta\end{pmatrix},\ \cos\theta=a .
\label{eq:exact}
\end{equation}
Hence $\Gam$ restricted to $P_{H}$ is the reflection $I-2\ket{\varphi}\bra{\varphi}$ about $\ket{\varphi}=M\ket{0}$, and $Q$ is a rotation by $2\theta$ with eigenvalues $e^{\pm2i\theta}$.
\end{proposition}

\begin{proof}
$M^{2}=\IP^{\dagger}A^{2}\IP=I$. The map $(x,r)\mapsto(x',r')$ realized by $\IP$ on basis states is a bijection in which $r'$ determines whether the complement was taken; since $A$ is diagonal, $\bra{0^{N},r}M\ket{0^{N},0}=2^{-K}\sum_{\tilde x,x}[x'(\tilde x,r)=x'(x,0)][r'(\tilde x,r)=r'(x,0)](\pm1)$ vanishes for $r\neq0$, so the component $\ket{\perp}$ of $M\ket{0}$ has no overlap with any $\inp=0^{N}$ state and $S_{0}\ket{\perp}=\ket{\perp}$. With $S_{0}\ket0=-\ket0$ and $M\ket{\perp}=b\ket0-a\ket{\perp}$ (from $M^{2}=I$), the matrices in \eqref{eq:exact} follow by direct multiplication. Numerically, $\|M^{2}-I\|_{\max}\le2.3\times10^{-15}$ and $|\bra0\Gam\ket0-(1-2a^{2})|\le2.2\times10^{-15}$ on every configuration used in this paper.
\end{proof}

For a clique, $a=-1$: $\Gam\ket{0}=-\ket{0}$ exactly, the ancillas are restored and $Q=I$ on $P_{H}$. For any other $H$ with $a\neq0$, $\Gam$ applies the phase $1-2a^{2}\in(0,1]$ to $\ket{H,0}$ \emph{and} transfers probability $4a^{2}(1-a^{2})$ into an ancilla state $\ket{\perp}$ on which the diffuser, which acts on $\idx$ alone, cannot act (Supplementary Fig.~S1a,b). $\Gam$ is therefore not a phase oracle, and the standard amplitude-amplification analysis \cite{Brassard2002} does not apply to it; this is the precise reason why, used directly, it is a heuristic.

\subsection{The spectral gap}\label{sec:gap}

The key to a rigorous algorithm is a uniform bound on $|a(H)|$ away from $1$ for all non-cliques. Let $F=\overline{H}$ be the complement of $H$ (inside $\hat G[H]$) with adjacency matrix $A_{F}$ over $\mathrm{GF}(2)$, and let $Q_{F}(x)=\sum_{uv\in F}x_{u}x_{v}$. Because $e(x)+e_{F}(x)=\binom{\hw(x)}{2}$, equation~\eqref{eq:aH} becomes
\begin{equation}
a(H)=-2^{-K}\big[F_{1}-\mathrm{Re}\big((1-i)F_{i}\big)\big],\qquad F_{z}:=\sum_{x\in\{0,1\}^{K}}z^{\hw(x)}(-1)^{Q_{F}(x)} .
\label{eq:gauss}
\end{equation}
$F_{1}$ is the Gauss sum of the $\mathrm{GF}(2)$ quadratic form $Q_{F}$, whose polar form has matrix $A_{F}$, and $F_{i}$ is the Gauss sum of the $\mathbb{Z}_{4}$-valued form $\hw(x)+2Q_{F}(x)$, whose polar form has matrix $I+A_{F}$ \cite{Brown1972,Lidl1997}. Both are $2^{K}$ times overlaps of stabilizer states \cite{Aaronson2004}, and a three-line argument (Supplementary Note~2) gives $|F_{1}|\in\{0,2^{K-r/2}\}$ and $|F_{i}|\in\{0,2^{K-r'/2}\}$ with $r=\rank_{2}A_{F}$ and $r'=\rank_{2}(I+A_{F})$, hence
\begin{equation}
|a(H)|\le2^{-r/2}+\sqrt{2}\cdot2^{-r'/2},\qquad r+r'\ge K,\ r\ge2,\ r'\ge2 .
\label{eq:rankbound}
\end{equation}

\begin{theorem}[Spectral gap]\label{thm:gap}
Let $K\equiv3\pmod4$, $K\ge7$, and let $H$ be a graph on $K$ vertices that is not complete and has at least one universal vertex. Then $|a(H)|\le\tfrac12+2^{-(K-1)/2}$, and the bound is attained for every such $K$: by $K_{K}$ minus an edge (and by the star $K_{1,K-1}$) when $K\equiv7\pmod8$, and by the disjoint union $K_{3}\sqcup K_{2}$ completed by $K-5$ universal vertices when $K\equiv3\pmod8$. In particular $|a(H)|\le5/8$, with equality only at $K=7$, and the worst case tends to $1/2$ as $K\to\infty$.
\end{theorem}

\begin{corollary}[No false positives]\label{cor:nofp}
$a(H)=-1$ if and only if $H$ is a clique; otherwise the phase applied by $\Gam$ to $\ket{H,0}$ satisfies $1-2a^{2}\ge7/32$. In particular no vertex set that is not a $k$-clique can ever be fully marked, for any $k$ and $n$.
\end{corollary}

The proof (Supplementary Note~2) combines the rank bound \eqref{eq:rankbound} with the parity of the Brown invariant of the $\mathbb Z_{4}$-valued form \cite{Brown1972,Wall1963}, which settles every rank class with $r,r'\ge4$ at $1/2$, and with closed-form evaluations of the three remaining low-rank classes: $r'=2$ forces a star, $r'=3$ forces the complement to be three disjoint cliques ($|a|\le\tfrac14$), and $r=2$ gives the complete multipartite complements that attain the bound. No enumeration enters the proof; exhaustive enumeration reproduces the maxima $5/8$ at $K=7$ (Fig.~\ref{fig:overview}b) and $17/32$ at $K=11$ exactly. Cliques are therefore always marked exactly, but a near-clique leaks most of its amplitude into the ancillas at every query ($95\%$ for $K_{7}$ minus an edge), so the plain oracle is a poor amplitude-amplification oracle (Supplementary Note~6): the filter of the next section is a necessity rather than a refinement.

\section{A one-sided phase oracle by phase estimation}\label{sec:filter}

Because $Q=\Gam S_{0}$ is an exact rotation whose angle vanishes for cliques and is at least $2\arccos(5/8)=102.6^{\circ}$ in modulus otherwise, phase estimation of $Q$ separates the two cases. Let $L=2^{t}$ and let $\est$ be a $t$-qubit register. Define $\mathrm{QPE}_{t}=(\mathrm{QFT}_{L}^{\dagger}\otimes I)\big(\sum_{j<L}\ket{j}\bra{j}\otimes Q^{j}\big)(H^{\otimes t}\otimes I)$ and
\begin{equation}
\Ot=\mathrm{QPE}_{t}^{\dagger}\,\big(I-2\ket{0}\bra{0}_{\est}\big)\,\mathrm{QPE}_{t}
\label{eq:Ot}
\end{equation}
(Fig.~\ref{fig:overview}c; Supplementary Fig.~S17b). Because $M^{2}=I$, a controlled $Q$ is $M(\mathrm{c}S_{0})M(\mathrm{c}S_{0})$: only the reflection $S_{0}$ acquires the control qubit (Supplementary Fig.~S17c), so no controlled version of Alpha or of $\IP$ is needed and the edge-coloring depth is preserved.

\begin{theorem}[Phase-estimation filter]\label{thm:filter}
On $\ket{y}\ket{0}_{\est}\ket{0}$ with $H=\supp(y)$ and $a=a(H)$,
\begin{align}
\Ot\ket{y,0,0}&=\ket{y}\big[(1-2\varepsilon)\ket{0,0}+\ket{g}\big],\qquad \braket{0,0}{g}=0,\qquad \|g\|^{2}=4\varepsilon(1-\varepsilon),\nonumber\\
\varepsilon&=\varepsilon_{t}(a)=\Big[\frac{\sin(L\arccos a)}{L\sin(\arccos a)}\Big]^{2}=\Big[\frac{U_{L-1}(a)}{L}\Big]^{2},
\label{eq:eps}
\end{align}
where $U_{L-1}$ is the Chebyshev polynomial of the second kind. Thus $\varepsilon=1$ and $\Ot\ket{y,0,0}=-\ket{y,0,0}$ exactly when $H$ is a clique, while for every non-clique $\varepsilon_{t}\le64/(39\cdot4^{t})$; the suprema over $|a|\le5/8$ are $2/27$, $0.023$ and $0.0062$ for $t=2,3,4$. $\Ot$ uses $2(L-1)$ applications of $Q$, i.e.\ $4(L-1)$ applications of $\IP^{\dagger}A\,\IP$ and as many controlled $S_{0}$, $t$ additional qubits, and depth $\Theta(n)$ for fixed $t$. With $s$ copies of the $\est$ register whose estimates are combined by AND, the error becomes $\varepsilon_{t}(a)^{s}$ at $s$ times the cost.
\end{theorem}

\begin{proof}
The $\est=0$ row of $\mathrm{QFT}_{L}^{\dagger}$ is uniform, so $\bra{0}_{\est}\mathrm{QPE}_{t}\ket{0}_{\est}=L^{-1}\sum_{j<L}Q^{j}=:A_{t}$, and $\Ot\ket{0_{\est},0}=\ket{0_{\est},0}-2\,\mathrm{QPE}_{t}^{\dagger}(\ket{0_{\est}}\otimes A_{t}\ket0)$, whose $\ket{0_{\est},0}$ component is $1-2\|A_{t}\ket0\|^{2}$. On $P_{H}$, $\ket0=(v_{+}+v_{-})/\sqrt2$ with $Qv_{\pm}=e^{\pm2i\theta}v_{\pm}$ and $A_{t}v_{\pm}=e^{\pm i(L-1)\theta}\frac{\sin L\theta}{L\sin\theta}v_{\pm}$, which gives \eqref{eq:eps}; $\sin^{2}\theta=1-a^{2}\ge39/64$ gives the bound. With $s$ registers the all-zero component after the $s$ estimations is $A_{t}^{s}\ket0$, and $A_{t}$ is normal with eigenvalues of modulus $\sqrt\varepsilon$, so the error is $\varepsilon^{s}$. The identity $\mathrm{c}Q=M(\mathrm{c}S_{0})M(\mathrm{c}S_{0})$ holds because the control-$0$ branch gives $M^{2}=I$.
\end{proof}

$\Ot$ is a one-sided bounded-error phase oracle: cliques are never missed, and a non-clique is marked with probability at most $2/27<1/3$ already for $t=2$ (Supplementary Note~3, Fig.~S3). 

The distinction from the faulty-oracle model of Regev and Schiff~\cite{Regev2008}, in which no speed-up survives, is essential. There each call silently applies the identity with some probability, a classical mixture whose hidden failure record decoheres the search. $\Ot$ is instead one fixed, known unitary whose deviation from the ideal phase oracle is a coherent, flagged, one-sided amplitude of norm at most $2\varepsilon^{s/2}$ that remains available to interference; this is the bounded-error model of H{\o}yer et al.~\cite{Hoyer2003} that Theorem~\ref{thm:search}(ii) invokes. The full comparison, including the modern picture of which faulty-oracle models retain a quantum advantage \cite{Lolck2024}, is in Supplementary Note~11. Equation~\eqref{eq:eps} is verified on the compiled circuits to $8\times10^{-14}$ (Sect.~\ref{sec:protocol}).

The filter is not free: once compiled, $\mathcal O_{2}$ costs about seven $\Gam$ queries ($6.98$ in depth, $6.71$ in non-Clifford count and $6.99$ in $T$ count) and two extra qubits (Supplementary Note~9). What is bought is a known iteration count and a guarantee: with $\mathcal O_{2}$ the optimum returns to the ideal $\lfloor\tfrac{\pi}{4}\sqrt{\mathcal N/\mathcal M}\rfloor$, whereas $\Gam$ alone needs $2.1$--$3.8$ times as many iterations to reach a maximum of only $0.32$--$0.52$ (Table~\ref{tab:bench}).

\section{Bounded-error $k$-clique search}\label{sec:search}

The search register is prepared in $\ket{D_{n,k}}\otimes\ket{1}^{\otimes q}$ with the Dicke circuit of B{\"a}rtschi and Eidenbenz~\cite{Bartschi2019} and B{\"a}rtschi and Eidenbenz~\cite{Bartschi2022}, so that the search space is the $\binom{n}{k}$ $k$-subsets of $V(G)$, each queried together with the $q$ dummies; one iteration applies $\Ot^{(s)}$ and the diffuser $I-2\ket{\sigma}\bra{\sigma}$ about the initial state; after $r$ iterations $\idx_{0..n-1}$ is measured and the outcome verified classically (complete circuit in Supplementary Fig.~S13). Let $\mathcal{N}=\binom{n}{k}$ and let $\mathcal{M}$ be the number of $k$-cliques of $G$.

\begin{theorem}[Bounded-error $k$-clique search]\label{thm:search}
(i) With $r=\lfloor\tfrac{\pi}{4}\sqrt{\mathcal{N}/\mathcal{M}}\rfloor$ iterations of $\Ot^{(s)}$ and the diffuser, the probability that the measurement returns a $k$-clique is at least $\sin^{2}\!\big((2r+1)\arcsin\sqrt{\mathcal{M}/\mathcal{N}}\big)-4r\,\varepsilon_{t}^{\,s/2}$, where $\varepsilon_{t}=\sup_{|a|\le5/8}\varepsilon_{t}(a)$; for $t=2$ the loss term is $4r(2/27)^{s/2}$, i.e.\ $1.09\,r$ for $s=1$ (vacuous for every $r\ge1$), $0.30\,r$ for $s=2$ and $0.081\,r$ for $s=3$. Hence $t=2$ and $s=O(\log(\mathcal{N}/\mathcal{M}))$ give constant success probability with $O(\sqrt{\mathcal{N}/\mathcal{M}}\,\log\mathcal{N})$ applications of $\Gam$. (ii) Since $\varepsilon_{2}<1/3$, the bounded-error search algorithm of H{\o}yer et al.~\cite{Hoyer2003} finds a $k$-clique, if one exists, with probability at least $2/3$ using $O(\sqrt{\mathcal{N}})$ applications of $\mathcal{O}_{2}$, i.e.\ $O(\sqrt{\binom{n}{k}})$ applications of $\Gam$, each of depth $O(\chi'(G)+\log n)\subseteq O(n)$ (Proposition~\ref{prop:logdepth}), $\Theta(n^{2})$ gates of which only $\Theta(n)$ are non-Clifford, and $3(n+q)+3$ qubits ($2(n+q)+4$ at depth $\Theta(n)$). (iii) For every $r$ and $s$, the success probability is at least $\sin^{2}\!\big((2r+1)\arcsin\sqrt{\mathcal{M}/\mathcal{N}}\big)-4\,\varepsilon_{t}^{\,s}\,r^{2}$, and $\varepsilon_{t}$ may be replaced by the largest $\varepsilon_{t}(a_{y})$ over the non-clique candidates of the instance; for $t=2$ the loss is $0.30\,r^{2}$ ($s{=}1$), $0.022\,r^{2}$ ($s{=}2$) and $0.0016\,r^{2}$ ($s{=}3$), sharper than (i) whenever $r<\varepsilon_{t}^{-s/2}$ and non-vacuous for $s=1$ at $r=1$.
\end{theorem}

Part (i) is a hybrid argument, part (iii) is proved by tracking the amplitude that leaks out of and back into the search sector rather than by a worst-case triangle inequality (Supplementary Note~14), which makes it first order in $\varepsilon$ at the price of a quadratic dependence on $r$, and part (ii) removes the logarithmic factor because the error reduction is folded into the amplitude amplification \cite{Hoyer2003}.

Which statement covers which experiment is spelled out once: \emph{the rigorous basis of every claim in this paper is the regime $s\ge2$}. There the better of (i) and (iii) is non-vacuous for $s=3$ at every $r^{*}$ that occurs and for $s=2$ on every configuration of the main protocol, and Table~\ref{tab:bench} lists it next to the observed median $P(r^{*})$ on the same instances, with $s=2$ and $s=3$ run on every instance: guaranteed $0.63$--$0.91$ against observed $0.92$--$0.99$ for $s=2$, and $0.92$--$0.99$ against $0.93$--$1.00$ for $s=3$, where the bound is attained to numerical precision. At the size ceiling ($r^{*}$ up to $11$) the $s=2$ bound becomes vacuous or weak while the observed success stays at $0.96$--$0.99$, and $s=3$ still guarantees $0.80$--$0.94$; the price of this regime is a factor $s$ in filter cost and $2s-2$ additional qubits. The cheaper single-register variant ($s=1$) is covered only at the smallest iteration counts, by part (iii), and is reported as an illustration of the same behavior rather than as evidence for a theorem (Supplementary Note~11).

Its observed loss is first order and linear in $r$ (at most $3.7$ times $r\bar\varepsilon$ over all $312$ instances with sweeps to $r=4r^{*}+10$, and at most $3.9$ times over $10\,000$ further random graphs; since the loss never exceeds $1$, these sweeps cover every $r$ at which the ratio could exceed $4$), whereas (iii) is quadratic in $r$; closing this gap is the remaining open problem, stated as Conjecture~S6 with the constant $4$, which all these data satisfy (Supplementary Note~11). When $\mathcal{M}$ is unknown the exponential schedule of Boyer et al.~\cite{Boyer1998} applies, and fixed-point search \cite{Yoder2014} avoids over-shooting.

Two points deserve emphasis. First, gate count and depth do not separate the constructions ($\Theta(n^{2})$ gates for every one, depth $O(\chi'(G)+\log n)$ for the parallel counter and for $\Gam$ alike); the non-Clifford count does, $\Theta(m\log k)$ against $\Theta(n)$ (Proposition~\ref{prop:tcount}), at the price of a one-sided error and a constant factor in depth for $\mathcal O_{2}$. Second, the total oracle depth $O(n\sqrt{\binom{n}{k}})=O(n^{k/2+1})$ does not beat the classical exponent $\omega k/3$ for $k=3$ and is only marginally below it for $k=4$; the contribution is the reduction of depth and of non-Clifford cost over previous quantum circuits and the rigorous error guarantee, not a claim of superiority over classical clique algorithms (Supplementary Note~10 places the result among the query, QRAG and classical models).

\section{Compiled resources and non-Clifford cost}\label{sec:resources}

All circuits are built from one- and two-qubit gates plus multi-controlled gates, with no unitary black boxes, and can therefore be transpiled (a gate-level rendering of $\Gam$ for the smallest valid instance is Supplementary Fig.~S9). Depth is one axis of the comparison; the number of non-Clifford gates is the other, because in the surface code and in every magic-state architecture Clifford gates are cheap while each $T$ gate consumes a distilled magic state, so that the $T$-count is the standard figure of merit for fault-tolerant cost \cite{Bravyi2005,Fowler2012,Babbush2018,Gidney2018,Litinski2019,Beverland2022}. In every construction of Part~I the graph is written into the circuit one Toffoli-class gate per edge. In $\Gam$ it is not:

\begin{proposition}[Non-Clifford cost]\label{prop:tcount}
Every gate of $\Gam$ that depends on $G$ is a CZ. The non-Clifford gates of $\Gam$ lie in the four applications of $\IP$ and $\IP^{\dagger}$ ($N$ controlled-Hadamards and $2N$ Toffolis each) and in $S_{0}$ (one $(N-1)$-controlled $Z$, $O(N)$ Toffolis with dirty ancillas; \cite{Barenco1995,Iten2016}), so $\Gam$ uses $O(N)=O(n)$ non-Clifford gates for every graph, and $\Ot$ uses about $2(2^{t}-1)$ times as many ($6.7$ times as compiled for $t=2$). The counting oracles of Chang et al.~\cite{Chang2018}, of Metwalli et al.~\cite{Metwalli2020} and of Proposition~\ref{prop:count} use $\Theta(m\log k)$.
\end{proposition}

The depth of $\Gam$ needs the same treatment. Compiled block by block, $\Gam$ has depth $\Theta(n)$ on every graph for three structural reasons (the dummies make $\chi'(\hat G)\ge N-1$, the mod-4 counter of $\IP$ is a chain of $N$ Toffolis, and $S_{0}$ is an $N$-controlled gate), and none of them is essential:

\begin{proposition}[Log-depth compilation of $\Gam$]\label{prop:logdepth}
Replace, in $\Gam=\mathrm{M}S_{0}\mathrm{M}$, the dummy CZ gates of $A$ by an in-place CNOT parity tree over the real $\inp$ qubits, $q$ CZ gates on its root and $\binom q2$ CZ gates among the dummies; the counter of $\IP$ by a tree of in-place two-bit mod-4 adders on $\lceil N/2\rceil$ ancillas whose carries are left in place and canceled by $\IP^{\dagger}$; the Toffolis of step~3 of $\IP$ by the same Toffolis controlled by $\rem_{1}$ and $\lfloor N/2\rfloor-1$ CNOT copies of it; and $S_{0}$ by an AND tree on $N-1$ ancillas. The result is the same unitary on the code space, with $3N+1$ qubits ($3N+1+t$ for $\Ot$), $\Theta(m)$ Clifford gates, the same $\Theta(N)$ non-Clifford gates, and depth $O(\chi'(G)+\log n)$, i.e.\ $O(\Delta+\log n)$ on sparse graphs and $O(n)$ on every graph.
\end{proposition}

\begin{figure}[t]
\centering
\includegraphics[width=\textwidth]{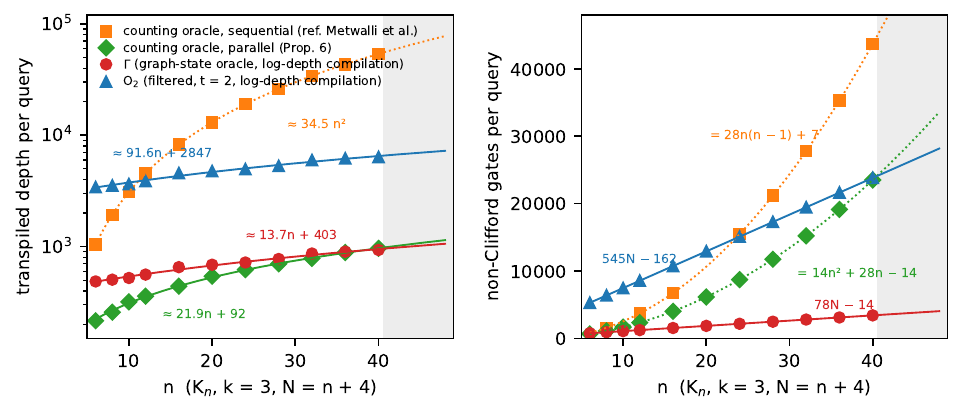}
\caption{\textbf{Compiled resources of one $k$-clique oracle call} (complete graph $K_{n}$, $k=3$, $N=n+4$; transpiled to $\{R_{z},\sqrt X,X,\mathrm{CX}\}$, optimization level 1, all-to-all coupling; shaded region: extrapolated fits). \textbf{a}, Transpiled depth: sequential counting oracle of Metwalli et al.~\cite{Metwalli2020} ($\approx34.5n^{2}$), parallel counting oracle of Proposition~\ref{prop:count} ($\approx21.9n+92$), and $\Gam$ ($\approx13.7n+403$) and the explicitly built $\mathcal O_{2}$ ($\approx91.6n+2\,847$) in the log-depth compilation of Proposition~\ref{prop:logdepth}. \textbf{b}, $T$ gates per call under the single normalization of the main text (pure Clifford$+T$ after Toffoli-only re-synthesis; a Toffoli contributes $7$ $T$, a controlled-Hadamard $2$; Sect.~\ref{sec:protocol}): $78N-14$ for $\Gam$ and $545N-162$ for $\mathcal O_{2}$ (Proposition~\ref{prop:tcount}), $14n^{2}+28n-14$ for the parallel counting oracle, and exactly $28n(n-1)+7$ for the re-synthesized sequential one (points at $n=8,12,16$). The mixed-basis circuit that the transpiler emits for the sequential counter ($\approx20n^{2}$ rotations, some $\pi/8$) is reported in Supplementary Note~12.}
\label{fig:oracles}
\end{figure}

The proof is in Supplementary Note~14. The compiled circuit equals the block-by-block one to $10^{-13}$, and the filtered oracle is verified directly, including full-width statevector checks at $24$--$27$ qubits (Supplementary Note~12). Fig.~\ref{fig:oracles} and Table~\ref{tab:tradeoff} give the compiled numbers on the complete graph ($k=3$) for the sequential counting oracle of Metwalli et al.~\cite{Metwalli2020}, the parallel counting oracle of Proposition~\ref{prop:count}, and $\Gam$ and $\mathcal O_{2}$ in the compilation of Proposition~\ref{prop:logdepth}, all under the same transpiler settings (Sect.~\ref{sec:protocol}; full data in Supplementary Tables~S6, S7 and~S18). On $K_{n}$, measured up to $n=40$, one $\Gam$ query has depth $\approx13.7n+403$ and one $\mathcal O_{2}$ query $\approx91.6n+2\,847$; on random graphs the depth of $\Gam$ follows the number of color classes of $G$ (Supplementary Fig.~S15c).

The synthesis-invariant statement comes first: the $T$-counts separate as $\Theta(n)$ for $\Gam$ and $\mathcal O_{2}$ against $\Theta(m\log k)$ for the sequential and $\Theta(m)$ for the parallel counting oracle (a constant number of Toffoli-class blocks per vertex against at least one per edge), and no choice of Toffoli synthesis moves a construction between these classes, only within its constants. For the constants we fix one normalization for every Clifford$+T$ number in the main text: pure Clifford$+T$ after Toffoli-only re-synthesis of all multi-controlled gates, counted at the transpiled level ($7$ $T$ per Toffoli, $2$ per controlled-Hadamard; Sect.~\ref{sec:protocol}). In it the counts are exactly $78N-14$ for $\Gam$ and $545N-162$ for $\mathcal O_{2}$, $14n^{2}+28n-14$ for the parallel counting oracle, and exactly $28n(n-1)+7=56m+7$ for the sequential one on $K_{n}$ (the mixed-basis circuit that the transpiler emits for it, $\approx20n^{2}$ rotations of which some are $\pi/8$, is kept in Supplementary Note~12 only).

The comparison is two-sided. In depth on $K_{n}$ the parallel counter ($\approx21.9n+92$) and $\Gam$ are within a factor $2.3$ of each other up to $n=36$, $\Gam$ becomes the shallower of the two at $n=40$ ($933$ against $965$ layers), and $\mathcal O_{2}$ is about nine times deeper than the parallel counter at $n=20$ (four times asymptotically). In non-Clifford cost the order reverses: $\Gam$ is below both counters from $n=8$, and $\mathcal O_{2}$ falls below the sequential counter from $n=24$ and below the parallel one just above $n=40$ (the exact counts $14n^{2}+28n-14$ and $545N-162$ cross between $n=40$ and $41$), beyond which the ratio grows linearly in $n$; all of these are measured on compiled circuits. Neither family dominates: the counting oracles are the constructions of choice when width is scarce or exactness is required at the smallest depth, the graph-state oracle when non-Clifford gates are. Supplementary Note~12 gives the full comparison, including the QAA oracle of Perriello~\cite{Perriello2025}.

Two qualifications make the fault-tolerant reading precise (Supplementary Note~12). The comparison assumes, as is standard \cite{Babbush2018,Gidney2018,Beverland2022}, that magic-state distillation dominates the space-time volume; the Clifford CZ gates of $A$ still cost $O(d)$ code cycles each in lattice surgery \cite{Litinski2019} but consume no magic states. And where magic-state factories are the bottleneck, the relevant quantity is the $T$-depth, in which the families separate even more sharply: every $n$-dependent layer of $\Gam$ is Clifford, so its $T$-depth is logarithmic ($145$--$193$ for $n=6$--$40$, rising in steps), against $\Theta(\chi'(G))$ for the parallel and $\Theta(m)$ for the sequential counter.

\begin{table}[t]
\centering\footnotesize
\caption{\textbf{Explicit $k$-clique oracles in the hard-wired circuit model, and the trade-off between the two families of this work.} Per-query costs; $b=\lceil\log_{2}k\rceil$, $N=n+q$ with $q\le4$, $\mathcal N=\binom nk$; ``non-Clifford'' counts Toffoli-class gates, each a constant number of $T$ gates; measured columns are transpiled numbers on $K_{16}$, $k=3$ (settings of Fig.~\ref{fig:oracles}; $\Gam$ and $\mathcal O_{2}$ in the compilation of Proposition~\ref{prop:logdepth}; full data, $T$-depths and measurement-based $T$-counts in Supplementary Tables~S6, S18 and~S19). The filtered oracle $\mathcal O_{t}^{(s)}$ uses $2(2^{t}-1)s$ applications of $Q=\Gam S_{0}$; the state-preparation depth per iteration, $O(k\log(n/k))$ with the Dicke circuits of B{\"a}rtschi and Eidenbenz~\cite{Bartschi2022}, is dominated by the oracle.}
\label{tab:tradeoff}
\setlength{\tabcolsep}{3.5pt}\resizebox{\textwidth}{!}{\begin{tabular}{@{}>{\raggedright\arraybackslash}p{2.7cm}>{\raggedright\arraybackslash}p{2.4cm}>{\raggedright\arraybackslash}p{1.6cm}>{\raggedright\arraybackslash}p{1.7cm}>{\raggedright\arraybackslash}p{1.9cm}>{\raggedright\arraybackslash}p{1.4cm}>{\raggedright\arraybackslash}p{1.0cm}>{\raggedright\arraybackslash}p{1.05cm}@{}}
\toprule
construction & depth / query & qubits & non-Clifford / query & error; queries for $k$-\textsc{clique} & wins when & depth $K_{16}$ & non-Cliff.\ $K_{16}$\\
\midrule
Chang et al.~\cite{Chang2018} (\textsc{max-clique}) & $O(n^{2})$ & $O(n^{2})$ & $O(n^{2})$ & exact; $O(2^{n/2})$ & -- & -- & --\\
Metwalli et al.~\cite{Metwalli2020} & $\Theta(m\log k)\subseteq O(n^{2})$ & $n+O(\log k)$ & $\Theta(m\log k)$ & exact; $O(\sqrt{\mathcal N})$ & width & $8\,248$ & $6\,727$\\
Perriello~\cite{Perriello2025}, QAA variant & $O(n\log n)$ (interlaced adders) & $n(\lceil\log_{2}n\rceil+2)$ & $\Theta(m\log n)$ (one adder per edge) & exact; $O(\sqrt{\mathcal N})$ & $T$-depth at moderate $n$ & -- & --\\
parallel counting, Prop.~\ref{prop:count} (this work) & $O(\chi'(G)\,\mathrm{poly}\log k$ ${}+\log n)$; $O(\Delta+\log n)$ for fixed $k$ & $n(4+b)-1$ & $\Theta(m\log k)$ (one per edge) & exact; $O(\sqrt{\mathcal N})$ & depth; sparse graphs & $441$ & $4\,018$\\
graph-state $\Gam$, Props.~\ref{prop:exact}, \ref{prop:tcount}, \ref{prop:logdepth} (this work) & $O(\chi'(G)+\log n)$, i.e.\ $O(\Delta+\log n)$ & $3N+1$ ($2N+2$ at depth $\Theta(n)$) & $\Theta(n)$, graph-independent & leaks amplitude (not a phase oracle); no bound & non-Clifford gates, $n\ge8$ & $654$ & $1\,546$\\
filtered $\mathcal O_{2}^{(s)}$, Thms.~\ref{thm:filter}, \ref{thm:search} (this work) & $O(\chi'(G)+\log n)$, about $7s$ times $\Gam$ & $3N+1+2s$ & $\Theta(n)$, graph-independent & one-sided, $\le(2/27)^{s}$; $O(\sqrt{\mathcal N})$ & non-Clifford gates, $n\gtrsim40$, with a guarantee & $4\,561$ ($s=1$) & $10\,738$ ($s=1$)\\
\bottomrule
\end{tabular}}
\end{table}

\section{Exact simulations on brain-network subgraphs}\label{sec:sims}

We tested the algorithm on induced subgraphs of two graphs from the Network Repository \cite{Rossi2015}, the macaque cerebral-cortex network ($91$ vertices, $1\,401$ edges, density $0.342$, clique number $17$) and the mouse-retina network ($1\,076$ vertices, $90\,811$ edges, density $0.157$, clique number $51$; Supplementary Table~S10), both taken as the simple graphs of their full vertex sets. For each $(n,k)$, $n$-subsets of the vertex set were drawn uniformly at random, kept if the induced subgraph contains a $k$-clique and is not degenerate ($\mathcal M/\mathcal N\ge1/2$), and de-duplicated up to graph isomorphism, giving $29$--$60$ pairwise non-isomorphic instances per configuration ($288$ in all, from $1.28\times10^{6}$ samples; Sect.~\ref{sec:protocol}). For every instance we computed the \emph{exact} success probability $P(r)$ (the probability that a measurement of $\idx_{0..n-1}$ after $r$ iterations returns a $k$-clique) with a statevector simulator that exploits the block structure of $\Gam$ (Sect.~\ref{sec:protocol}); no shot sampling enters. We compare the ideal phase oracle, $P_{\mathrm{ideal}}(r)=\sin^{2}((2r+1)\theta)$ with $\sin^{2}\theta=\mathcal{M}/\mathcal{N}$, the plain oracle $\Gam$, and the filtered oracles $\mathcal{O}_{2}$ and $\mathcal{O}_{3}$, at the ideal iteration count $r^{*}=\lfloor\pi/(4\theta)\rfloor$ and at their own maximum over $r$.

\begin{figure}[t]
\centering
\includegraphics[width=0.75\textwidth]{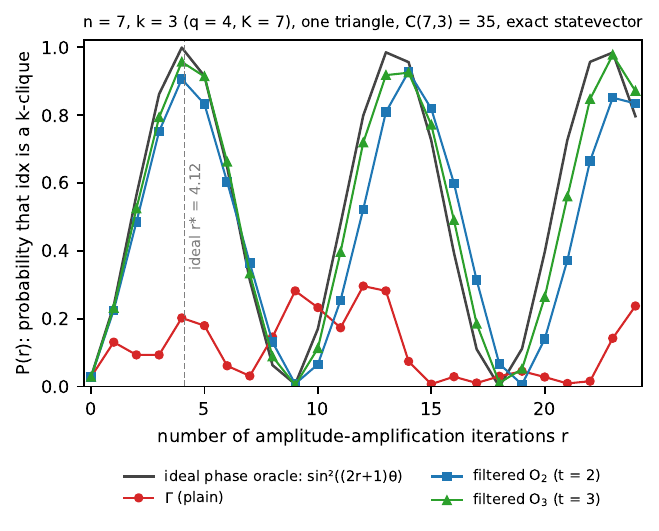}
\caption{\textbf{Amplitude-amplification trajectories on one test instance.} Exact probability $P(r)$ that a measurement of $\idx$ returns a $k$-clique after $r$ iterations, for an induced subgraph of the macaque network with $n=7$, $k=3$ ($q=4$, $K=7$), $m=7$ edges and exactly one triangle ($\mathcal N=35$, ideal $r^{*}=4.12$). Ideal phase oracle: $\sin^{2}((2r+1)\theta)$, $\theta=\arcsin\sqrt{1/35}$. Plain $\Gam$ peaks at $0.30$ ($r=12$) and collapses to $0.008$ at $r=15$; the filtered oracles $\mathcal{O}_{2}$ and $\mathcal{O}_{3}$ reach $0.91$ and $0.96$ at $r=4$ and follow the ideal curve (single estimation register, $s=1$: the illustrative regime; the guaranteed regime $s\ge2$ is quantified in Table~\ref{tab:bench}).}
\label{fig:trajectory}
\end{figure}

Fig.~\ref{fig:trajectory} shows a representative instance ($n=7$, $k=3$, one triangle, $\mathcal{N}=35$, ideal optimum $r^{*}=4$). With $\Gam$ alone the trajectory is not a sinusoid: it peaks at $0.30$ at $r=12$, about three times the ideal iteration count, and collapses to $0.008$ at $r=15$, an aggravated form of the over-shooting known from Grover search \cite{Brassard1997}, because leaked amplitude is never recovered; with $\mathcal{O}_{2}$ and $\mathcal{O}_{3}$ the success probability at $r=4$ is $0.91$ and $0.96$, against $0.998$ for the ideal oracle. Aggregated over instances (Table~\ref{tab:bench}), the median of $P(r^{*})$ for $\Gam$ alone is $0.05$--$0.29$ and its maximum over $r$ is $0.32$--$0.52$, reached at $2.1$--$3.8$ times $r^{*}$, whereas the filtered oracle in its guaranteed regime ($s=2,3$) reaches an observed $P(r^{*})$ of $0.92$--$0.99$ and $0.93$--$1.00$ on every instance, above the proven $0.63$--$0.91$ and $0.92$--$0.99$. The single-register variant reaches $0.83$--$0.90$ at $r^{*}$, against $0.94$--$1.00$ for the ideal oracle, and two size-ceiling rows ($(10,6)$ and $(12,10)$, $12$ instances each, $r^{*}$ up to $11$; $312$ instances in total) show no width dependence; there the filtered oracle reaches $0.96$--$0.99$ with $s=2$ and $1.00$ with $s=3$. Inter-quartile ranges, bootstrap confidence intervals and the sampling details are in Supplementary Note~9.

\begin{table}[t]
\centering\small
\caption{\textbf{Exact success probabilities at the ideal iteration count, observed and guaranteed} (induced subgraphs of the two host graphs, full vertex sets; $\#$ = number of pairwise non-isomorphic clique-bearing, non-degenerate instances; $r^{*}=\lfloor\pi/(4\theta)\rfloor$; $\langle r^{*}\rangle$ = mean $r^{*}$). Entries are medians over instances (inter-quartile range in brackets) of $P(r^{*})$ for the ideal oracle, the plain oracle $\Gam$ and the filtered oracle $\mathcal O_{2}^{(s)}$ with $s=1,2,3$ estimation registers ($2(n+q)+2+2s$ qubits; $s=2,3$ were run on every instance, and the number of instances is in parentheses). ``guar.'' is the better of the lower bounds of Theorem~\ref{thm:search}(i) and (iii) at the same $r^{*}$ with the worst-case constant, $P_{\mathrm{ideal}}(r^{*})-\min\!\big(4r^{*}(2/27)^{s/2},\,4(2/27)^{s}r^{*2}\big)$, taken over the same instances as the observed column; negative medians are shown as $<0$. The $s\ge2$ columns constitute the guaranteed regime on which the claims of this paper rest; the $s=1$ columns are the illustrative half-cost variant. The last two rows probe the size ceiling ($r^{*}$ up to $11$; $s=1$ sweeps to $r^{*}+2$); at such $r^{*}$ the $s=2$ bound is vacuous or weak, while the $s=3$ bound remains non-vacuous. Bootstrap $95\%$ confidence intervals on every median are in Supplementary Table~S22; best success over $r$, $\mathcal O_{3}$, and the instance-wise bounds are in Supplementary Tables~S11 and~S17.}
\label{tab:bench}
\resizebox{\textwidth}{!}{\begin{tabular}{@{}ccclccccccccccc@{}}
\toprule
$n$&$k$&$q$&host&$\#$&$\langle r^{*}\rangle$&ideal&$\Gam$&$\mathcal O_{2}^{(1)}$&guar.\ $s{=}1$&$\mathcal O_{2}^{(2)}$&guar.\ $s{=}2$&$\mathcal O_{2}^{(3)}$&guar.\ $s{=}3$\\
\midrule
6&3&4&macaque&60&1.6&1.00&0.29 [0.07, 0.35]&0.90 [0.87, 0.93]&0.57&0.99 (60)&0.91&1.00 (60)&0.99\\
6&4&3&mouse&34&2.4&0.94&0.05 [0.03, 0.10]&0.83 [0.80, 0.86]&$<0$&0.92 (34)&0.74&0.93 (34)&0.92\\
7&3&4&macaque&60&2.3&0.99&0.07 [0.05, 0.17]&0.88 [0.86, 0.91]&$<0$&0.98 (60)&0.79&0.99 (60)&0.97\\
7&4&3&macaque&30&2.6&0.99&0.09 [0.04, 0.13]&0.84 [0.82, 0.86]&$<0$&0.98 (30)&0.79&0.99 (30)&0.97\\
7&5&2&mouse&45&2.6&1.00&0.07 [0.04, 0.13]&0.84 [0.82, 0.87]&$<0$&0.99 (45)&0.80&1.00 (45)&0.98\\
8&3&4&mouse&30&3.7&0.99&0.08 [0.06, 0.15]&0.88 [0.87, 0.90]&$<0$&0.99 (30)&0.63&0.99 (30)&0.95\\
8&6&1&mouse&29&3.6&0.98&0.13 [0.11, 0.15]&0.88 [0.86, 0.90]&$<0$&0.97 (29)&0.63&0.98 (29)&0.95\\
10&6&1&mouse&12&9.8&1.00&0.09 [0.06, 0.12]&0.64 [0.62, 0.66]&$<0$&0.96 (12)&$<0$&1.00 (12)&0.80\\
12&10&1&macaque&12&5.5&1.00&0.28 [0.23, 0.31]&0.87 [0.86, 0.89]&$<0$&0.99 (12)&0.21&1.00 (12)&0.94\\
\bottomrule
\end{tabular}}
\end{table}

\subsection{Protocol}\label{sec:protocol}

Because $\idx$ enters $\Gam$ and $\Ot$ only as a control, both are block-diagonal in the $\idx$ basis, while the diffuser acts on $\idx$ alone; the simulator therefore evolves one ancilla vector per candidate and applies the diffuser as a rank-one update. It reproduces a gate-by-gate Qiskit statevector simulation \cite{Qiskit2024} of the full circuits to $|\Delta P|<8.1\times10^{-14}$ at about $2000$ times the speed, and Theorem~\ref{thm:filter} is verified on the compiled filtered oracle to $8\times10^{-14}$. The host graphs are used as simple graphs, with duplicate edges and self-loops removed. For each configuration $(n,k)$, $n$-subsets are drawn uniformly with a fixed seed; degenerate instances ($\mathcal M/\mathcal N\ge1/2$, for which the ideal iteration count is zero) are excluded, the remaining ones are de-duplicated up to graph isomorphism, and clique-free subgraphs are kept for the false-positive check of Corollary~\ref{cor:nofp}. Resource counts use Qiskit's transpiler with basis $\{R_{z},\sqrt X,X,\mathrm{CX}\}$, optimization level~1 and a fixed seed. Every Clifford$+T$ number in this paper uses one normalization: all multi-controlled gates are re-synthesized into Toffolis, and each non-Clifford rotation left after transpilation is an odd multiple of $\pi/4$, i.e.\ one $T$ gate ($7$ per Toffoli and $2$ per controlled-Hadamard). Supplementary Notes~5 and~6 also show why the search register must be initialized as in Sect.~\ref{sec:search}: preparing the Dicke state on all $N$ index qubits instead queries the oracle on a different candidate set, so that a shot-sampled benchmark with that initialization does not test the algorithm. The complete protocols are given in Supplementary Note~15: sweep ranges, sampling counts, the synthesis of multi-controlled gates with borrowed ancillas, and the re-implementation of the oracle of Metwalli et al.~\cite{Metwalli2020}.

\section{Discussion}\label{sec:discussion}

We have separated four statements that the literature on explicit quantum clique circuits tends to merge. The first is a statement about scheduling: an adjacency-type oracle for an $n$-vertex graph can be compiled to depth $O(\chi'(G)+\log n)$, $\chi'(G)\le\Delta+1\le n$, with $\Theta(n)$ qubits and no random-access memory, within a logarithmic factor of the depth--width lower bound, and this applies to counting oracles and to the graph-state oracle alike, so that an exact $k$-clique oracle of depth and width $\Theta(n)$ exists for every fixed $k$ (Proposition~\ref{prop:count}); the quadratic depth of the published circuits was a property of the circuits, not of the problem. The second is a statement about non-Clifford cost: every counting oracle writes the graph into the circuit one Toffoli-class gate per edge and therefore uses $\Theta(m)$ non-Clifford gates, whereas the graph-state oracle carries the graph in a Clifford unitary and uses $\Theta(n)$ (Proposition~\ref{prop:tcount}). The third is a statement about that oracle: the graph-state unitary, tested against the symmetric state \eqref{eq:psiH}, is an exact reflection controlled by an overlap that equals $-1$ for cliques and is at most $5/8$ in modulus otherwise, for every clique size and every graph. The fourth is a statement about an algorithm: a two-qubit phase-estimation filter converts that reflection into a one-sided bounded-error phase oracle, and the resulting search is a rigorous $O(\sqrt{\binom nk})$-query algorithm at $\Theta(n)$ depth and $\Theta(n)$ non-Clifford gates per query.

Where the method works, and where it does not, is a measured statement. The depth reduction is unconditional and is confirmed on compiled circuits for all three linear-depth constructions. The exact parallel counting oracle is the shallowest construction up to $n=36$ and has no error; it is the right choice whenever depth is the binding constraint and Toffolis are cheap, as on unencoded hardware. The graph-state oracle never produces a false positive, and with the filter it recovers the ideal amplitude-amplification behavior at a constant factor of about seven per query, so that its non-Clifford advantage over an exact counting oracle appears just above $n=40$ and then grows linearly; over a full search it is a factor $n$ in the dominant fault-tolerant resource (Table~\ref{tab:tradeoff}). The remaining deficits are the worst-case constant $2/27$, the quadratic $r$-dependence of Theorem~\ref{thm:search}(iii) for $s=1$ (Conjecture~S6), and the absence of a time bound for triangle finding. On the instance sizes we can simulate, classical solvers answer instantly; the contribution is the reduction of depth and of non-Clifford cost over previous quantum circuits and the rigorous error guarantee, not a claim of practical superiority.

What these results imply depends on the resources one is prepared to assume: in the query model the oracle is free and our constructions are invisible; in the QRAG model, whose time-efficient algorithmics is an active subject \cite{Buhrman2022,Allcock2023}, already a plain Grover search over the $\binom n3$ triples runs in $\Otil(n^{3/2})$ time, which no compiled oracle can match; in the hard-wired circuit model, the model of every explicit quantum clique circuit published so far \cite{Chang2018,Metwalli2020}, the per-query depth sets the coherence time of a run and the per-query non-Clifford count sets its fault-tolerant cost (Supplementary Note~10). For fault-tolerant execution, where Clifford gates are cheap (not free, but without magic states) and every $T$ gate consumes a distilled one \cite{Bravyi2005,Fowler2012,Litinski2019}, the non-Clifford part of $\Gam$ is the same circuit for every graph: the graph enters through the Clifford unitary $A$, and a factor-$n$ reduction of the magic-state budget, at a $T$-depth that is logarithmic rather than linear in $n$, is the relevant improvement. For hardware whose native resource is the graph state (measurement-based, photonic and neutral-atom platforms; \cite{Raussendorf2001,Hein2006,Bartolucci2023,Bluvstein2022}), $A$ is a state preparation rather than a circuit, the graph-independent blocks $\IP$ and $S_{0}$ are the only operations that require non-Clifford resources, and the depth accounting of this paper becomes a statement about the number of entangling rounds.

Three open questions stand out (a longer list is Supplementary Note~10). The counting argument of Proposition~\ref{prop:lower} bounds all gates, not the non-Clifford ones, and $\Gam$ shows that the graph can be carried entirely by Clifford gates; whether an \emph{exact} clique oracle can use $o(m)$ non-Clifford gates is open. Theorem~\ref{thm:search}(iii) is first order in the error but quadratic in the iteration count, where the observed loss is linear (Conjecture~S6); removing the quadratic dependence, i.e.\ controlling the interference of the amplitude that returns from the ancilla sectors, would make the single-register guarantee match the demonstrations. And the reduction of the oracle overlap to Gauss sums of $\mathbb Z_{4}$-valued quadratic forms is not specific to cliques: replacing the complete pattern by another graph changes the quadratic form and the rank conditions, and the same technique should yield exact error bounds for graph-state tests of other induced-subgraph patterns.

\section*{Acknowledgments}
No funding was used in this project.

\section*{Data availability}
The data that produce the results of this paper are available upon reasonable request.

\section*{Author contributions}
A.H.M. designed the edge-detection, Alpha, Input Preparator and $\Gam$ oracles and the edge-coloring layering, wrote the initial Qiskit implementation, performed the initial experiments and wrote the initial manuscript. P.K. developed the analysis of $\Gam$ (Proposition~\ref{prop:exact}), the spectral-gap theorem (Theorem~\ref{thm:gap}), the phase-estimation filter and the complexity results (Theorems~\ref{thm:filter} and \ref{thm:search}), the depth--width bound (Proposition~\ref{prop:lower}), the simulations, the compiled-resource analysis, the audit of the initial experiments, and the text and figures of this manuscript. H.A. supervised and advised the project and verified the results. All authors reviewed and approved the manuscript.

\section*{Competing interests}
The authors declare no competing interests.

\bibliographystyle{quantum}
\bibliography{refs}

\clearpage
\setcounter{figure}{0}\setcounter{table}{0}\setcounter{equation}{0}\setcounter{theorem}{0}
\renewcommand{\thefigure}{S\arabic{figure}}\renewcommand{\thetable}{S\arabic{table}}
\renewcommand{\theequation}{S\arabic{equation}}\renewcommand{\thetheorem}{S\arabic{theorem}}
\renewcommand{\theHfigure}{S\arabic{figure}}\renewcommand{\theHtable}{S\arabic{table}}
\renewcommand{\theHequation}{S\arabic{equation}}\renewcommand{\theHtheorem}{S\arabic{theorem}}
\captionsetup{font=small,labelfont=bf}
\setlength{\emergencystretch}{3em}
\begin{center}{\Large\bfseries Supplementary Material}\\[6pt]
{\large Linear-depth quantum oracles for clique problems from edge colorings and graph states, with linear non-Clifford cost and a provably bounded-error $k$-clique search}\end{center}
\vspace{4pt}
\noindent\textbf{Contents}\par\smallskip
{\small
{\hypersetup{linkcolor=black}\noindent\begin{tabular}{@{}b{0.92\linewidth}@{\hspace{0.6em}}r@{}}
Supplementary Note~1: the round-robin 1-factorization\dotfill & \pageref{S:note:1}\\
Supplementary Note~2: Gauss sums, low-rank cases and the proof of Theorem 8\dotfill & \pageref{S:note:2}\\
Supplementary Note~3: the phase-estimation filter\dotfill & \pageref{S:note:3}\\
Supplementary Note~4: the exact parallel counting oracle\dotfill & \pageref{S:note:4}\\
Supplementary Note~5: search-register initialization and the host graphs\dotfill & \pageref{S:note:5}\\
Supplementary Note~6: trajectories of the plain oracle\dotfill & \pageref{S:note:6}\\
Supplementary Note~7: compiled circuit resources and validation\dotfill & \pageref{S:note:7}\\
Supplementary Note~8: triangle finding with the edge-detection oracle\dotfill & \pageref{S:note:8}\\
Supplementary Note~9: the complete circuit, the cost of the filter and the aggregated benchmarks\dotfill & \pageref{S:note:9}\\
Supplementary Note~10: comparison across models and further directions\dotfill & \pageref{S:note:10}\\
Supplementary Note~11: guarantees versus experiments for $s=1,2,3$ estimation registers\dotfill & \pageref{S:note:11}\\
Supplementary Note~12: the log-depth compilation of $\Gam$ and the fault-tolerant accounting\dotfill & \pageref{S:note:12}\\
Supplementary Note~13: complementary compiled details (the edge detector, the counting oracle and the size ceiling)\dotfill & \pageref{S:note:13}\\
Supplementary Note~14: proofs of Theorem~11 and of Propositions~4, 6 and~13\dotfill & \pageref{S:note:14}\\
Supplementary Note~15: computational protocols\dotfill & \pageref{S:note:15}\\
Supplementary Note~16: circuit diagrams of the constructions\dotfill & \pageref{S:note:16}\\
\end{tabular}}
}
\medskip
\begin{figure}[!htbp]
\centering
\includegraphics[width=\textwidth]{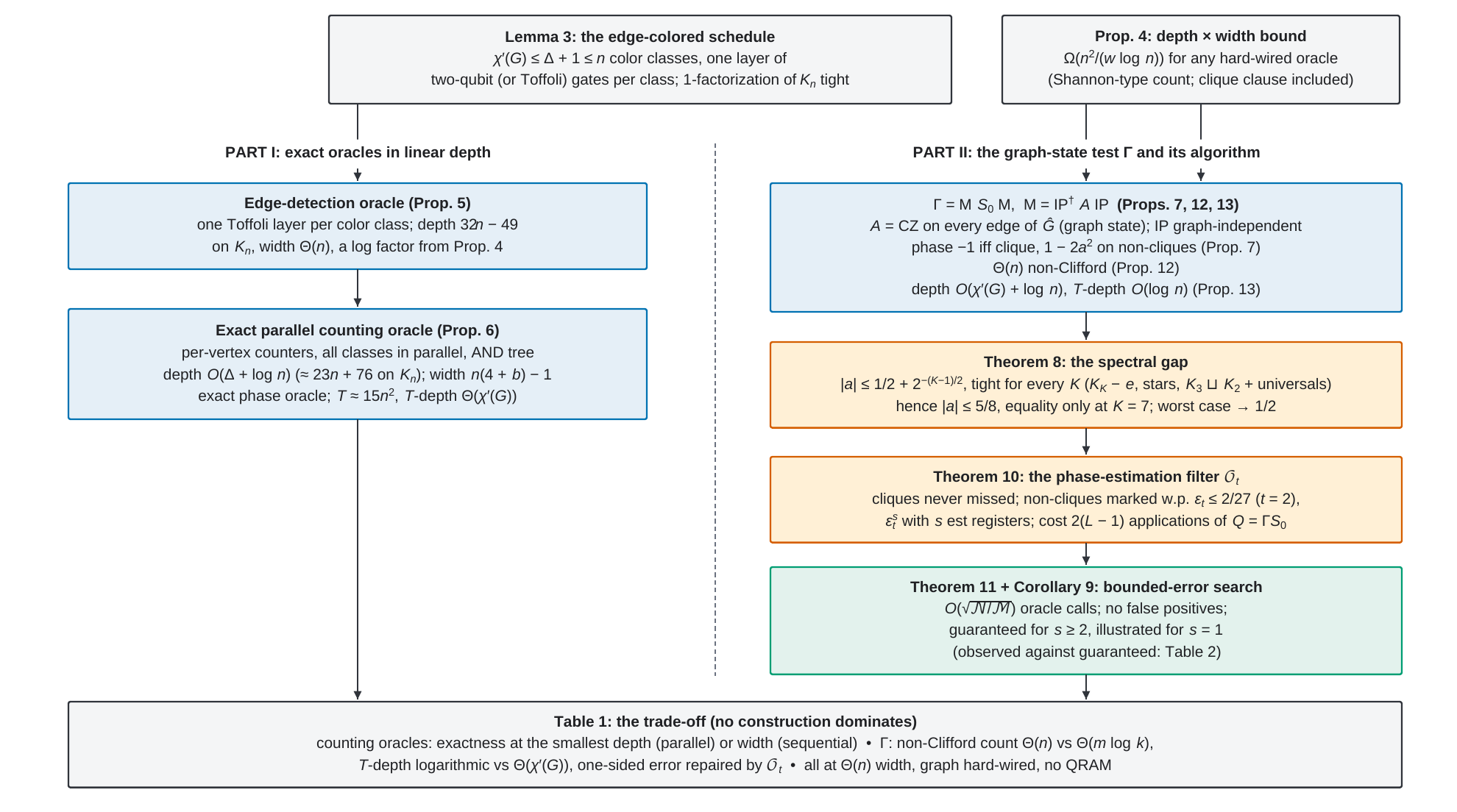}
\caption*{\textbf{Map of the main-text results.} Both parts rest on the edge-coloring lemma (main-text Lemma~3). Part~I: the edge-detection oracle (Proposition~5), the depth--width trade-off (Proposition~4) and the exact parallel counting oracle (Proposition~6), which shares the depth of Part~II but not its non-Clifford cost (Proposition~12). Part~II: the graph-state gate Alpha and the graph-independent Input Preparator combine into the clique oracle $\Gam$, whose action is an exact reflection controlled by the overlap $a(H)$ (Proposition~7); Theorem~8 bounds $a(H)$; the phase-estimation filter $\mathcal{O}_{t}$ (Theorem~10) makes $\Gam$ a bounded-error phase oracle; amplitude amplification gives Theorem~11 and Corollary~9. The compilation of Proposition~13 gives both parts the same asymptotic depth, $O(\chi'(G)+\log n)$. The Input Preparator uses neither Alpha nor the edge-detection oracle, and the edge-detection oracle is not part of the $k$-clique pipeline; its use in triangle finding is Supplementary Note~8.}
\end{figure}

\FloatBarrier
\section*{Supplementary Note 1: the round-robin 1-factorization}\label{S:note:1}

For even $n$, label the vertices $0,\dots,n-1$ and define, for $i=0,\dots,n-2$,
\begin{equation}
S_{i}=\big\{\{n-1,\,i\}\big\}\cup\big\{\{(i-j)\bmod(n-1),\ (i+j)\bmod(n-1)\}:\ j=1,\dots,\tfrac n2-1\big\}.
\end{equation}
Each $S_{i}$ is a perfect matching: the pairs are disjoint because $(i-j)+(i+j)\equiv2i$ and the map $j\mapsto(i\pm j)$ is injective on $\{1,\dots,n/2-1\}$ modulo $n-1$. If $\{(i-j),(i+j)\}=\{(h-l),(h+l)\}$ then $2i\equiv2h\pmod{n-1}$ and, since $n-1$ is odd, $i=h$; the pairs $\{n-1,i\}$ are distinct for distinct $i$. Hence the $n-1$ matchings are pairwise disjoint, and since they contain $(n-1)\cdot n/2=\binom n2$ edges in total they partition $E(K_{n})$. For odd $n$ the construction is applied to $n+1$ and the pairs touching the extra vertex are dropped, giving $n$ near-perfect matchings of $(n-1)/2$ edges. This is the classical construction of Kirkman (1847) and Lucas (1883) \citesupp{S-Wallis1997}; it is the \texttt{alpha()} routine of the accompanying code, and it was verified for $3\le n\le16$ to cover every edge exactly once with disjoint rounds, emitting $\chi'(K_{n})=n-1$ (even $n$) or $n$ (odd $n$) rounds (Table~\ref{S:tab:onefact}).

\begin{table}[!htbp]
\centering\small
\caption{Verification of the round-robin layering used by \texttt{alpha()}.}
\label{S:tab:onefact}
\resizebox{\textwidth}{!}{\begin{tabular}{@{}cccccc@{}}
\toprule
$n$ & rounds emitted & max round size & covers every edge once & disjoint within round & $\chi'(K_{n})$\\
\midrule
3 & 3 & 1 & yes & yes & 3\\ 4 & 3 & 2 & yes & yes & 3\\ 5 & 5 & 2 & yes & yes & 5\\ 6 & 5 & 3 & yes & yes & 5\\
7 & 7 & 3 & yes & yes & 7\\ 8 & 7 & 4 & yes & yes & 7\\ 9 & 9 & 4 & yes & yes & 9\\ 10 & 9 & 5 & yes & yes & 9\\
11 & 11 & 5 & yes & yes & 11\\ 12 & 11 & 6 & yes & yes & 11\\ 13--16 & $n$ or $n-1$ & $\lfloor n/2\rfloor$ & yes & yes & $n$ or $n-1$\\
\bottomrule
\end{tabular}}
\end{table}

\paragraph{The edge partition in context.} The first statement is Vizing's theorem with the constructive proof of Misra and Gries \citesupp{S-Vizing1964,S-Misra1992}; the value of $\chi'(K_{n})$ is attained by the classical round-robin 1-factorization of Kirkman and Lucas \citesupp{S-Wallis1997}, which we use as the explicit worst-case schedule (Supplementary Note~1). Only matchings are needed, not perfect matchings, which exist only for even $n$. Neither the partition nor its use as a depth schedule is claimed as new: scheduling commuting gates by an edge coloring is standard in graph-state preparation \citesupp{S-Hein2004}, in swap networks for fermionic simulation \citesupp{S-Kivlichan2018} and in general circuit schedulers \citesupp{S-Guerreschi2018}. The lemma is stated for general graphs because every cost in this paper is charged for it: a sparse graph costs $\Delta+1$ layers rather than the $n-1$ of the complete graph (Supplementary Fig.~S15c), $K_{n}$ is the tight worst case, and in every circuit below a whole color class, in the counting oracle including all of its controlled counter updates, is a single layer (main-text Propositions~5 and~6).

\FloatBarrier
\section*{Supplementary Note 2: Gauss sums, low-rank cases and the proof of Theorem 8}\label{S:note:2}

\paragraph{Complement form.} With $F=\overline H$ the complement of $H$ inside $\hat G[H]$ and $e_{F}(x)$ the number of edges of $F$ inside $\supp(x)$, $e(x)+e_{F}(x)=\binom{\hw(x)}{2}$, and $(-1)^{\binom h2}=-1$ exactly when $h\equiv2,3\pmod4$. Hence
\begin{equation}
a(H)=-2^{-(K-1)}\sum_{\hw(x)\equiv2,3\,(4)}(-1)^{e_{F}(x)}.
\label{S:eq:aF}
\end{equation}
The indicator of $h\equiv2,3\pmod4$ is $(1-\mathrm{Re}[(1-i)i^{h}])/2$ (check $h=0,1,2,3$), so with $F_{z}=\sum_{x}z^{\hw(x)}(-1)^{Q_{F}(x)}$,
\begin{equation}
a(H)=-2^{-K}\big[F_{1}-\mathrm{Re}((1-i)F_{i})\big].
\label{S:eq:aFz}
\end{equation}
Equation \eqref{S:eq:aFz} agrees with the definition of $a(H)$ to $0.0$ on all graphs with $K=7$ and on random graphs with $K=11,15$ (script \texttt{08\_gap\_theory.py}).

\paragraph{Gauss-sum lemma.}
\begin{lemma}\label{S:lem:gauss}
Let $\phi:\mathrm{GF}(2)^{K}\to\mathbb Z_{4}$ satisfy $\phi(x\oplus y)=\phi(x)+\phi(y)+2B(x,y)\pmod4$ for a symmetric bilinear form $B$ over $\mathrm{GF}(2)$. Then $\big|\sum_{x}i^{\phi(x)}\big|\in\{0,\,2^{K-\rank_{2}(B)/2}\}$.
\end{lemma}
\begin{proof}
$|\sum_{x}i^{\phi(x)}|^{2}=\sum_{y,z}i^{\phi(y\oplus z)-\phi(y)}=\sum_{z}i^{\phi(z)}\sum_{y}(-1)^{B(y,z)}=2^{K}\sum_{z\in\mathrm{Rad}B}i^{\phi(z)}$. On $\mathrm{Rad}B$ the function $\phi$ is additive, and $2\phi(z)=\phi(0)=0$ gives $\phi(z)\in\{0,2\}$, so $z\mapsto\phi(z)/2$ is a linear functional on $\mathrm{Rad}B$; the sum is $|\mathrm{Rad}B|=2^{K-\rank B}$ if the functional vanishes and $0$ otherwise.
\end{proof}
For $\phi=2Q_{F}$ one has $Q_{F}(x\oplus y)=Q_{F}(x)+Q_{F}(y)+x^{\top}A_{F}y$ over $\mathrm{GF}(2)$, so $B=A_{F}$; for $\phi=\hw+2Q_{F}$, $\hw(x\oplus y)=\hw(x)+\hw(y)-2x\cdot y$ and $-2\equiv2\pmod4$ give $B=I+A_{F}$. Hence $|F_{1}|\in\{0,2^{K-r/2}\}$ and $|F_{i}|\in\{0,2^{K-r'/2}\}$ with $r=\rank_{2}A_{F}$, $r'=\rank_{2}(I+A_{F})$, and $|a(H)|\le2^{-r/2}+\sqrt2\,2^{-r'/2}$. Both magnitude statements hold exactly on every graph tested (Table~\ref{S:tab:gapcheck}).

\paragraph{Facts about the ranks.} $A_{F}$ is symmetric with zero diagonal (alternating), so $r$ is even and $r\ge2$ as soon as $F$ has an edge. $r+r'\ge\rank(I)=K$ by subadditivity. $r'=1$ means $I+A_{F}=vv^{\top}$ with $v_{u}^{2}=1$ for all $u$, i.e.\ $I+A_{F}=J$ and $F=K_{K}$, so $H$ is empty; a universal vertex excludes this.

\paragraph{The case $r'=2$.} $S=I+A_{F}$ is symmetric with unit diagonal, hence not alternating, hence congruent over $\mathrm{GF}(2)$ to a diagonal matrix (Albert's theorem \citesupp{S-Lidl1997}); rank $2$ gives $S=vv^{\top}+ww^{\top}$ with $v,w$ independent. The diagonal condition $v_{u}+w_{u}=1$ forces $w=v+\mathbf1$ and $S=v\mathbf1^{\top}+\mathbf1v^{\top}+J$, i.e.\ $S_{uv}=v_{u}+v_{v}+1$. Off the diagonal $(A_{F})_{uv}=1$ iff $v_{u}=v_{v}$, so $F$ is the disjoint union of two cliques on $P=\supp(v)$ and its complement (both non-empty, else the rank is $1$) and $H=K_{|P|,K-|P|}$. A universal vertex of $K_{a,b}$ requires $a=1$ or $b=1$, so $H=K_{1,K-1}$, $q=1$ and $R$ is the empty graph on $K-1$ vertices.

\paragraph{Closed forms.} For the star, $F=K_{K-1}\cup\{\text{isolated center}\}$:
\[F_{1}=2\sum_{y\in\mathrm{GF}(2)^{K-1}}(-1)^{\binom{\hw y}{2}}=2\,\mathrm{Re}[(1-i)(1+i)^{K-1}]=4\,\mathrm{Re}(1+i)^{K-2}\] and, since $(1+i)^{2}=2i$ and $(K-3)/2$ is even, $(1+i)^{K-2}=(-1)^{(K-3)/4}2^{(K-3)/2}(1+i)$, so $F_{1}=(-1)^{(K-3)/4}2^{(K+1)/2}$;
\[F_{i}=(1+i)\sum_{y}i^{\hw(y)}(-1)^{\binom{\hw y}2}=(1+i)\sum_{y}i^{\hw(y)^{2}}=(1+i)\cdot2^{K-2}(1+i)=2^{K-1}i\] (because $h^{2}\equiv0$ or $1\pmod4$ according to the parity of $h$ and even and odd subsets are equinumerous). Then $\mathrm{Re}((1-i)F_{i})=2^{K-1}$ and
\begin{equation}
a(K_{1,K-1})=\tfrac12-(-1)^{(K-3)/4}\,2^{-(K-1)/2},
\end{equation}
i.e.\ $|a|=\tfrac12+2^{-(K-1)/2}$ for $K\equiv7\pmod8$ and $\tfrac12-2^{-(K-1)/2}$ for $K\equiv3\pmod8$ (numerically $0.625$, $0.46875$, $0.50781$, $0.49805$, $0.50049$ for $K=7,11,15,19,23$). For $F$ a single edge plus isolated vertices ($H=K_{K}-e$): $F_{1}=2^{K-1}$ and $F_{i}=(1+i)^{K-2}(1+i+i+1)=2(1+i)^{K-1}$, giving the same magnitudes $\tfrac12\pm2^{-(K-1)/2}$.

\paragraph{Proof of Theorem 8 of the main text.} By the parity refinement of Lemma~\ref{S:lem:parity} below, $|a|\le2^{-r/2}+2^{-\lfloor r'/2\rfloor}$, which settles every class with $r,r'\ge4$ at $\tfrac12<\tfrac12+2^{-(K-1)/2}$. The three remaining classes are evaluated exactly: $r'=2$ is the star computed above, with $|a|=\tfrac12\pm2^{-(K-1)/2}$; $r'=3$ is Lemma~\ref{S:lem:threeclique}, with $|a|\le2^{-(K-3)/2}\le\tfrac14$; and $r=2$ is Lemma~\ref{S:lem:rank2}, with $|a|\le\tfrac12+2^{-(K-1)/2}$, attained by $K_{K}-e$ (and the star) for $K\equiv7\pmod8$ and by parts $(3,2)$ plus $K-5$ universal vertices for $K\equiv3\pmod8$. The maximum over the four cases is $\tfrac12+2^{-(K-1)/2}$, for every $K\ge7$. \qed

\begin{lemma}[Parity of the Brown invariant]\label{S:lem:parity}
Let $\phi=\hw+2Q_{F}$ and $r'=\rank_{2}(I+A_{F})$. Then $F_{i}=\sum_{x}i^{\phi(x)}$ is $0$ or $2^{K-r'/2}\zeta_{8}^{\beta}$ with $\beta\equiv r'\pmod2$; consequently $\mathrm{Re}((1-i)F_{i})\le2^{K-r'/2}$ if $r'$ is even and $\le2^{K-(r'-1)/2}$ if $r'$ is odd, and $|a(H)|\le2^{-r/2}+2^{-\lfloor r'/2\rfloor}$.
\end{lemma}
\emph{Proof.} On the radical of $B=I+A_{F}$ the form $\phi$ takes values in $\{0,2\}$ and is either identically $0$, in which case the sum factorizes as $2^{K-r'}$ times the Gauss sum of the induced non-degenerate form on a space of dimension $r'$, or not, in which case $F_{i}=0$. A non-degenerate $\mathbb Z_{4}$-valued quadratic form is an orthogonal sum of rank-one forms $x\mapsto\pm x$ and of rank-two forms whose bilinear part is alternating \citesupp{S-Brown1972,S-Wall1963}; the Gauss sum of a rank-one form is $1\pm i=\sqrt2\,\zeta_{8}^{\pm1}$ and that of a rank-two alternating form is $\pm2=2\zeta_{8}^{0\text{ or }4}$. The Gauss sum of the orthogonal sum is the product, so $|F_{i}|=2^{K-r'}\cdot2^{r'/2}=2^{K-r'/2}$ and $\beta\equiv(\text{number of rank-one summands})\equiv r'\pmod2$. Since $(1-i)=\sqrt2\,\zeta_{8}^{-1}$, $\mathrm{Re}((1-i)F_{i})=\sqrt2\cos(\pi(\beta-1)/4)|F_{i}|$, which is $\pm|F_{i}|$ for $\beta$ even and $0$ or $\pm\sqrt2|F_{i}|$ for $\beta$ odd. \qed

\begin{lemma}[The rank-$2$ family in closed form]\label{S:lem:rank2}
Let $F=K_{\alpha,\beta,\gamma}$ plus $\delta\ge1$ isolated vertices ($\alpha\ge\beta\ge\gamma\ge0$, $\beta\ge1$, $\alpha+\beta+\gamma+\delta=K$), i.e.\ $H$ a disjoint union of at most three cliques completed by $\delta$ universal vertices, and set $u=i^{-\alpha}$, $v=i^{-\beta}$, $w=i^{-\gamma}$, $\sigma_{K}=(-1)^{(K-3)/4}$. Then $F_{1}=2^{K-1}$ if $\gamma=0$ and $F_{1}=0$ if $\gamma\ge1$; $F_{i}=2^{K/2-1}\zeta_{8}^{K}\,(u+v+w-uvw)$; and
\begin{gather*}
a(H)=-\tfrac12-\sigma_{K}\,\mathrm{Im}(S)\,2^{-(K+1)/2}\quad(\gamma=0,\ S=1+u+v-uv),\\
a(H)=-\sigma_{K}\,\mathrm{Im}(S)\,2^{-(K+1)/2}\quad(\gamma\ge1,\ S=u+v+w-uvw).
\end{gather*}
Since $\mathrm{Im}(S)\in\{0,\pm2\}$ in the bipartite case and $|\mathrm{Im}(S)|\le4$ in the tripartite case, $|a(H)|\le\tfrac12+2^{-(K-1)/2}$ throughout the family, with equality iff $\gamma=0$ and $\sigma_{K}\mathrm{Im}(S)=2$: at $(\alpha,\beta)\equiv(1,1)\pmod4$ for $K\equiv7\pmod8$ (e.g.\ $K_{K}-e$) and at $(\alpha,\beta)\equiv(3,2)\pmod4$ for $K\equiv3\pmod8$. In particular $a(K_{K}-e)=-\tfrac12+(-1)^{(K-3)/4}2^{-(K-1)/2}$.
\end{lemma}

\begin{proof}
$Q_{F}(x)=ab+ac+bc\bmod2$, with $(a,b,c)$ the numbers of chosen vertices in the three parts. Summing $(-1)^{a(b+c)}$ over $a$ kills every term with $b+c$ odd and doubles the rest, which gives $F_{1}$ (positive for $\gamma=0$; the even--even and odd--odd sectors cancel for $\gamma\ge1$). For $F_{i}$, split each part sum by parity, $\sum_{a}\binom{\alpha}{a}i^{a}(-1)^{a(b+c)}=(1+i(-1)^{b+c})^{\alpha}$, and collect the four parity sectors of $(b,c)$:
\begin{equation*}
F_{i}=\tfrac12\big[(1+i)^{K-\gamma}(1-i)^{\gamma}+(1+i)^{K-\beta}(1-i)^{\beta}+(1+i)^{K-\alpha}(1-i)^{\alpha}-(1+i)^{\delta}(1-i)^{K-\delta}\big],
\end{equation*}
and $(1+i)^{p}(1-i)^{q}=2^{(p+q)/2}\zeta_{8}^{p-q}$ turns this into the stated four-term form. Finally $\zeta_{8}^{K-1}=i^{(K-1)/2}=\sigma_{K}\,i$ for $K\equiv3\pmod4$, so $\mathrm{Re}((1-i)F_{i})=2^{(K-1)/2}\,\mathrm{Re}(\zeta_{8}^{K-1}S')=-\sigma_{K}2^{(K-1)/2}\,\mathrm{Im}(S')$ with $S'=u+v+w-uvw$ ($=S$ when $\gamma=0$, where $w=1$), and $a=-2^{-K}[F_{1}-\mathrm{Re}((1-i)F_{i})]$ gives the two displays. Over fourth roots $u,v$ the imaginary part of $1+u+v-uv$ takes only the values $0,\pm2$ (sixteen cases; the constant $1$ blocks $\pm4$), and $|S'|\le4$ always.
\end{proof}

\begin{lemma}[The class $r'=3$ in closed form]\label{S:lem:threeclique}
If $r'=\mathrm{rank}(I+A_{F})=3$ and $H$ has a universal vertex, then $F=K_{n_{1}}\sqcup K_{n_{2}}\sqcup K_{n_{3}}$ with $n_{1}+n_{2}+n_{3}=K$ and at least one $n_{j}=1$, i.e.\ $H$ is complete tripartite. Then $F_{i}=2^{K-2}(-1+i)$, so $\mathrm{Re}((1-i)F_{i})=0$ identically, and
\begin{equation*}
a(H)=-2^{-K}\,T(n_{1})\,T(n_{2})\,T(n_{3}),\qquad T(m)=2^{(m+1)/2}\cos\!\big(\pi(1-m)/4\big),
\end{equation*}
which vanishes iff some $n_{j}\equiv3\pmod4$ and otherwise satisfies $|a(H)|\le2^{-(K-3)/2}\le\tfrac14$.
\end{lemma}

\begin{proof}
$I+A_{F}$ is symmetric with unit diagonal; a rank-$3$ such matrix over $\mathrm{GF}(2)$ is congruent to $I_{3}\oplus0$, hence equals $WW^{\top}$ with rows $w_{i}\in\mathrm{GF}(2)^{3}$ of odd weight, so every vertex carries one of the four types $e_{1},e_{2},e_{3},(1,1,1)$ and $u\sim_{F}v$ iff $w_{u}\cdot w_{v}=1$: vertices of the same $e_{j}$-type are $F$-adjacent, different $e_{j}$-types are not, and the type $(1,1,1)$ is $F$-adjacent to everything. A universal vertex of $H$ is $F$-isolated, which excludes the type $(1,1,1)$ and makes the universal vertex's own type class a singleton; all three $e_{j}$-types must occur for the rank, so $F$ is three disjoint cliques, one of them a singleton. Then $e_{F}(x)=\sum_{j}\binom{a_{j}}{2}$ factorizes both Gauss sums: $F_{i}=\prod_{j}\sum_{a}\binom{n_{j}}{a}i^{a^{2}}=\prod_{j}2^{n_{j}-1}(1+i)=2^{K-3}(1+i)^{3}=2^{K-2}(-1+i)$, and $(1-i)(-1+i)=2i$ is purely imaginary; $F_{1}=\prod_{j}T(n_{j})$ with $T(m)=\sum_{a}\binom{m}{a}(-1)^{\binom a2}=\sqrt2\,\mathrm{Re}\big(\zeta_{8}(1-i)^{m}\big)=2^{(m+1)/2}\cos(\pi(1-m)/4)$, and $|T(m)|\le2^{(m+1)/2}$ gives the bound.
\end{proof}

Together with the star ($r'=2$) and the parity bound ($r,r'\ge4$), Lemmas~\ref{S:lem:rank2} and~\ref{S:lem:threeclique} prove main-text Theorem~8 in its tight form. Tables~\ref{S:tab:rankclasses} and~\ref{S:tab:rank2} are retained as verification: every class maximum equals the value the lemmas predict, and the rank-$2$ table reproduces Lemma~\ref{S:lem:rank2} in exact rational arithmetic for every $K\le31$ (script \texttt{25} checks the closed forms of both lemmas against direct summation for every composition at $K=7,11,15,19$).

\paragraph{The rank classes and the evidence.} The sharpened bound $|a(H)|\le2^{-r/2}+2^{-\lfloor r'/2\rfloor}$ was verified on all $32\,767$ graphs at $K=7$ and on $22\,945$ random graphs at $K=11$ and $15$ (the parity factor is $\pm1$ exactly when $r'$ is even; script \texttt{23}); the magnitude statements, the rank bound, the closed forms of Lemmas~\ref{S:lem:rank2}--\ref{S:lem:threeclique} and the class-by-class maxima were checked with zero violations on all $33\,864$ augmented graphs at $K=7$, on $12\,000$ random graphs at $K=11$ and on $2\,400$ at $K=15$ (scripts \texttt{23} and \texttt{25}). Script \texttt{23} computes $r$, $r'$, $F_{1}$, $F_{i}$ and the factor $c(F)=\mathrm{Re}((1-i)F_{i})/|F_{i}|$ for all $32\,767$ admissible graphs at $K=7$ and for $19\,942$ and $3\,003$ random admissible graphs at $K=11$ and $15$ (edge density uniform in $[0.05,0.95]$, plus the structured families); every graph satisfies $c(F)=\pm1$ exactly when $r'$ is even and $c(F)\in\{0,\pm\sqrt2\}$ when $r'$ is odd, and the parity-refined bound. Table~\ref{S:tab:rankclasses} lists, for $K=7$, the maximum of $|a|$ and the values of $c(F)$ in each rank class. The two classes that the rank bound alone does not settle, $(r,r')=(2,K-2)$ and $(K-3,3)$, both have maximum $1/4$ at $K=7$, exactly the cancellation now proved: in $(K-3,3)$ the second Gauss sum drops out identically (Lemma~\ref{S:lem:threeclique}), and in $(2,K-2)$ the bipartite four-term sum caps $|\mathrm{Im}(S)|$ at $2$ (Lemma~\ref{S:lem:rank2}). Table~\ref{S:tab:rank2} gives the exact maximum of $|a|$ over the whole rank-$2$ family ($F$ complete multipartite with at most three parts plus isolated vertices, i.e.\ $H$ a disjoint union of at most three cliques plus universal vertices) for $K=7,\dots,31$, computed in rational arithmetic from $a(H)=-2^{1-K}\sum_{\hw\equiv2,3}(-1)^{ab+ac+bc}$ with $(a,b,c)$ the numbers of chosen vertices in the parts; it equals $\tfrac12+2^{-(K-1)/2}$ for every $K$, attained exactly at the residues of Lemma~\ref{S:lem:rank2}, and $a(K_{K}-e)=-\tfrac12+(-1)^{(K-3)/4}2^{-(K-1)/2}$.

\begin{table}[!htbp]
\centering\small
\caption{Numerical verification of the Gauss-sum magnitudes and of the rank bound (script \texttt{08}). For $K=7$ all graphs $R$ on $k$ real vertices were enumerated; for $K=11,15$ random graphs with edge density uniform in $(0.05,0.95)$.}
\label{S:tab:gapcheck}
\resizebox{\textwidth}{!}{\begin{tabular}{@{}ccccccl@{}}
\toprule
$k$ & $q$ & $K$ & graphs & violations of $|F_{1}|,|F_{i}|$ formulas & violations of the bound & max $|a|$ (non-clique), attained at\\
\midrule
3 & 4 & 7 & 8 (all) & 0 & 0 & 0.625, $R$ = one edge ($r=2$, $r'=7$)\\
4 & 3 & 7 & 64 (all) & 0 & 0 & 0.625 ($r=2$, $r'=7$)\\
5 & 2 & 7 & 1\,024 (all) & 0 & 0 & 0.625 ($r=2$, $r'=7$)\\
6 & 1 & 7 & 32\,768 (all) & 0 & 0 & 0.625, $R$ empty (star; $r=6$, $r'=2$)\\
7 & 4 & 11 & 2\,097\,152 (all) & -- & -- & 0.53125\\
7--10 & 4--1 & 11 & 3\,000 random each & 0 & 0 & 0.53125 ($q=4$), 0.500\\
11--14 & 4--1 & 15 & 600 random each & 0 & 0 & 0.5078\\
\bottomrule
\end{tabular}}
\end{table}

\begin{table}[!htbp]
\centering\small
\caption{Exact $a(H)$ for structured families (vertex $K-1$ universal), with $r$, $r'$ and the rank bound in brackets for $K=7$.}
\label{S:tab:families}
\resizebox{\textwidth}{!}{\begin{tabular}{@{}lccccc@{}}
\toprule
family & $K=7$ & $K=11$ & $K=15$ & $K=19$ & $K=23$\\
\midrule
$K_{K}-e$ & $-0.625$ (2, 6; 0.677) & $-0.469$ & $-0.508$ & $-0.498$ & $-0.500$\\
$K_{K}-$2 disjoint edges & $-0.500$ (4, 5; 0.500) & $-0.188$ & $-0.266$ & $-0.246$ & $-0.251$\\
$K_{K}-$3 disjoint edges & $-0.375$ (6, 4; 0.479) & $-0.063$ & $-0.141$ & $-0.121$ & $-0.126$\\
$K_{K}-P_{3}$ & $-0.625$ (2, 7; 0.625) & $-0.469$ & $-0.508$ & $-0.498$ & $-0.500$\\
$K_{K}-$triangle & $-0.250$ (2, 5; 0.750) & $+0.063$ & $-0.016$ & $+0.004$ & $-0.001$\\
$K_{K}-K_{4}$ & $0.000$ (4, 4; 0.604; parity-refined 0.500) & $+0.313$ & $+0.234$ & $+0.254$ & $+0.249$\\
$K_{1}+(K_{a}\cup K_{b})$ & $-0.375$ (2, 6; 0.677) & $-0.469$ & $-0.492$ & $-0.498$ & $-0.500$\\
star $K_{1,K-1}$ & $+0.625$ (6, 2; 0.832) & $+0.469$ & $+0.508$ & $+0.498$ & $+0.500$\\
$K_{1}+$perfect matching & $0.000$ (2, 7; 0.625) & $+0.250$ & $0.000$ & $-0.063$ & $0.000$\\
\bottomrule
\end{tabular}}
\end{table}

\begin{table}[!htbp]
\centering\small
\caption{Distribution of $|a(H)|$ over all non-clique graphs on $k$ real vertices plus $q$ universal dummies (exhaustive). $\lambda=1-2a^{2}$ is the phase applied by $\Gam$; the leaked probability is $4a^{2}(1-a^{2})$.}
\label{S:tab:adist}
\begin{tabular}{@{}cccccccccc@{}}
\toprule
$k$ & $q$ & $K$ & graphs & $a$(clique) & max $|a|$ & mean $|a|$ & frac$(a=0)$ & min $\lambda$ & mean leaked\\
\midrule
3 & 4 & 7 & 8 & $-1$ & 0.625 & 0.571 & 0.000 & $+0.219$ & 0.850\\
4 & 3 & 7 & 64 & $-1$ & 0.625 & 0.429 & 0.016 & $+0.219$ & 0.596\\
5 & 2 & 7 & 1\,024 & $-1$ & 0.625 & 0.255 & 0.137 & $+0.219$ & 0.294\\
6 & 1 & 7 & 32\,768 & $-1$ & 0.625 & 0.134 & 0.288 & $+0.219$ & 0.111\\
7 & 4 & 11 & 2\,097\,152 & $-1$ & 0.531 & 0.091 & 0.202 & $+0.436$ & 0.053\\
\bottomrule
\end{tabular}
\end{table}

\begin{figure}[!htbp]
\centering
\includegraphics[width=\textwidth]{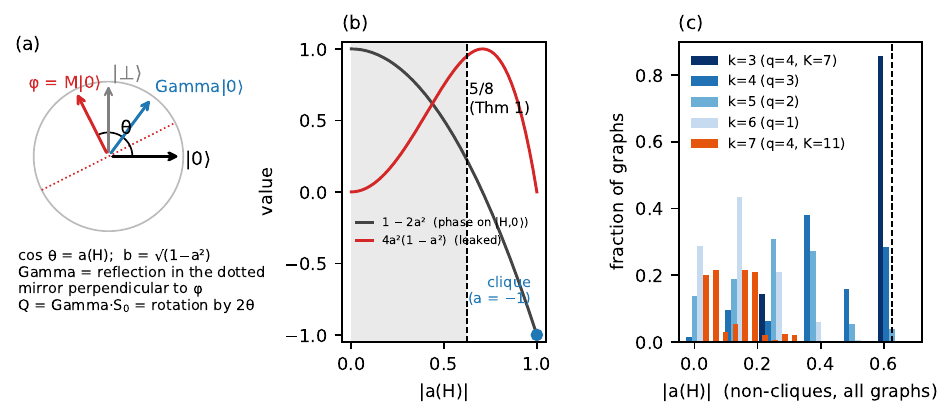}
\caption{\textbf{Exact action of $\Gam$ on a queried vertex set.} \textbf{a}, Geometry on the invariant plane spanned by the ancilla reference state $\ket{0}$ and $\varphi=M\ket{0}=a\ket{0}+b\ket{\perp}$, $\cos\theta=a(H)$: $\Gam$ reflects in the mirror perpendicular to $\varphi$, so $\Gam\ket{0}=(1-2a^{2})\ket{0}-2ab\ket{\perp}$, and $Q=\Gam S_{0}$ rotates by $2\theta$. \textbf{b}, The phase $1-2a^{2}$ applied to $\ket{H,0}$ and the probability $4a^{2}(1-a^{2})$ transferred to $\ket{\perp}$ as functions of $|a(H)|$; Theorem~8 confines non-cliques to the shaded region, while a clique has $a=-1$. \textbf{c}, Distribution of $|a(H)|$ over all non-clique graphs on $k$ real vertices plus $q$ universal dummies (exhaustive: $8$, $64$, $1\,024$, $32\,768$ and $2\,097\,152$ graphs for $k=3,\dots,7$); all values are multiples of $2^{-(K-1)}$ and none reaches the dashed line at $5/8$.}
\label{S:fig:theory}
\end{figure}

\begin{figure}[!htbp]
\centering
\includegraphics[width=0.62\textwidth]{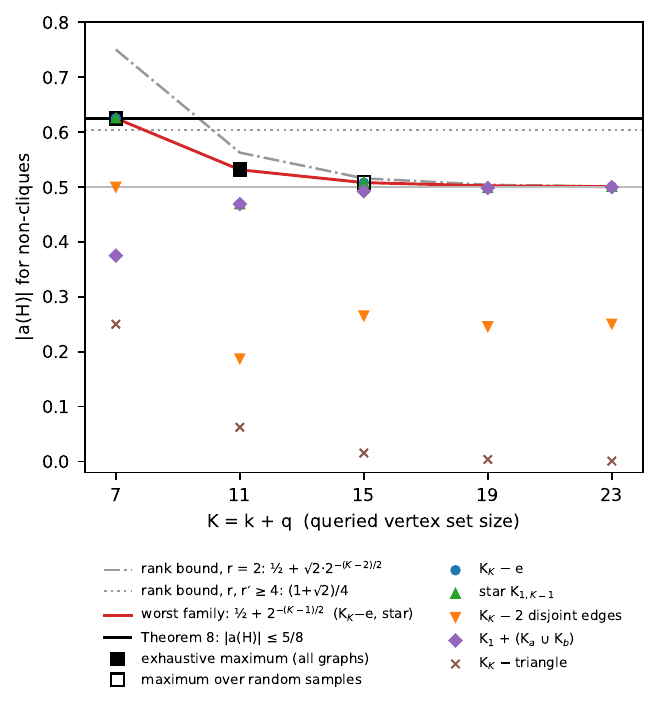}
\caption{\textbf{The spectral gap (Theorem~8).} Largest $|a(H)|$ over non-cliques as a function of $K=k+q$: exhaustive maxima at $K=7$ and $K=11$ (filled squares), the maximum over random graphs at $K=15$ (open square), exact values for structured families, the analytic worst family $\tfrac12+2^{-(K-1)/2}$, the rank bound (main-text equation~(5)) at $r=2$ with $r'\ge K-2$, and its value $(1+\sqrt2)/4$ when both ranks are at least $4$; Theorem~8 in its tight form gives $\tfrac12+2^{-(K-1)/2}$ for every $K$ (the curve labeled the analytic worst family) and $\tfrac12$ for the class $r,r'\ge4$.}
\label{S:fig:gap}
\end{figure}

\paragraph{Distribution of the overlap and the plain oracle.} The proof above combines the rank bound (main-text equation~(5)) with the observation that the two low-rank cases are highly structured: $r'=2$ forces $H$ to be a complete bipartite graph, hence a star, whose Gauss sums evaluate in closed form to $|a|=\tfrac12\pm2^{-(K-1)/2}$; $r=2$ forces $r'\ge K-2$; and $K=7$ is settled by enumerating all $2^{15}$ graphs on six vertices plus one universal vertex. Fig.~\ref{S:fig:gap} summarizes the bound, the exhaustive maxima ($5/8$ at $K=7$ over $33\,864$ augmented graphs, $17/32$ at $K=11$ over $2\,097\,152$ graphs) and the structured families; the distribution of $|a|$ over all non-cliques (Fig.~\ref{S:fig:theory}c) shows that the typical value is far smaller than the worst case (mean $0.09$ at $K=11$), and that the dominant error source of the plain oracle is the number of dummies: at fixed $K=7$ the mean leaked probability $4a^{2}(1-a^{2})$ falls from $0.85$ for $k=3$ to $0.11$ for $k=6$.

Main-text Theorem~8 and Fig.~\ref{S:fig:theory}c delimit the regime in which $\Gam$ can be used on its own. Cliques are always marked exactly, so the algorithm never mis-marks in the sense of main-text Corollary~9; the only failure mode is the erosion of the amplitude of \emph{non}-cliques into the ancilla space, which spoils the interference on which amplitude amplification relies. The erosion is largest for \emph{near}-cliques: $K_{K}$ minus one edge has $|a|=5/8$, receives the phase $1-2a^{2}=0.22$ instead of $+1$ and leaks $95\%$ of its amplitude per query, so the hardest instances for $\Gam$ are those in which many candidate sets miss a clique by one or two edges, i.e.\ dense host graphs. One might expect from the mean leakage that the case $q=1$ ($k\equiv2\bmod4$) makes $\Gam$ nearly exact. The exact trajectories do not bear this out: at $q=1$ the success probability at the ideal iteration count is $0.16$ and the best achievable $0.41$, no better than at $q=4$ (Supplementary Table~S7). The reason is that the relevant quantity is not the mean of $|a|$ over all graphs but its value on the non-cliques that carry most of the amplitude in a real instance, and near-cliques with $|a|\approx1/2$ exist for every $q$. In practice, therefore, the plain oracle is inadequate for all $k$, and the filter of the next section is a necessity rather than a refinement.

\paragraph{The proof in brief.} The proof (Supplementary Note~2) combines the rank bound main-text equation~(5) with the parity of the Brown invariant of the $\mathbb Z_{4}$-valued form \citesupp{S-Brown1972,S-Wall1963}, which sharpens it to $|a(H)|\le2^{-r/2}+2^{-\lfloor r'/2\rfloor}$ and settles every class with $r,r'\ge4$ at $1/2$, and with closed-form evaluations of the three remaining low-rank classes, which are rigid. $r'=2$ forces $H$ to be a star, with $|a|=\tfrac12\pm2^{-(K-1)/2}$. $r'=3$ forces the complement $F$ to be three disjoint cliques, where $\mathrm{Re}((1-i)F_{i})$ vanishes identically and $|a|\le2^{-(K-3)/2}\le\tfrac14$ (Lemma~S4). And $r=2$ makes $F$ complete multipartite with at most three parts plus isolated vertices, where both Gauss sums evaluate in closed form ($F_{i}=2^{K/2-1}\zeta_{8}^{K}(u+v+w-uvw)$ with $u,v,w$ fourth roots of unity set by the part sizes mod $4$), giving $|a|\le\tfrac12+2^{-(K-1)/2}$ with the attainment pattern of the theorem, and $a(K_{K}-e)=-\tfrac12+(-1)^{(K-3)/4}2^{-(K-1)/2}$ exactly (Lemma~S3). No enumeration enters the proof; the enumerations serve as verification: the maxima are $5/8$ at $K=7$ (all $33\,864$ augmented graphs) and $17/32$ at $K=11$ (all $2\,097\,152$), both equal to the bound, a random sample attains $\tfrac12+2^{-7}$ at $K=15$, and the closed forms reproduce every rank class (script \texttt{25}). In every structured family and every random sample the typical value is far smaller than the worst case (mean $|a|=0.09$ at $K=11$; Supplementary Figs.~S1c and S2). Cliques are always marked exactly, so by main-text Corollary~9 the oracle never mis-marks; its only failure mode is the erosion of \emph{non}-clique amplitude into the ancilla space, largest for near-cliques ($K_{K}$ minus one edge has $|a|=5/8$ at $K=7$ and leaks $95\%$ of its amplitude per query), and since induced subgraphs of real networks are full of near-cliques for every $q$, the plain oracle is a poor amplitude-amplification oracle for every $k$ (Supplementary Note~6). Nothing downstream tightens or loosens with the exact constant: the filter error $2/27$ of the next section is attained at $|a|=1/\sqrt6\approx0.408$, in the interior of the gap for every $K$. The filter is a necessity rather than a refinement.

\FloatBarrier
\section*{Supplementary Note 3: the phase-estimation filter}\label{S:note:3}

\begin{figure}[!htbp]
\centering\includegraphics[width=0.95\textwidth]{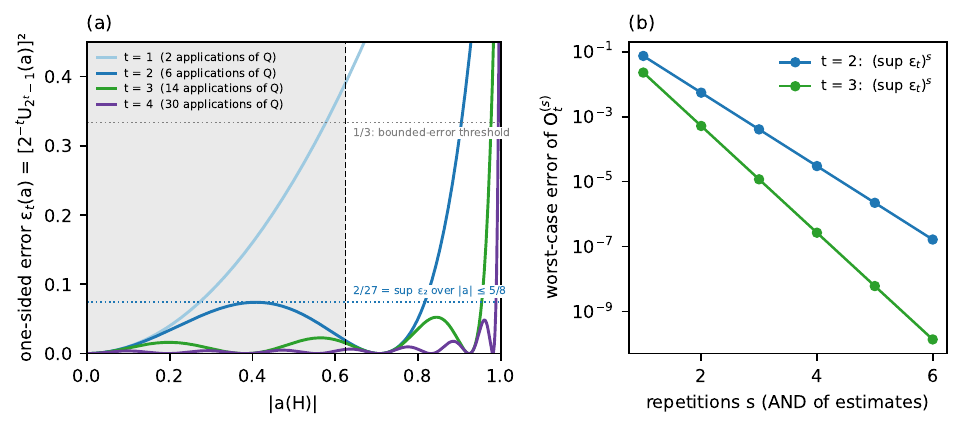}
\caption{\textbf{One-sided error of the phase-estimation filter.} \textbf{a}, $\varepsilon_{t}(a)=[2^{-t}U_{2^{t}-1}(a)]^{2}$ for $t=1,\dots,4$, the probability that a $t$-bit estimate of the eigenphase of $Q$ returns $0$ for a non-clique with overlap $a$; the number of applications of $Q$ per oracle call is $2(2^{t}-1)$. On $|a|\le5/8$ (shaded) the suprema are $0.39$, $2/27$, $0.023$ and $0.006$; $t=2$ already lies below the $1/3$ threshold of the bounded-error search model. \textbf{b}, Worst-case error of $s$ AND-combined repetitions, $(\sup\varepsilon_{t})^{s}$.}
\label{S:fig:eps}
\end{figure}

\paragraph{Derivation of the error formula.} Write $\mathrm{QPE}_{t}=(\mathrm{QFT}_{L}^{\dagger}\otimes I)\,\mathrm{c}Q\,(H^{\otimes t}\otimes I)$ with $\mathrm{c}Q=\sum_{j<L}\ket j\bra j\otimes Q^{j}$. Since the $\est=0$ row of $\mathrm{QFT}_{L}^{\dagger}$ is uniform, $\bra{0_{\est}}\mathrm{QPE}_{t}\ket{0_{\est}}=L^{-1}\sum_{j<L}Q^{j}=A_{t}$, and
\begin{equation}
\mathcal O_{t}\ket{0_{\est},0}=\ket{0_{\est},0}-2\,\mathrm{QPE}_{t}^{\dagger}\big(\ket{0_{\est}}\otimes A_{t}\ket0\big),
\end{equation}
whose $\ket{0_{\est},0}$ component is $1-2\|A_{t}\ket0\|^{2}$; the remaining component is orthogonal and has squared norm $4\varepsilon(1-\varepsilon)$ with $\varepsilon=\|A_{t}\ket0\|^{2}$. On the invariant plane, $\ket0=(v_{+}+v_{-})/\sqrt2$ with $Qv_{\pm}=e^{\pm2i\theta}v_{\pm}$, and $\sum_{j<L}e^{\pm2ij\theta}=e^{\pm i(L-1)\theta}\sin(L\theta)/\sin\theta$, so $\varepsilon_{t}(a)=[\sin(L\theta)/(L\sin\theta)]^{2}=[U_{L-1}(a)/L]^{2}$ with $\cos\theta=a$. For $t=2$, $\varepsilon_{2}(a)=a^{2}(2a^{2}-1)^{2}$, whose maximum on $|a|\le5/8$ is $2/27$ at $a^{2}=1/6$. For general $t$ and $|a|\le5/8$, $\sin^{2}\theta\ge39/64$ gives $\varepsilon_{t}\le64/(39\cdot4^{t})$. Table~\ref{S:tab:eps} lists exact values; Fig.~\ref{S:fig:eps} plots them.

\paragraph{Repetition.} With $s$ independent $\est$ registers, applying $\mathrm{QPE}_{t}$ $s$ times on the same data register and flipping the phase iff all estimates are $0$ gives the all-zero component $A_{t}^{s}\ket0$. $A_{t}$ is a polynomial in the unitary $Q$, hence normal, with eigenvalues of modulus $\sqrt\varepsilon$ on $v_{\pm}$, so $\|A_{t}^{s}\ket0\|^{2}=\varepsilon^{s}$; cliques ($A_{t}=I$) still receive the exact phase $-1$.

\paragraph{Cost.} $\mathrm{QPE}_{t}$ needs $\sum_{m<t}2^{m}=L-1$ controlled applications of $Q$; with un-computation, $\mathcal O_{t}$ needs $2(L-1)$ controlled $Q$'s, i.e.\ $4(L-1)$ applications of $M=\IP^{\dagger}A\,\IP$ and $4(L-1)$ controlled $S_{0}$'s (each an $(N+1)$-controlled Z). Since $M^{2}=I$, $\mathrm{c}Q=M(\mathrm cS_{0})M(\mathrm cS_{0})$, so neither $A$ nor $\IP$ is ever controlled and the edge-coloring depth of $A$ is preserved. For $t=2$, $\mathcal O_{2}$ costs six applications of $Q$, i.e.\ about seven times the depth, gate count and non-Clifford count of $\Gam$ once the controlled reflections and the estimation register are compiled (Table~\ref{S:tab:nonclifford}), plus two qubits.

\paragraph{Hybrid argument (Theorem 11(i)).} Let $O^{*}$ be the ideal phase oracle and $D$ the diffuser. Because $D$ acts on $\idx$ alone, $(DO^{*})^{j}\ket{\sigma,0}$ is of the form $\sum_{y}c_{y}\ket{y,0}$, and $(\mathcal O_{t}^{(s)}-O^{*})$ maps $\ket{y,0}$ to $0$ for cliques and to $-2\varepsilon_{y}\ket{y,0}+\ket{g_{y}}$, of norm $2\sqrt{\varepsilon_{y}}$, otherwise; these images lie in orthogonal $\idx$-sectors, so $\|(\mathcal O_{t}^{(s)}-O^{*})\sum_{y}c_{y}\ket{y,0}\|\le2\sqrt{\varepsilon}$ with $\varepsilon=\sup\varepsilon_{y}$. Summing over the $r$ hybrids, $\|(D\mathcal O_{t}^{(s)})^{r}\ket{\sigma,0}-(DO^{*})^{r}\ket{\sigma,0}\|\le2r\sqrt\varepsilon$, and $|\,\|\Pi\psi\|^{2}-\|\Pi\psi^{*}\|^{2}|\le2\|\psi-\psi^{*}\|$ for any projector $\Pi$. With $\varepsilon=(2/27)^{s}$ and $r\approx\tfrac\pi4\sqrt{\mathcal N/\mathcal M}$, the choice $s=\lceil2\log(4r/\delta)/\log(27/2)\rceil$ makes the loss at most $\delta$, at $6rs=O(\sqrt{\mathcal N/\mathcal M}\log(\mathcal N/\mathcal M))$ applications of $Q$.

\begin{table}[!htbp]
\centering\small
\caption{One-sided error $\varepsilon_{t}(a)$ of the $t$-bit filter and its supremum over $|a|\le5/8$ (script \texttt{08}, part C).}
\label{S:tab:eps}
\begin{tabular}{@{}lcccc@{}}
\toprule
$|a|$ & $t=1$ & $t=2$ & $t=3$ & $t=4$\\
\midrule
0.125 & 0.0156 & 0.0147 & 0.0113 & 0.0033\\
0.250 & 0.0625 & 0.0479 & 0.0135 & 0.0026\\
0.375 & 0.1406 & 0.0726 & 0.0001 & 0.0001\\
0.408 & 0.1665 & 0.0741 & 0.0009 & 0.0009\\
0.500 & 0.2500 & 0.0625 & 0.0156 & 0.0039\\
0.625 & 0.3906 & 0.0187 & 0.0153 & 0.0062\\
\midrule
$\sup_{|a|\le5/8}$ & 0.391 & $2/27=0.0741$ & 0.0228 & 0.0062\\
applications of $Q$ in $\mathcal O_{t}$ & 2 & 6 & 14 & 30\\
\bottomrule
\end{tabular}
\end{table}

\paragraph{Circuit-level checks.} On the $n=7$, $k=3$ instance of main-text Fig.~3 (35 queries), the diagonal element $\bra{y,0,0}\mathcal O_{t}\ket{y,0,0}$ computed on the simulated circuit agrees with $1-2\varepsilon_{t}(a_{y})$ to $2.7\times10^{-15}$, $1.6\times10^{-14}$ and $7.7\times10^{-14}$ for $t=1,2,3$; for the clique $\varepsilon=1$ exactly.

\FloatBarrier
\section*{Supplementary Note 4: the exact parallel counting oracle}\label{S:note:4}

\paragraph{Construction.} A vertex set $H=\supp(y)$ of weight $k$ is a clique iff every $v\in H$ has exactly $k-1$ neighbors in $H$. The oracle of main-text Proposition~6 (Fig.~\ref{S:fig:counting}a) computes all these degrees at once. Registers: $\idx$ ($n$ qubits, one-hot candidate set); one counter $\mathtt{cnt}_{v}$ of $b=\lceil\log_{2}k\rceil$ qubits, one comparison bit $e_{v}$ and one flag $f_{v}$ per vertex; and $n-1$ tree ancillas $t$, in total $n(4+b)-1$ qubits. The edges of $G$ are partitioned by a proper edge coloring (Misra--Gries, $\chi'(G)\le\Delta+1$; the accompanying code uses the round-robin 1-factorization on $K_{n}$ and a greedy coloring with at most $2\Delta-1$ classes otherwise). For every color class, and for every edge $uv$ in it, the two controlled increments $\mathtt{cnt}_{v}\mathrel{+}=\idx_{u}$ and $\mathtt{cnt}_{u}\mathrel{+}=\idx_{v}$ are applied; they act on the disjoint qubit sets $\{\idx_{u}\}\cup\mathtt{cnt}_{v}$ and $\{\idx_{v}\}\cup\mathtt{cnt}_{u}$, and two edges of the same class share no vertex, so a whole class is one layer of controlled increments. A controlled increment of a $b$-bit register is the sequence of multi-controlled X gates on bits $b-1,\dots,0$ with controls $\idx_{u}$ and all lower bits (top bit first); for $b=3$ the top gate has three controls and is realized as three Toffolis with the idle flag $f_{v}$ as a clean ancilla, so that the gates of different vertices never share a borrowed qubit and a round remains one layer. After the last class, $e_{v}\leftarrow[\mathtt{cnt}_{v}=k-1]$ (a $b$-controlled X with the appropriate inverted controls) and $f_{v}\leftarrow\idx_{v}\wedge\neg e_{v}$ (a CNOT and a Toffoli). The inverted flags are combined by a binary AND tree of Toffolis on $t$ ($\lceil\log_{2}n\rceil$ layers), a $Z$ on the root applies $-1$ iff all flags are $0$, and the tree, the flags and the counting layers are un-computed in reverse order. A narrower variant replaces the tree by an $n$-controlled $Z$ on the flags synthesized with the idle counters as dirty ancillas \citesupp{S-Iten2016}: $n(3+b)$ qubits, depth $O(n)$ for the final test instead of $O(\log n)$ (Table~\ref{S:tab:nonclifford}, rows ``$n$-controlled $Z$ variant'').

\paragraph{Correctness and cost.} Increments commute, so $\mathtt{cnt}_{v}\equiv|N(v)\cap H|\pmod{2^{b}}$ after all classes. For $v\in H$ the count is at most $k-1<2^{b}$, so $e_{v}=1$ iff $v$ is adjacent to all of $H\setminus\{v\}$; for $v\notin H$ the flag is masked by $\idx_{v}$, and a counter that wrapped around for such a $v$ is un-wrapped exactly by the reverse sequence, because every step is a permutation of the computational basis. Hence the phase is $-1$ exactly on $k$-cliques and all ancillas are restored. The oracle needs $2m$ controlled increments ($\Theta(m\log k)$ Toffoli-class gates) in $\chi'(G)$ layers of constant Toffoli depth for $b\le3$, $O(n)$ Toffolis for the comparison bits, flags and tree, and depth $O(\chi'(G)\,\mathrm{poly}\log k+\log n)$. Both variants were verified by statevector simulation for $n=4$ and $k\in\{3,4\}$ on $K_{4}$ (four triangles, one 4-clique), $K_{4}$ minus an edge, a 4-cycle with a chord and two random graphs (script \texttt{12}, 20 cases): on every weight-$k$ basis input the output amplitude is $\pm1$ on the input state with the correct sign and no leakage, to $3\times10^{-13}$.

\paragraph{Compiled cost.} Table~\ref{S:tab:nonclifford} lists the transpiled resources of the oracle for $k=3$--$6$ on $K_{n}$, alongside the sequential counter of \citesupp{S-Metwalli2020}, the graph-state oracle $\Gam$ and the explicitly built $\mathcal O_{2}$; Table~\ref{S:tab:sparse} gives the depth on random graphs $G(n,p)$ for the two schedules (Fig.~\ref{S:fig:counting}c). On $K_{n}$ the depth for $k=3$ is $216$, $258$, $319$, $359$, $441$, $542$ for $n=6,8,10,12,16,20$ ($\approx23n+76$), and the number of non-Clifford gates is exactly $14n^{2}+28n-14$ (measured up to $n=40$); on sparse graphs the depth is $\approx21$ layers per color class plus $\approx150$ for flags and tree at $n=20$, so that a graph with $\Delta=6$ ($p=0.15$) costs $294$ layers against $542$ on $K_{20}$.

\paragraph{Relation to previous constructions.} The counting oracle of \citesupp{S-Metwalli2020} increments a single $\lceil\log_{2}\binom k2\rceil$-bit counter once per edge and compares it with $\binom k2$; its depth is $\Theta(m\log k)$ because all increments share the counter. Per-vertex counters have been used in Grover oracles for domination problems \citesupp{S-Jiang2023}; the edge-colored schedule above is what makes all $n$ counters advance simultaneously. An alternative exact oracle tests the non-edges of $G$ directly (a vertex set is a clique iff no non-edge lies inside it): one flag per color class of the complement graph, each computed by a logarithmic-depth OR tree over the at most $\lfloor n/2\rfloor$ pair indicators of the class, gives depth $\Theta(n\log n)$ at width $\Theta(n)$, or depth $\Theta(n)$ at width $\Theta(n\log n)$ if the OR trees of successive classes are pipelined; it uses one Toffoli per non-edge and is dominated by the counting oracle on dense graphs.

\paragraph{Non-Clifford cost.} All constructions of this note, like the sequential counter and the edge detectors of Fig.~\ref{S:fig:primitivesSI}, write the graph into the circuit one Toffoli-class gate per edge (or per non-edge); their non-Clifford count is therefore $\Theta(m)$ on every input graph, whereas the graph-state oracle $\Gam$ of the main text carries the graph in the Clifford unitary $A$ and uses $O(n)$ non-Clifford gates (main-text Proposition~12). The fitted forms for the block-by-block compilation are $89N-1$ (depth of $\Gam$, even $N$), $622N-13$ (depth of $\mathcal O_{2}$), $72N-10$ and $480N-34$ (non-Clifford gates of $\Gam$ and $\mathcal O_{2}$); the log-depth compilation of main-text Proposition~13 (Table~\ref{S:tab:logdepth}) has depth $\approx13.7n+403$ and $\approx91.6n+2\,847$ (fits over $n=6$--$40$) and $T$-counts $78N-14$ and $545N-162$; $\approx23n+76$ (depth of the parallel counter, $k=3$), $\approx34.5n^{2}$ (depth of the sequential counter) and $14n^{2}+28n-14$, $\approx21n^{2}$ (non-Clifford gates of the parallel and sequential counters). For $\Gam$, $\mathcal O_{2}$ and the parallel counter every non-Clifford rotation is an odd multiple of $\pi/4$, so these counts are $T$-counts and already sit in the single normalization of the main text. The sequential counter is the one construction with two accountings, and this note is the home of the mixed-basis one: as compiled it contains $\pi/8$ rotations from Qiskit's ancilla-free synthesis of its three-controlled gates ($\approx20n^{2}$ non-Clifford rotations in total), while the re-synthesis with three Toffolis on a clean ancilla, which the main text uses throughout, gives exactly $28n(n-1)+7=56m+7$ $T$ gates on $K_{n}$ ($1\,575$, $3\,703$, $6\,727$ at $n=8,12,16$; script \texttt{13}). Under the main-text normalization the non-Clifford crossover of $\mathcal O_{2}$ with the sequential counter is at $n=24$ (measured: $15\,098$ against $15\,463$); counting the mixed-basis rotations of the as-compiled circuit instead (a lower bound for the competitor, since each $\pi/8$ costs many $T$ gates; $21n^{2}-19n+5$ measured up to $n=40$) would put it between $n=28$ and $32$. Both are illustrative; the $\Theta(n)$-against-$\Theta(m\log k)$ separation is synthesis-invariant.

\begin{table}[p]
\centering\small
\caption{Measured resources of one oracle call on $K_{n}$ (script \texttt{11}; basis $\{R_{z},\sqrt X,X,\mathrm{CX}\}$, optimization level 1, seed 7, all-to-all coupling). Non-Clifford = $R_{z}$ rotations with angle not a multiple of $\pi/2$ after transpilation; Toffoli-class = gates with at least two controls plus controlled-Hadamards before transpilation. $\mathcal O_{2}$ was built explicitly ($t=2$: three controlled $Q$ and three controlled $Q^{-1}$, each $M\,\mathrm{c}S_{0}\,M\,\mathrm{c}S_{0}$). For $k=3$ and $k=4$ the parallel counting oracle is the same circuit ($b=2$), and so is it for $k=5$ and $k=6$ ($b=3$).}
\label{S:tab:nonclifford}
\renewcommand{\arraystretch}{0.9}\setlength{\tabcolsep}{4pt}\scriptsize\adjustbox{max width=\textwidth, max totalheight=0.88\textheight}{\begin{tabular}{@{}cclccccccc@{}}
\toprule
$n$ & $k$ & oracle & qubits & depth & CX & gates & non-Clifford & Toffoli-class\\
\midrule
6 & 3 & counting, sequential~\citesupp{S-Metwalli2020} & 9 & 1\,043 & 576 & 1\,448 & 647 & 61\\
6 & 3 & counting, parallel (main-text Prop.~6) & 35 & 216 & 576 & 1\,472 & 658 & 94\\
6 & 3 & counting, parallel, $n$-controlled $Z$ variant & 30 & 239 & 550 & 1\,391 & 626 & 85\\
6 & 3 & $\Gam$ & 22 & 889 & 676 & 1\,863 & 710 & 121\\
6 & 3 & $\mathcal O_{2}$ (explicit) & 24 & 6\,207 & 4\,519 & 12\,565 & 4\,766 & 732\\
8 & 3 & counting, sequential~\citesupp{S-Metwalli2020} & 11 & 1\,932 & 1\,070 & 2\,669 & 1\,197 & 113\\
8 & 3 & counting, parallel (main-text Prop.~6) & 47 & 258 & 964 & 2\,392 & 1\,106 & 158\\
8 & 3 & counting, parallel, $n$-controlled $Z$ variant & 40 & 327 & 930 & 2\,287 & 1\,062 & 145\\
8 & 3 & $\Gam$ & 26 & 1\,067 & 838 & 2\,283 & 854 & 145\\
8 & 3 & $\mathcal O_{2}$ (explicit) & 28 & 7\,451 & 5\,579 & 15\,365 & 5\,726 & 876\\
10 & 3 & counting, sequential~\citesupp{S-Metwalli2020} & 13 & 3\,097 & 1\,716 & 4\,262 & 1\,915 & 181\\
10 & 3 & counting, parallel (main-text Prop.~6) & 59 & 319 & 1\,448 & 3\,520 & 1\,666 & 238\\
10 & 3 & counting, parallel, $n$-controlled $Z$ variant & 50 & 415 & 1\,406 & 3\,391 & 1\,610 & 221\\
10 & 3 & $\Gam$ & 30 & 1\,245 & 1\,008 & 2\,711 & 998 & 169\\
10 & 3 & $\mathcal O_{2}$ (explicit) & 32 & 8\,695 & 6\,687 & 18\,213 & 6\,686 & 1020\\
12 & 3 & counting, sequential~\citesupp{S-Metwalli2020} & 15 & 4\,538 & 2\,514 & 6\,227 & 2\,801 & 265\\
12 & 3 & counting, parallel (main-text Prop.~6) & 71 & 359 & 2\,028 & 4\,856 & 2\,338 & 334\\
12 & 3 & counting, parallel, $n$-controlled $Z$ variant & 60 & 503 & 1\,978 & 4\,703 & 2\,270 & 313\\
12 & 3 & $\Gam$ & 34 & 1\,423 & 1\,186 & 3\,147 & 1\,142 & 193\\
12 & 3 & $\mathcal O_{2}$ (explicit) & 36 & 9\,939 & 7\,843 & 21\,109 & 7\,646 & 1164\\
16 & 3 & counting, sequential~\citesupp{S-Metwalli2020} & 19 & 8\,248 & 4\,566 & 11\,273 & 5\,077 & 481\\
16 & 3 & counting, parallel (main-text Prop.~6) & 95 & 441 & 3\,476 & 8\,152 & 4\,018 & 574\\
16 & 3 & counting, parallel, $n$-controlled $Z$ variant & 80 & 679 & 3\,410 & 7\,951 & 3\,926 & 545\\
16 & 3 & $\Gam$ & 42 & 1\,779 & 1\,566 & 4\,043 & 1\,430 & 241\\
16 & 3 & $\mathcal O_{2}$ (explicit) & 44 & 12\,427 & 10\,299 & 27\,045 & 9\,566 & 1452\\
20 & 3 & counting, sequential~\citesupp{S-Metwalli2020} & 23 & 13\,062 & 7\,226 & 17\,807 & 8\,025 & 761\\
20 & 3 & counting, parallel (main-text Prop.~6) & 119 & 542 & 5\,308 & 12\,280 & 6\,146 & 878\\
20 & 3 & counting, parallel, $n$-controlled $Z$ variant & 100 & 855 & 5\,226 & 12\,031 & 6\,030 & 841\\
20 & 3 & $\Gam$ & 50 & 2\,135 & 1\,978 & 4\,971 & 1\,718 & 289\\
20 & 3 & $\mathcal O_{2}$ (explicit) & 52 & 14\,915 & 12\,947 & 33\,173 & 11\,486 & 1740\\
8 & 4 & counting, sequential~\citesupp{S-Metwalli2020} & 12 & 4\,643 & 2\,142 & 5\,958 & 2\,464 & 169\\
8 & 4 & counting, parallel (main-text Prop.~6) & 47 & 258 & 964 & 2\,376 & 1\,106 & 158\\
8 & 4 & counting, parallel, $n$-controlled $Z$ variant & 40 & 327 & 930 & 2\,271 & 1\,062 & 145\\
8 & 4 & $\Gam$ & 24 & 961 & 756 & 2\,077 & 782 & 133\\
8 & 4 & $\mathcal O_{2}$ (explicit) & 26 & 6\,731 & 5\,043 & 13\,989 & 5\,246 & 804\\
12 & 4 & counting, sequential~\citesupp{S-Metwalli2020} & 16 & 10\,919 & 5\,030 & 13\,962 & 5\,778 & 397\\
12 & 4 & counting, parallel (main-text Prop.~6) & 71 & 359 & 2\,028 & 4\,832 & 2\,338 & 334\\
12 & 4 & counting, parallel, $n$-controlled $Z$ variant & 60 & 503 & 1\,978 & 4\,679 & 2\,270 & 313\\
12 & 4 & $\Gam$ & 32 & 1\,313 & 1\,096 & 2\,933 & 1\,070 & 181\\
12 & 4 & $\mathcal O_{2}$ (explicit) & 34 & 9\,195 & 7\,259 & 19\,685 & 7\,166 & 1092\\
16 & 4 & counting, sequential~\citesupp{S-Metwalli2020} & 20 & 19\,851 & 9\,134 & 25\,326 & 10\,484 & 721\\
16 & 4 & counting, parallel (main-text Prop.~6) & 95 & 441 & 3\,476 & 8\,120 & 4\,018 & 574\\
16 & 4 & counting, parallel, $n$-controlled $Z$ variant & 80 & 679 & 3\,410 & 7\,919 & 3\,926 & 545\\
16 & 4 & $\Gam$ & 40 & 1\,665 & 1\,468 & 3\,821 & 1\,358 & 229\\
8 & 5 & counting, sequential~\citesupp{S-Metwalli2020} & 13 & 7\,463 & 3\,478 & 10\,048 & 4\,000 & 225\\
8 & 5 & counting, parallel (main-text Prop.~6) & 55 & 739 & 3\,172 & 8\,456 & 3\,682 & 526\\
8 & 5 & counting, parallel, $n$-controlled $Z$ variant & 48 & 808 & 3\,138 & 8\,351 & 3\,638 & 513\\
8 & 5 & $\Gam$ & 22 & 889 & 676 & 1\,863 & 710 & 121\\
8 & 5 & $\mathcal O_{2}$ (explicit) & 24 & 6\,207 & 4\,519 & 12\,565 & 4\,766 & 732\\
10 & 5 & counting, sequential~\citesupp{S-Metwalli2020} & 15 & 11\,923 & 5\,564 & 16\,059 & 6\,396 & 361\\
10 & 5 & counting, parallel (main-text Prop.~6) & 69 & 924 & 4\,928 & 13\,060 & 5\,726 & 818\\
10 & 5 & counting, parallel, $n$-controlled $Z$ variant & 60 & 1\,020 & 4\,886 & 12\,931 & 5\,670 & 801\\
10 & 5 & $\Gam$ & 26 & 1\,067 & 838 & 2\,283 & 854 & 145\\
10 & 5 & $\mathcal O_{2}$ (explicit) & 28 & 7\,451 & 5\,579 & 15\,365 & 5\,726 & 876\\
12 & 5 & counting, sequential~\citesupp{S-Metwalli2020} & 17 & 17\,431 & 8\,138 & 23\,474 & 9\,352 & 529\\
12 & 5 & counting, parallel (main-text Prop.~6) & 83 & 1\,088 & 7\,068 & 18\,656 & 8\,218 & 1174\\
12 & 5 & counting, parallel, $n$-controlled $Z$ variant & 72 & 1\,232 & 7\,018 & 18\,503 & 8\,150 & 1153\\
12 & 5 & $\Gam$ & 30 & 1\,245 & 1\,008 & 2\,711 & 998 & 169\\
12 & 5 & $\mathcal O_{2}$ (explicit) & 32 & 8\,695 & 6\,687 & 18\,213 & 6\,686 & 1020\\
16 & 5 & counting, sequential~\citesupp{S-Metwalli2020} & 21 & 31\,591 & 14\,750 & 42\,516 & 16\,944 & 961\\
16 & 5 & counting, parallel (main-text Prop.~6) & 111 & 1\,418 & 12\,500 & 32\,824 & 14\,546 & 2078\\
16 & 5 & counting, parallel, $n$-controlled $Z$ variant & 96 & 1\,656 & 12\,434 & 32\,623 & 14\,454 & 2049\\
16 & 5 & $\Gam$ & 38 & 1\,601 & 1\,372 & 3\,591 & 1\,286 & 217\\
8 & 6 & counting, sequential~\citesupp{S-Metwalli2020} & 13 & 7\,463 & 3\,478 & 10\,048 & 4\,000 & 225\\
8 & 6 & counting, parallel (main-text Prop.~6) & 55 & 739 & 3\,172 & 8\,440 & 3\,682 & 526\\
8 & 6 & counting, parallel, $n$-controlled $Z$ variant & 48 & 808 & 3\,138 & 8\,335 & 3\,638 & 513\\
8 & 6 & $\Gam$ & 20 & 785 & 598 & 1\,661 & 638 & 109\\
8 & 6 & $\mathcal O_{2}$ (explicit) & 22 & 5\,499 & 4\,007 & 11\,213 & 4\,286 & 660\\
12 & 6 & counting, sequential~\citesupp{S-Metwalli2020} & 17 & 17\,431 & 8\,138 & 23\,474 & 9\,352 & 529\\
12 & 6 & counting, parallel (main-text Prop.~6) & 83 & 1\,088 & 7\,068 & 18\,632 & 8\,218 & 1174\\
12 & 6 & counting, parallel, $n$-controlled $Z$ variant & 72 & 1\,232 & 7\,018 & 18\,479 & 8\,150 & 1153\\
12 & 6 & $\Gam$ & 28 & 1\,137 & 922 & 2\,501 & 926 & 157\\
\bottomrule
\end{tabular}}
\end{table}

\begin{table}[!htbp]
\centering\small
\caption{Depth of the parallel counting oracle ($k=3$) and of the edge-detection oracle on random graphs $G(n,p)$ (script \texttt{11 sparse}; same settings), with the rounds given by a greedy proper edge coloring of $G$ (``coloring'') or by the $K_{n}$ 1-factorization (``1-fact.''). ``tree'': AND-tree final test (main-text Proposition~6); ``$n$-cZ'': $n$-controlled $Z$ variant. The last two columns are the edge detector of Fig.~\ref{S:fig:primitivesSI}b.}
\label{S:tab:sparse}
\resizebox{\textwidth}{!}{\begin{tabular}{@{}ccccccccccc@{}}
\toprule
$n$ & $p$ & $m$ & $\Delta$ & rounds & tree, coloring & tree, 1-fact. & $n$-cZ, coloring & $n$-cZ, 1-fact. & edge det., coloring & edge det., 1-fact.\\
\midrule
12 & 0.15 & 8 & 3 & 3 & 197 & 197 & 339 & 313 & 106 & 160\\
12 & 0.40 & 26 & 8 & 8 & 297 & 345 & 421 & 490 & 200 & 280\\
12 & 1.00 & 66 & 11 & 11 & 359 & 359 & 503 & 503 & 388 & 335\\
16 & 0.15 & 15 & 4 & 5 & 239 & 289 & 464 & 529 & 138 & 288\\
16 & 0.40 & 47 & 8 & 10 & 337 & 353 & 568 & 573 & 284 & 389\\
16 & 1.00 & 120 & 15 & 15 & 441 & 441 & 679 & 679 & 461 & 463\\
20 & 0.15 & 39 & 6 & 7 & 294 & 366 & 613 & 685 & 261 & 465\\
20 & 0.40 & 84 & 13 & 14 & 436 & 488 & 738 & 794 & 393 & 504\\
20 & 1.00 & 190 & 19 & 19 & 542 & 542 & 855 & 855 & 756 & 591\\
\bottomrule
\end{tabular}}
\end{table}

\paragraph{Multi-controlled gates in the compiled circuits.} Qiskit's default high-level synthesis selects, for each multi-controlled X, the cheapest available method given the qubits of the circuit that are idle at that point, which it may borrow as clean or dirty ancillas \citesupp{S-Iten2016,S-Khattar2025}. In $\Gam$ the $N$ $\idx$ qubits and the two $\rem$ qubits are idle during $S_{0}$ and are used as dirty ancillas; in the edge detector the $\idx$ qubits are idle during the OR block. This is what makes the compiled depths of Table~\ref{S:tab:gammares} and Table~\ref{S:tab:edge} linear in $N$ and $n$; a standalone $c$-controlled X with no idle qubits (Fig.~\ref{S:fig:mcx}, ancilla-free curve) is quadratic in $c$. Borrowed ancillas are also why the parallel counting oracle avoids multi-controlled gates on the vertex registers: gates of different vertices that borrow the same idle qubit are serialized by the transpiler, which would otherwise make the flag stage, rather than the counting rounds, dominate the depth.

\paragraph{The construction step by step.} The layering applies equally to the edge-counting $k$-clique oracles of Chang et al.~\citesupp{S-Chang2018} and Metwalli et al.~\citesupp{S-Metwalli2020}. Their depth $\Theta(m)$ has the same origin as that of the naive edge detector: a single counter that every edge increments, so that the $m$ increments are serialized whatever the graph looks like. Distributing the count over the vertices removes the bottleneck (Supplementary Fig.~S15a). Give every vertex $v$ a $b=\lceil\log_{2}k\rceil$-bit counter $\mathtt{cnt}_{v}$ and process the edges color class by color class: for each edge $uv$ of a class, add $\idx_{u}$ to $\mathtt{cnt}_{v}$ and $\idx_{v}$ to $\mathtt{cnt}_{u}$. The two controlled increments act on the disjoint qubit sets $\{\idx_{u}\}\cup\mathtt{cnt}_{v}$ and $\{\idx_{v}\}\cup\mathtt{cnt}_{u}$, and two edges of the same class share no vertex, so a whole class is one layer; on $K_{n}$ the round-robin 1-factorization gives $n-1$ rounds, on a sparse graph the Misra--Gries coloring gives $\Delta+1$. After the last class $\mathtt{cnt}_{v}=|N(v)\cap H|$ for every $v$, where $H=\supp(y)$. A comparison bit $e_{v}=[\mathtt{cnt}_{v}=k-1]$ and a flag $f_{v}=\idx_{v}\wedge\neg e_{v}$ mark a vertex of $H$ that misses a neighbor in $H$; the flags are combined by an AND tree of Toffolis on $n-1$ clean ancillas, a $Z$ on the root applies the phase $-1$ exactly when no flag is set, and the tree, the flags and the rounds are un-computed in mirror image. Every gate that acts on a single vertex has at most two controls (for $b=3$ the top bit of an increment is a three-controlled X, realized as three Toffolis with the idle flag $f_{v}$ as a clean ancilla), so no multi-controlled gate other than the tree is ever needed, and the tree has depth $\lceil\log_{2}n\rceil$. Per-vertex counters have been used before in Grover oracles for domination problems \citesupp{S-Jiang2023}, and the schedule itself is the standard one of main-text Lemma~3; what is added here is their combination on an arbitrary graph, which advances all $n$ counters in parallel in $\chi'(G)$ rounds.

\FloatBarrier
\section*{Supplementary Note 5: search-register initialization and the host graphs}\label{S:note:5}

\begin{figure}[!htbp]
\centering\includegraphics[width=\textwidth]{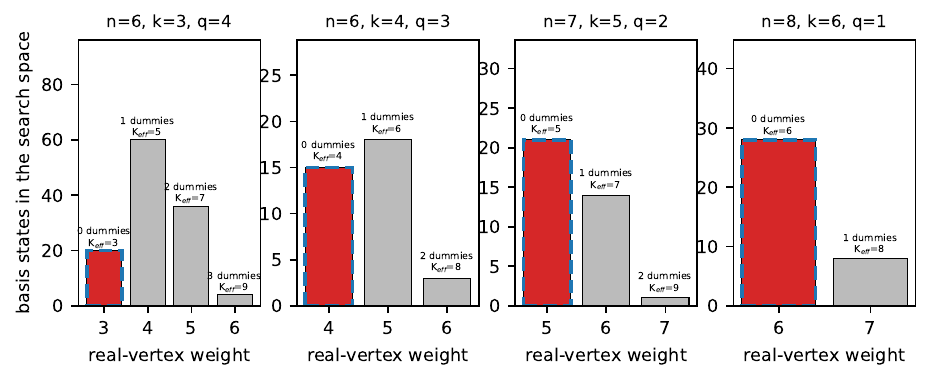}
\caption{\textbf{Search space of the initial driver versus the algorithm of the main text.} Number of $\idx$ basis states by real-vertex weight in the initial driver's search space ($\ket{D_{N,k+q}}$ on all $N$ qubits followed by X on the dummies) for four configurations, with the number of dummy bits set and the effective $K$ annotated; every candidate with the correct weight $k$ (red) carries no dummy vertex. Dashed: the search space of Fig.~\ref{S:fig:aa}, $\binom nk$ states all carrying the $q$ dummies.}
\label{S:fig:init}
\end{figure}

\paragraph{Initialization.} The driver of the initial shot-sampled benchmark prepared $\ket{D_{N,k+q}}$ on all $N=n+q$ $\idx$ qubits and then applied X to the $q$ dummy qubits, whereas the algorithm of Fig.~\ref{S:fig:aa} prepares $\ket{D_{n,k}}$ on $\idx_{0..n-1}$ and sets the dummies to $\ket1$. A weight-$K$ string with $r$ real vertices has $K-r$ dummy bits set before the flip and $r-k$ after it, so every candidate with the correct real-vertex weight $k$ carries \emph{no} dummy vertex (Fig.~\ref{S:fig:init}): the oracle was queried on $K_{\mathrm{eff}}=k$ vertices, for which $k=3$ marks independent triples as well as triangles and $k\not\equiv3\pmod4$ does not produce the state \eqref{S:eq:aF} at all. That benchmark is therefore not a test of the algorithm studied here; the exact simulations of the main text replace it. Exact trajectories for both initializations (Table~\ref{S:tab:init}) show that neither approaches the ideal oracle without the filter.

\begin{table}[!htbp]
\centering\small
\caption{Exact trajectories of the plain $\Gam$ oracle for the two initializations (8 instances per row).}
\label{S:tab:init}
\begin{tabular}{@{}cclcccc@{}}
\toprule
$n$ & $k$ & host & initialization & ideal $P(r^{*})$ & $P(r^{*})$ & $\max_{r}P$\\
\midrule
6 & 3 & macaque & main text & 0.750 & 0.309 & 0.417\\
6 & 3 & macaque & earlier driver & 0.757 & 0.225 & 0.267\\
6 & 4 & mouse & main text & 0.882 & 0.256 & 0.550\\
6 & 4 & mouse & earlier driver & 0.882 & 0.257 & 0.263\\
7 & 3 & macaque & main text & 0.895 & 0.238 & 0.427\\
7 & 3 & macaque & earlier driver & 0.821 & 0.159 & 0.185\\
\bottomrule
\end{tabular}
\end{table}

\begin{figure}[!htbp]
\centering\includegraphics[width=\textwidth]{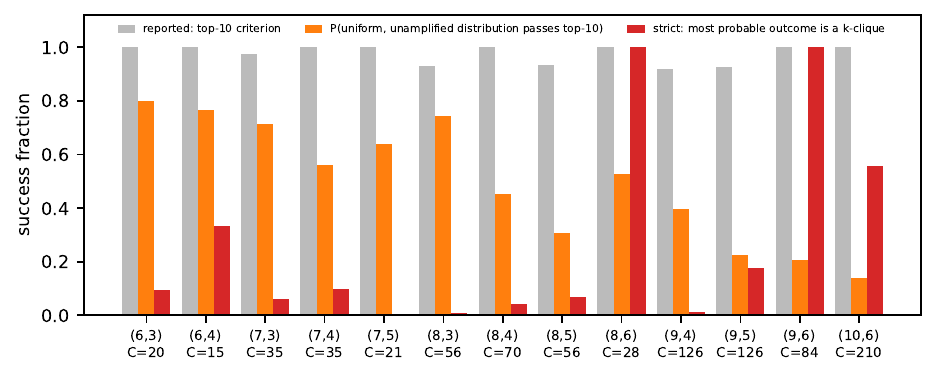}
\caption{\textbf{Audit of the initial benchmark.} For each $(n,k)$ ($C=\binom nk$): the reported success fraction under the top-ten criterion, the probability that an unamplified uniform distribution satisfies that criterion (computed per instance from its number of $k$-cliques), and the fraction of instances in which the single most probable outcome was a $k$-clique. $805$ clique-bearing instances.}
\label{S:fig:audit}
\end{figure}

\paragraph{Success criterion.} The initial benchmark scored a run as a success when one of the ten most probable outcomes of $1000$ shots was a $k$-clique, within $\lfloor2\cdot\tfrac\pi4\sqrt{\binom nk}\rfloor$ iterations. For search spaces of size $15$--$210$ this criterion is nearly vacuous: the probability that an unamplified uniform distribution already satisfies it is $0.14$--$0.80$ (Fig.~\ref{S:fig:audit}, Table~\ref{S:tab:audit}). Under the strict criterion that the single most probable outcome be a $k$-clique, its stored data give $0$--$56\%$ success (typically $1$--$10\%$), against $94$--$100\%$ for the exact counting oracle of Metwalli \emph{et al.} on the same instances. The $338$ clique-free subgraphs that were run produced no false positive, as Corollary~9 of the main text requires.

\begin{table}[!htbp]
\centering\small
\caption{Audit of the earlier stored result files (script \texttt{10}); a duplicate file for $(8,3)$ was excluded and 20 records of that configuration have no stored edge list.}
\label{S:tab:audit}
\resizebox{\textwidth}{!}{\begin{tabular}{@{}ccccccccc@{}}
\toprule
$n$ & $k$ & $\binom nk$ & instances & mean $\mathcal M$ & top-10 (reported) & top-1 (strict) & mean rank of first clique & P(uniform passes)\\
\midrule
6&3&20&139&3.5&100.0\%&9.4\%&2.78&0.800\\ 6&4&15&45&1.7&100.0\%&33.3\%&0.84&0.767\\
7&3&35&161&5.5&97.5\%&6.2\%&4.75&0.714\\ 7&4&35&81&3.2&100.0\%&9.9\%&2.86&0.559\\
7&5&21&16&2.6&100.0\%&0.0\%&5.31&0.640\\ 8&3&56&97&10.0&92.8\%&1.0\%&5.64&0.743\\
8&4&70&74&5.5&100.0\%&4.1\%&4.74&0.452\\ 8&5&56&30&2.1&93.3\%&6.7\%&4.39&0.307\\
8&6&28&5&2.4&100.0\%&100.0\%&0.00&0.528\\ 9&4&126&86&8.0&91.9\%&1.2\%&5.47&0.395\\
9&5&126&40&3.6&92.5\%&17.5\%&3.78&0.226\\ 9&6&84&13&2.0&100.0\%&100.0\%&0.00&0.208\\
10&6&210&18&3.4&100.0\%&55.6\%&1.50&0.140\\
\midrule
total & & & 805 & & 97.1\% & 10.9\% & & \\
\bottomrule
\end{tabular}}
\end{table}

\begin{figure}[!htbp]
\centering\includegraphics[width=\textwidth]{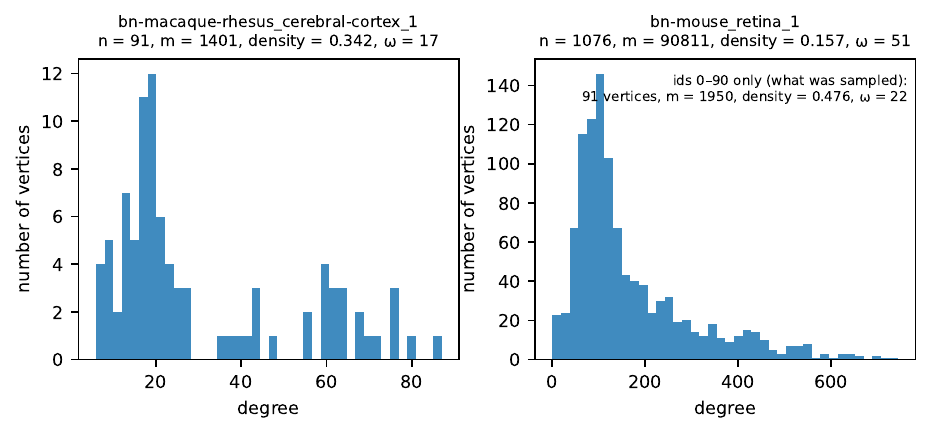}
\caption{\textbf{The two host graphs.} Degree distributions after removal of duplicate lines. Inset: the induced subgraph on vertex ids $0$--$90$ from which the initial benchmark's mouse-retina subgraphs were drawn.}
\label{S:fig:hosts}
\end{figure}

\paragraph{Host graphs.} Statistics recomputed from the raw edge lists are in Table~\ref{S:tab:hosts}; degree distributions are shown in Fig.~\ref{S:fig:hosts}. The repository's densities ($0.486$, $0.998$) are reproduced to three decimals by counting duplicate lines. The repository's ``at least 21'' for the macaque clique number is not reproduced by the simple graph ($\omega=17$); ``at least 42'' for the mouse graph is consistent ($\omega=51$). The preprocessing script of the initial benchmark hard-coded 91 vertices for both graphs, so its mouse-retina subgraphs were drawn from vertex ids $0$--$90$ only (induced subgraph: $1\,950$ edges, density $0.476$, $\omega=22$); the exact simulations of the main text (main-text Table~2, Table~\ref{S:tab:gammatraj}, Fig.~\ref{S:fig:aggregated}) use a new sample drawn from the full vertex sets of both graphs with the protocol of main-text Sect.~8.1 and Supplementary Note~15 (script \texttt{15}); only the instance of main-text Fig.~3 and Table~\ref{S:tab:single}, an induced subgraph of the macaque graph, is taken from the earlier sample.

\begin{table}[!htbp]
\centering\small
\caption{\textbf{The two host graphs} \citesupp{S-Rossi2015}, recomputed from the raw edge lists.}
\label{S:tab:hosts}
\begin{tabular}{@{}lcc@{}}
\toprule
Property & bn-macaque-rhesus\_cerebral-cortex\_1 & bn-mouse\_retina\_1\\
\midrule
lines in the edge file (with duplicates) & 1\,989 & 577\,350\\
distinct vertices & 91 & 1\,076\\
distinct simple edges & 1\,401 & 90\,811\\
density (simple graph) & 0.342 & 0.157\\
degree min / mean / max & 6 / 30.8 / 87 & 1 / 168.8 / 744\\
triangles & 10\,336 & 3\,289\,057\\
exact clique number $\omega$ & 17 & 51\\
\bottomrule
\end{tabular}
\end{table}

\FloatBarrier
\section*{Supplementary Note 6: trajectories of the plain oracle}\label{S:note:6}

\begin{figure}[!htbp]
\centering\includegraphics[width=\textwidth]{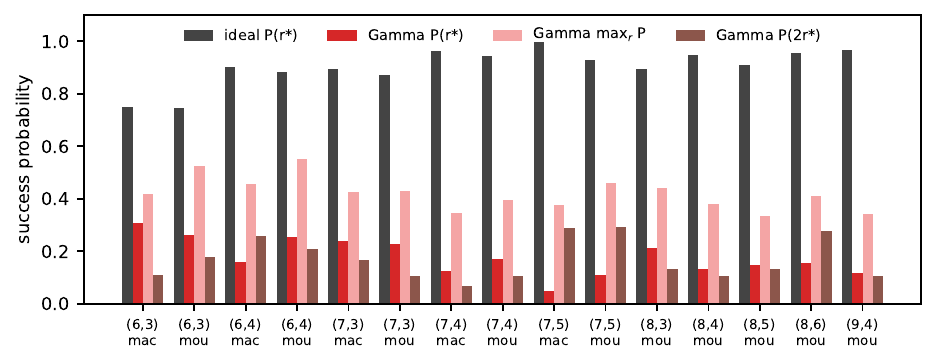}
\caption{\textbf{Plain $\Gam$ oracle: success at $r^{*}$, best achievable success and over-shooting.} Exact probabilities aggregated over the stored instances (main-text initialization): ideal $P(r^{*})$, $\Gam$ $P(r^{*})$, $\Gam$ $\max_{r}P(r)$ and $\Gam$ $P(2r^{*})$.}
\label{S:fig:overshoot}
\end{figure}

Table~\ref{S:tab:gammatraj} gives the exact trajectories of the plain $\Gam$ oracle on the instances of the main-text set (main-text initialization, up to $60$ pairwise non-isomorphic instances per row; Fig.~\ref{S:fig:overshoot} shows the corresponding quantities for a smaller exploratory sample). At the ideal iteration count $r^{*}$, $\Gam$ delivers $0.05$--$0.31$ where the ideal oracle delivers $0.75$--$1.00$; its own maximum lies at $\approx3$--$5\,r^{*}$ and is only $0.34$--$0.55$; doubling the iteration count halves the probability or worse. Table~\ref{S:tab:single} lists the single instance of main-text Fig.~3.

Two features of Table~\ref{S:tab:gammatraj} deserve comment. First, the deficit at $r^{*}$ is not monotone in $q$: the rows with $q=1$ or $2$ (mean leakage $0.11$--$0.29$ over all graphs, Table~\ref{S:tab:adist}) are no better than those with $q=4$ (mean leakage $0.85$). The mean over all graphs is therefore not the right predictor; what matters is the overlap of the particular non-cliques that carry most of the amplitude, and in induced subgraphs of dense networks these are near-cliques with $|a|\approx1/2$, which exist for every $q$. Second, the location of the maximum ($3$--$5\,r^{*}$) and its height ($0.34$--$0.55$) are remarkably uniform across configurations, which is consistent with the picture of Proposition~7 of the main text: each query removes a roughly constant fraction of the non-clique amplitude from the coherent search subspace, so that the amplification proceeds more slowly and saturates rather than oscillates.

\begin{table}[!htbp]
\centering\small
\caption{Plain $\Gam$ oracle on the instances of the main-text set (scripts \texttt{15}--\texttt{17}): number of pairwise non-isomorphic non-degenerate instances, mean $r^{*}=\lfloor\pi/(4\theta)\rfloor$, mean ideal $P(r^{*})$, mean and median of $P(r^{*})$, mean location of the maximum, mean and worst maximum, mean $P(2r^{*})$, and the false-positive check on the stored clique-free subgraphs of the same configuration (count, $\min_{y}\lambda_{y}$, $\max_{y}|a_{y}|$; Corollary~9 requires $\lambda\ge7/32=0.21875$, Theorem~8 $|a|\le5/8$).}
\label{S:tab:gammatraj}
\resizebox{\textwidth}{!}{\begin{tabular}{@{}ccclcccccccccccc@{}}
\toprule
$n$&$k$&$q$&host&$\#$&$\langle r^{*}\rangle$&ideal $P(r^{*})$&$\Gam$ $P(r^{*})$ mean&median&argmax $r$&$\max_r P$ mean&worst&$P(2r^{*})$&clique-free&$\min\lambda$&$\max|a|$\\
\midrule
6&3&4&macaque&60&1.6&0.960&0.222&0.293&5.1&0.470&0.253&0.144&300&+0.21875&0.6250\\
6&3&4&mouse&60&1.9&0.961&0.178&0.086&4.8&0.438&0.253&0.171&300&+0.21875&0.6250\\
6&4&3&macaque&34&2.4&0.929&0.115&0.048&5.8&0.532&0.389&0.239&300&+0.21875&0.6250\\
6&4&3&mouse&34&2.4&0.929&0.115&0.048&5.8&0.532&0.389&0.239&300&+0.21875&0.6250\\
7&3&4&macaque&60&2.3&0.955&0.139&0.074&6.8&0.391&0.235&0.188&300&+0.21875&0.6250\\
7&3&4&mouse&60&2.9&0.976&0.123&0.086&8.6&0.349&0.238&0.153&300&+0.21875&0.6250\\
7&4&3&macaque&30&2.6&0.972&0.123&0.089&5.4&0.408&0.243&0.143&300&+0.21875&0.6250\\
7&4&3&mouse&60&3.3&0.976&0.061&0.053&5.9&0.386&0.247&0.169&300&+0.21875&0.6250\\
7&5&2&macaque&56&2.6&0.992&0.105&0.065&5.8&0.432&0.288&0.282&300&+0.21875&0.6250\\
7&5&2&mouse&45&2.6&0.994&0.110&0.069&6.0&0.430&0.288&0.274&300&+0.21875&0.6250\\
8&3&4&mouse&30&3.7&0.978&0.094&0.080&10.9&0.343&0.268&0.118&300&+0.21875&0.6250\\
8&4&3&mouse&60&4.8&0.992&0.098&0.112&11.0&0.299&0.197&0.134&300&+0.21875&0.6250\\
8&5&2&mouse&60&4.7&0.988&0.080&0.074&10.1&0.321&0.216&0.081&300&+0.21875&0.6250\\
8&6&1&mouse&29&3.6&0.976&0.135&0.129&7.0&0.345&0.266&0.110&300&+0.21875&0.6250\\
9&4&3&mouse&40&6.5&0.993&0.085&0.094&15.6&0.225&0.130&0.118&300&+0.21875&0.6250\\
\bottomrule
\end{tabular}}
\end{table}

\begin{table}[!htbp]
\centering\small
\caption{The instance of main-text Fig.~3 (an induced subgraph of the macaque graph from the exploratory sample): $n=7$, $k=3$, $E=\{03,12,13,23,34,35,36\}$, one triangle, $\mathcal N=35$; exact $P(r)$ for $r=0$--$24$ from one run (script \texttt{14}).}
\label{S:tab:single}
\begin{tabular}{@{}ccccc|ccccc@{}}
\toprule
$r$&ideal&$\Gam$&$\mathcal O_{2}$&$\mathcal O_{3}$&$r$&ideal&$\Gam$&$\mathcal O_{2}$&$\mathcal O_{3}$\\
\midrule
0&0.029&0.029&0.029&0.029&13&0.984&0.282&0.808&0.917\\
1&0.238&0.131&0.225&0.231&14&0.955&0.074&0.927&0.924\\
2&0.564&0.094&0.484&0.524&15&0.724&0.007&0.818&0.771\\
3&0.861&0.093&0.751&0.793&16&0.394&0.029&0.599&0.490\\
4&0.998&0.202&0.906&0.956&17&0.110&0.011&0.314&0.185\\
5&0.914&0.180&0.832&0.913&18&0.000&0.031&0.067&0.011\\
6&0.646&0.061&0.603&0.662&19&0.112&0.046&0.007&0.052\\
7&0.313&0.032&0.364&0.332&20&0.396&0.028&0.140&0.263\\
8&0.063&0.146&0.131&0.088&21&0.726&0.009&0.372&0.559\\
9&0.007&0.282&0.008&0.004&22&0.956&0.016&0.664&0.846\\
10&0.170&0.233&0.066&0.113&23&0.984&0.143&0.851&0.977\\
11&0.479&0.173&0.254&0.396&24&0.796&0.238&0.834&0.870\\
12&0.798&0.296&0.522&0.718&&&&&\\
\bottomrule
\end{tabular}
\end{table}

\FloatBarrier
\section*{Supplementary Note 7: compiled circuit resources and validation}\label{S:note:7}

\begin{table}[!htbp]
\centering\small
\caption{One $\Gam$ query on the complete host graph, transpiled to $\{R_{z},\sqrt X,X,\mathrm{CX}\}$ (optimization level 1); scripts \texttt{05}, \texttt{11} and \texttt{14}.}
\label{S:tab:gammares}
\resizebox{\textwidth}{!}{\begin{tabular}{@{}ccccccccccc@{}}
\toprule
$n$&$k$&$q$&$N$&qubits&depth (all-to-all)&$89N-1$&CX&gates&depth (linear chain)&CX (linear)\\
\midrule
6&3&4&10&22&889&889&676&1\,863&2\,796&4\,123\\
6&4&3&9&20&785&(odd)&598&1\,661&2\,396&3\,448\\
7&3&4&11&24&961&(odd)&756&2\,077&3\,144&4\,986\\
8&3&4&12&26&1\,067&1\,067&838&2\,283&3\,580&5\,794\\
8&4&3&11&24&961&(odd)&756&2\,077&--&--\\
10&3&4&14&30&1\,245&1\,245&1\,008&2\,711&4\,582&7\,929\\
12&3&4&16&34&1\,423&1\,423&1\,186&3\,147&5\,402&10\,147\\
14&3&4&18&38&1\,601&1\,601&1\,372&3\,591&--&--\\
16&3&4&20&42&1\,779&1\,779&1\,566&4\,043&--&--\\
20&3&4&24&50&2\,135&2\,135&1\,978&4\,971&--&--\\
\bottomrule
\end{tabular}}
\end{table}

\begin{table}[!htbp]
\centering\small
\caption{$\Gam$ versus the checking-based oracle of Metwalli \emph{et al.}, same complete graphs, all-to-all coupling.}
\label{S:tab:vsmet}
\begin{tabular}{@{}ccccccccc@{}}
\toprule
$n$&$k$&$\Gam$ qubits&$\Gam$ depth&$\Gam$ CX&ref.\ qubits&ref.\ depth&ref.\ CX&depth ratio\\
\midrule
6&3&22&889&676&9&1\,043&576&1.17\\
7&3&24&961&756&10&1\,453&804&1.51\\
8&3&26&1\,067&838&11&1\,932&1\,070&1.81\\
10&3&30&1\,245&1\,008&13&3\,097&1\,716&2.49\\
12&3&34&1\,423&1\,186&15&4\,538&2\,514&3.19\\
16&3&42&1\,779&1\,566&19&8\,248&4\,566&4.64\\
20&3&50&2\,135&1\,978&23&13\,062&7\,226&6.12\\
8&4&24&961&756&12&4\,643&2\,142&4.83\\
16&4&40&1\,665&1\,468&20&19\,851&9\,134&11.92\\
\bottomrule
\end{tabular}
\end{table}

\begin{table}[!htbp]
\centering\small
\caption{Edge-detection oracle on $K_{n}$: naive versus matching-partitioned circuit.}
\label{S:tab:edge}
\begin{tabular}{@{}cccccccc@{}}
\toprule
$n$&$m$&naive depth&naive CX&partitioned depth&partitioned CX&qubits&$32n-49$\\
\midrule
4&6&55&36&76&78&7&79\\ 6&15&127&90&129&194&10&143\\ 8&28&231&168&207&362&13&207\\
10&45&367&270&271&574&16&271\\ 12&66&535&396&335&834&19&335\\ 16&120&967&720&463&1\,498&25&463\\
20&190&1\,527&1\,140&591&2\,354&31&591\\ 32&496&3\,975&2\,976&975&6\,074&49&975\\
\bottomrule
\end{tabular}
\end{table}

\begin{figure}[!htbp]
\centering
\includegraphics[width=0.72\textwidth]{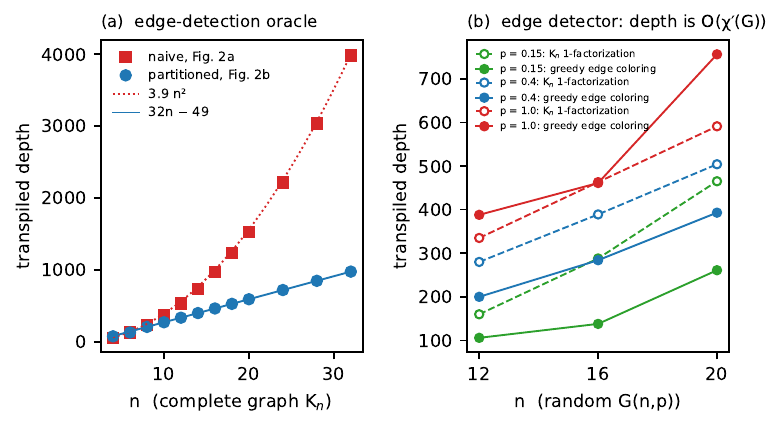}
\caption{\textbf{Compiled depth of the edge-detection oracle} (settings of Fig.~5). \textbf{a}, On $K_{n}$: naive (Fig.~\ref{S:fig:primitivesSI}a) $\approx3.9n^{2}$ versus partitioned (Fig.~\ref{S:fig:primitivesSI}b) $32n-49$. \textbf{b}, On random $G(n,p)$: layering by a greedy proper edge coloring of $G$ (solid) versus the $K_{n}$ 1-factorization (dashed); for sparse graphs the depth follows $\chi'(G)\le\Delta+1$.}
\label{S:fig:resources}
\end{figure}

\paragraph{Feasibility on hardware.} The counts in Tables~\ref{S:tab:gammares}--\ref{S:tab:vsmet} put the circuits well beyond present unencoded devices: one $\Gam$ query at $N=10$ needs $676$ CX gates in $889$ layers under all-to-all coupling, and $4\,123$ CX gates in $2\,796$ layers on a linear chain; a filtered query costs about six times as much. The comparison with the counting oracle of Metwalli \emph{et al.} shows that the depth advantage is asymptotic (a factor $1.2$ at $n=6$, $3.2$ at $n=12$, $6.1$ at $n=20$) while the CX counts are of the same order and the width is about twice as large. These numbers are the honest basis for the claim of the main text: a $\Theta(n)$ versus $\Theta(n^{2})$ reduction of per-query depth at $\Theta(n^{2})$ gates, not a near-term demonstration.

\paragraph{Validation of the simulator.} The B\"artschi--Eidenbenz Dicke circuit (Fig.~\ref{S:fig:dicke}) prepares $\ket{D_{n,k}}$ with amplitude error $\le1.9\times10^{-16}$ for $(n,k)=(4,2),(5,3),(6,3),(7,3),(9,7)$. The fast simulator agrees with a gate-for-gate Qiskit statevector run of the full amplitude-amplification circuits on all output distributions, with $\max|\Delta P|=3.3\times10^{-14}$ and $8.1\times10^{-14}$ ($n=4$, $k=3$, $r=1,2$, initial driver's initialization), $2.0\times10^{-14}$ and $3.2\times10^{-14}$ (main-text initialization), and $1.9$--$7.9\times10^{-14}$ for two graphs at $n=5$; it is about $2000$ times faster ($0.04$\,s versus $91$\,s per data point at $n=5$).

\begin{figure}[!htbp]
\centering\includegraphics[width=\textwidth]{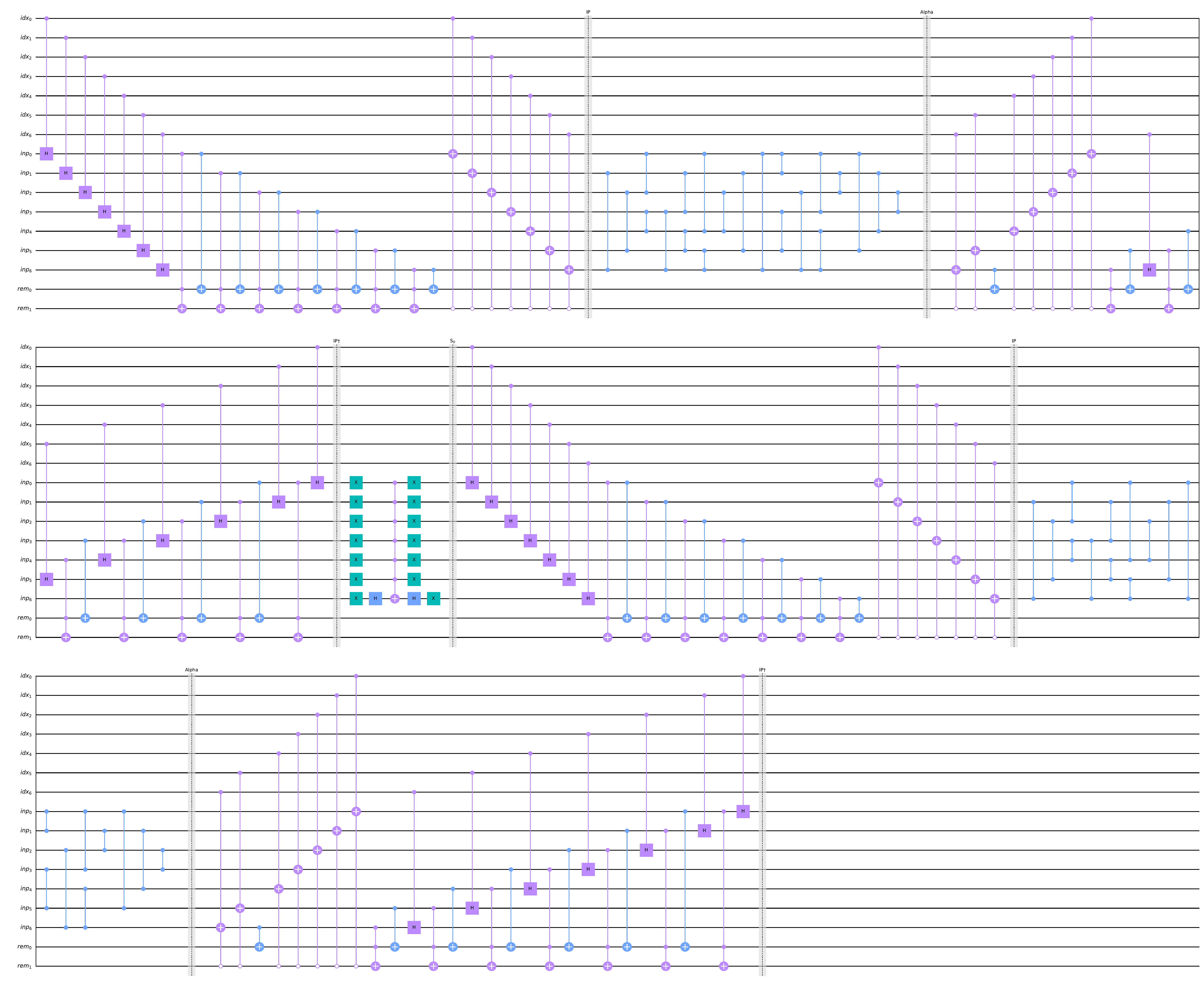}
\caption{\textbf{Gate-level $\Gam$ oracle for the smallest valid instance} ($n=3$, $k=3$; $q=4$, $K=7$, $N=7$, 16 qubits). The blocks $\IP$, Alpha, $\IP^{\dagger}$, $S_{0}$, $\IP$, Alpha, $\IP^{\dagger}$ are separated by labeled barriers; 171 gates, depth 103 before transpilation. The host graph is the triangle, so the augmented graph is $K_{7}$ and Alpha contains all 21 CZ gates in 7 layers.}
\label{S:fig:gatelevel}
\end{figure}

\begin{figure}[!htbp]
\centering\includegraphics[width=0.9\textwidth]{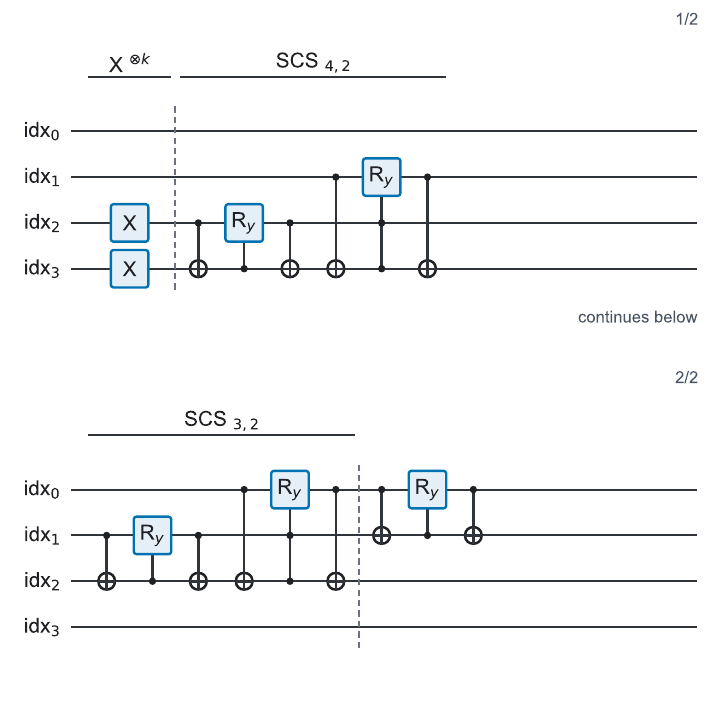}
\caption{\textbf{B\"artschi--Eidenbenz Dicke-state preparation} for $n=4$, $k=2$: $X^{\otimes k}$ followed by the split-and-cyclic-shift blocks $\mathrm{SCS}_{l,k}$. It is used for state preparation and twice per iteration inside the diffuser; the 2022 short-depth construction lowers this part to $O(k\log(n/k))$ depth.}
\label{S:fig:dicke}
\end{figure}

\begin{figure}[!htbp]
\centering\includegraphics[width=0.55\textwidth]{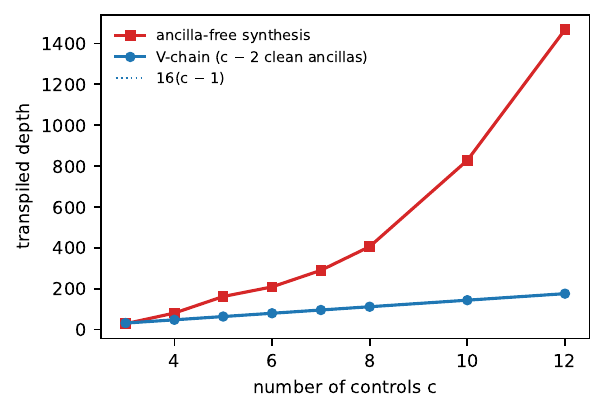}
\caption{\textbf{Cost of the multi-controlled X} that appears in $S_{0}$, in the OR block of the edge detector and in the diffuser: transpiled depth of a $c$-controlled X with the ancilla-free synthesis (quadratic in $c$) and with the V-chain using $c-2$ clean ancillas (depth $16(c-1)$). In $\Gam$ the idle $\idx$ qubits can serve as ancillas, so linear depth is available at no extra width.}
\label{S:fig:mcx}
\end{figure}

\begin{figure}[!htbp]
\centering\includegraphics[width=0.9\textwidth]{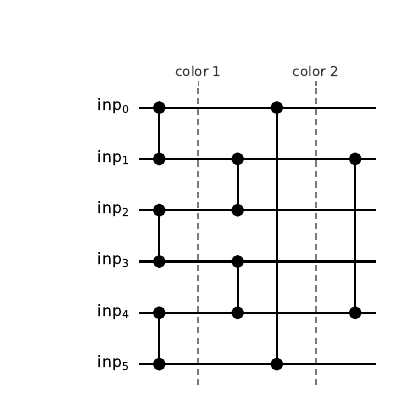}
\caption{\textbf{Alpha for a sparse graph} on six vertices (a 6-cycle with one chord, $m=7$, $\Delta=3$): a greedy proper edge coloring uses $\chi'(G)\le\Delta+1$ color classes, so the depth is $O(\Delta)$ rather than $O(n)$.}
\label{S:fig:sparse}
\end{figure}

\FloatBarrier
\section*{Supplementary Note 8: triangle finding with the edge-detection oracle}\label{S:note:8}

\begin{corollary}[Triangle finding]\label{S:cor:triangle}
Le Gall's algorithm \citesupp{S-LeGall2014} run with the oracle of Proposition~5 uses $\Otil(n^{5/4})$ oracle calls of depth $O(n)$ each, i.e.\ $\Otil(n^{9/4})$ total oracle depth in the hard-wired circuit model.
\end{corollary}

We state Corollary~\ref{S:cor:triangle} as a bound on oracle depth and \emph{not} as a time complexity: the non-query cost of the nested quantum walk in \citesupp{S-LeGall2014} has not been analyzed in the literature, and no claim is made here of improving on the classical $O(n^{\omega})$ bound. Within our model the comparison that is defensible is with Grover search over the $\binom{n}{3}$ triples using the same oracle, which costs $\Otil(n^{5/2})$ total oracle depth; in the QRAG model an adjacency query costs $O(1)$ and Grover search alone runs in $\Otil(n^{3/2})$ time \citesupp{S-Allcock2023}.

A binary-to-one-hot decoder on $2\lceil\log_{2}n\rceil$ address qubits costs $O(\log n)$ depth and $O(n)$ ancillas, so the oracle of main-text Proposition~5 also accepts the binary-encoded queries of the query-model algorithms at no asymptotic cost in depth. If the non-query operations of Le Gall's walk can be implemented in polylogarithmic depth per step, Corollary~\ref{S:cor:triangle} becomes an $\Otil(n^{9/4})$ time bound in the hard-wired circuit model; the same question determines whether the $\Otil(n^{3/2})$ time of a plain Grover search over triples in the QRAG model (Table~\ref{S:tab:models}) can be matched without random access.

\FloatBarrier
\section*{Supplementary Note 9: the complete circuit, the cost of the filter and the aggregated benchmarks}\label{S:note:9}

\begin{figure}[!htbp]
\centering
\includegraphics[width=\textwidth]{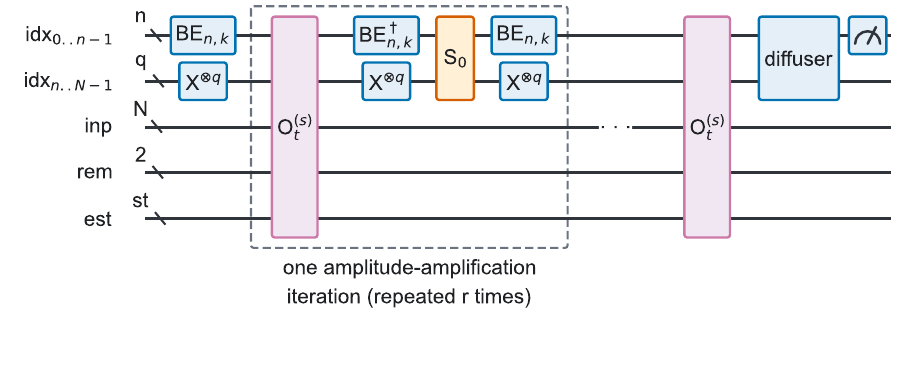}
\caption{\textbf{Complete amplitude-amplification circuit for $k$-\textsc{clique}.} State preparation: the B\"artschi--Eidenbenz unitary $\mathrm{BE}_{n,k}$ creates the Dicke state of weight $k$ on $\idx_{0..n-1}$ and X gates set the $q$ dummy qubits to $\ket1$. One iteration (shaded) applies the filtered oracle $\Ot^{(s)}$ ($s$ repetitions of $\Ot$ combined by AND; $s=1$ with the schedule of \citesupp{S-Hoyer2003}) followed by the diffuser $I-2\ket\sigma\bra\sigma$, implemented as $\mathrm{BE}^{\dagger}$, $X^{\otimes q}$, $S_{0}$ on all $N$ $\idx$ qubits, $X^{\otimes q}$, $\mathrm{BE}$. After $r$ iterations $\idx_{0..n-1}$ is measured. Registers: $\idx$ ($N$), $\inp$ ($N$), $\rem$ (2), $\est$ ($st$).}
\label{S:fig:aa}
\end{figure}

\paragraph{The complete circuit.} Fig.~\ref{S:fig:aa} shows the amplitude-amplification circuit of main-text Theorem~11: the B\"artschi--Eidenbenz unitary prepares $\ket{D_{n,k}}$ on $\idx_{0..n-1}$ and X gates set the $q$ dummy qubits to $\ket1$; one iteration applies the filtered oracle $\mathcal O_{t}^{(s)}$ and the diffuser $I-2\ket\sigma\bra\sigma$, implemented as $\mathrm{BE}^{\dagger}$, $X^{\otimes q}$, $S_{0}$ on all $N$ $\idx$ qubits, $X^{\otimes q}$, $\mathrm{BE}$; after $r$ iterations $\idx_{0..n-1}$ is measured and the outcome verified classically.

\paragraph{The error model, formally.} The impossibility theorem of Regev and Schiff \citesupp{S-Regev2008} concerns a specific channel: each call, independently, applies the ideal oracle with probability $1-p$ and the identity with probability $p$, i.e.\ the mixture $\mathcal E(\rho)=(1-p)\,O_{\mathrm{id}}\,\rho\,O_{\mathrm{id}}^{\dagger}+p\,\rho$, with no record available to the algorithm of which occurred. The failure is classical, silent and incoherent (formally the algorithm interacts with an environment that stores the failure pattern) and the hybrid argument over that hidden record forces $\Omega(\mathcal N)$ queries; repetition does not help, because a repeated silent failure is still silent. The filtered oracle is outside this model on every count. $\mathcal O_{t}^{(s)}$ is one fixed, known unitary; on a candidate $\ket y\ket0$ its action is $\sigma_{y}(1-2\varepsilon_{y}^{\,s})\ket y\ket 0$ plus an amplitude of norm $2\sqrt{\varepsilon_{y}^{\,s}(1-\varepsilon_{y}^{\,s})}\le2\varepsilon^{s/2}$ on states whose estimation register is orthogonal to $\ket0$ (main-text Theorem~10): coherent, amplitude-bounded, carried on a flag that the algorithm can read out as a definite ``non-clique'' verdict or un-compute, and one-sided, since $\varepsilon_{y}=0$ on cliques (Corollary~9). No environment records a failure, so the Regev--Schiff hybrid has nothing to average over, and the interference between the marked and the leaked branches remains available to the algorithm; the phase-estimation filter is precisely a use of that interference, and the first-order accounting of Theorem~11(iii) tracks it explicitly. This is the bounded-error model of H{\o}yer, Mosca and de Wolf \citesupp{S-Hoyer2003} (a unitary that computes the marking bit with bounded error and can be inverted), whose search theorem gives $\Theta(\sqrt{\mathcal N})$ queries with no logarithmic overhead and underlies main-text Theorem~11(ii); the modern picture of which faulty-oracle models preserve a quantum advantage is given in \citesupp{S-Lolck2024}: even in the Regev--Schiff model a roughly quadratic robustification blow-up preserves any super-cubic advantage, while the quadratic advantage of search does not survive.

\begin{figure}[!htbp]
\centering
\includegraphics[width=\textwidth]{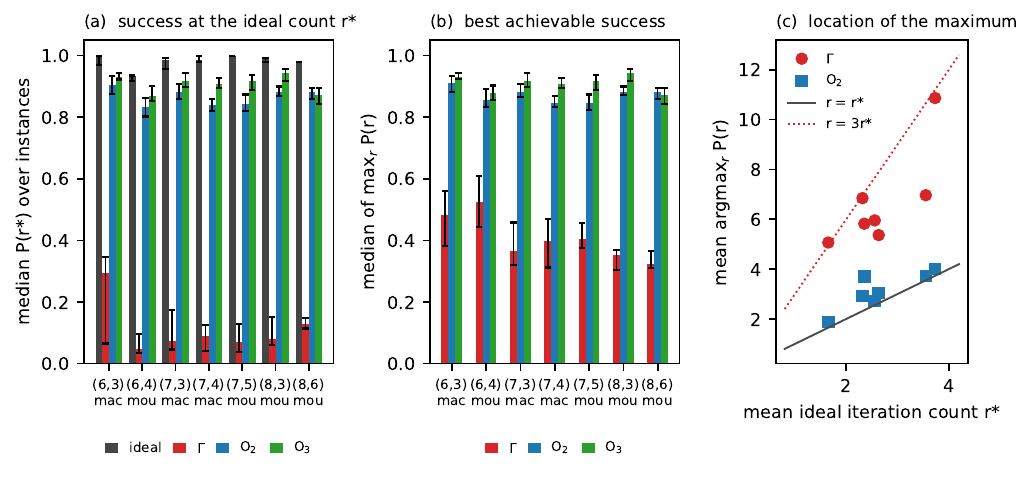}
\caption{\textbf{Aggregated benchmarks on induced subgraphs of the two brain networks} (exact probabilities; instances of the main-text set, full vertex sets). Configurations $(n,k)$ with host graph (mac: macaque cortex, mou: mouse retina); the number of instances per bar is given in main-text Table~2; $r^{*}=\lfloor\pi/(4\theta)\rfloor$ with $\sin^{2}\theta=\mathcal M/\mathcal N$. \textbf{a}, Median success probability at $r^{*}$ for the ideal oracle, plain $\Gam$, $\mathcal O_{2}$ and $\mathcal O_{3}$; error bars are inter-quartile ranges. \textbf{b}, Median of the best success probability over $r$. \textbf{c}, Mean iteration count at which the maximum occurs versus mean $r^{*}$: plain $\Gam$ peaks at about $3$--$5\,r^{*}$, the filtered oracle at $r^{*}$ or its neighbor.}
\label{S:fig:aggregated}
\end{figure}

\paragraph{The cost of the filter.} The filter is not free. $\mathcal{O}_{2}$ applies $Q$ six times (three in $\mathrm{QPE}_{2}$ and three in its inverse), so one filtered query costs $6.98$ times the depth and $6.71$ times the non-Clifford count of one $\Gam$ query once the controlled reflections and the estimation register are compiled: $6\,207$ transpiled layers, $4\,519$ CX gates and $4\,766$ non-Clifford gates at $N=10$, against $889$ and $710$, for the explicitly built circuit (main-text Fig.~2), and two more qubits; $\mathcal{O}_{3}$ costs $14$ applications of $Q$. What is bought is a known iteration count and a guarantee: with $\mathcal{O}_{2}$ the optimum returns to the ideal $\lfloor\tfrac{\pi}{4}\sqrt{\mathcal N/\mathcal M}\rfloor$ and the success probability there is within $0.03$--$0.15$ of the ideal oracle's, whereas $\Gam$ alone needs $3$--$5$ times as many iterations to reach a maximum of only $0.34$--$0.55$ (Fig.~\ref{S:fig:aggregated}). Counting the oracle depth spent per successful run, the two are comparable (about $12$ applications of $\Gam$ per success for $\Gam$ alone against about $11$ $\Gam$-equivalents for $\mathcal{O}_{2}$ at $r^{*}\approx1.5$), so the filter costs essentially nothing in total depth while removing the heuristic character. The worst-case bound $4r\sqrt{\varepsilon}$ of main-text Theorem~11(i) is loose at small $r$ (it exceeds $1$ for $r=1$); the observed losses are much smaller because $|a(H)|$ is typically far below $5/8$ (mean $0.09$--$0.57$, Fig.~\ref{S:fig:theory}c) and $\varepsilon_{2}(a)$ is correspondingly small ($0.048$ at $|a|=0.25$).

\paragraph{Aggregated benchmarks.} Throughout this note the $s\ge2$ columns are the guaranteed regime of main-text Theorem~11, on which the paper's claims rest; the $s=1$ columns are the illustrative half-cost variant. Fig.~\ref{S:fig:aggregated} visualizes main-text Table~2 (medians with inter-quartile ranges over the pairwise non-isomorphic instances of the main-text set) together with the location of the maximum; the plain-oracle data for all configurations are in Table~\ref{S:tab:gammatraj}. The numerical evidence has clear limits, which we state so that the theorems and the simulations are not confused. The instances are small ($n\le8$, at most $28$ qubits), $29$--$60$ pairwise non-isomorphic subgraphs per configuration; the host graphs are two specific brain networks, chosen because their induced subgraphs have the degree heterogeneity and clique structure of real data rather than of Erd\H{o}s--R\'enyi graphs; and the asymptotic statements rest on Theorems~8, 10 and~11, not on main-text Table~2. The simulations are noiseless. One $\Gam$ query at $n=6$--$8$ already contains $680$--$840$ CX gates after transpilation and a filtered query about six times as many, so these circuits are far beyond present unencoded hardware; at these sizes the exact counting oracles are cheaper than $\mathcal O_{2}$ in every metric, and the non-Clifford advantage of the filtered oracle only appears for $n\gtrsim30$ (main-text Fig.~2). Finally, the $\mathcal O_{3}$ column shows that a third estimation bit buys $0.01$--$0.08$ in success probability for $2.3$ times the cost of $\mathcal O_{2}$, so $t=2$ is the sensible operating point.

\paragraph{Summary of the exact simulations.} Fig.~\ref{S:fig:gammaSI} shows a representative instance ($n=7$, $k=3$, one triangle, $\mathcal{N}=35$, ideal optimum $r^{*}=4$). With $\Gam$ alone the trajectory is not a sinusoid: it rises slowly, peaks at $0.30$ at $r=12$, about three times the ideal iteration count, and collapses to $0.008$ at $r=15$, an aggravated form of the over-shooting known from Grover search \citesupp{S-Brassard1997}, because leaked amplitude is never recovered. With $\mathcal{O}_{2}$ the success probability at $r=4$ is $0.91$, and with $\mathcal{O}_{3}$ it is $0.96$, against $0.998$ for the ideal oracle; both follow the ideal curve over three full periods. Aggregated over instances (main-text Table~2; Supplementary Note~9), the median of $P(r^{*})$ for $\Gam$ alone is $0.05$--$0.29$ and its maximum over $r$ is $0.32$--$0.52$, reached at $2.1$--$3.8$ times $r^{*}$, whereas the filtered oracle in its guaranteed regime (two and three estimation registers, the regime on which the claims of this paper rest; main-text Theorem~11) reaches an observed $P(r^{*})$ of $0.92$--$0.99$ and $0.93$--$1.00$ against guaranteed $0.63$--$0.91$ and $0.92$--$0.99$ on the same instances. The illustrative single-register variant, at half the filter cost, reaches $0.83$--$0.90$ at $r^{*}$ and $0.85$--$0.91$ at its maximum, which lies at $r^{*}$ or its neighbor, against $0.94$--$1.00$ for the ideal oracle; $\mathcal{O}_{3}$ at $s=1$ reaches $0.87$--$0.95$. The inter-quartile ranges in main-text Table~2 show that these are properties of the configurations, not of a few instances, and the samples are uniform on the guaranteed regime: the four cheapest rows carry two- and three-register runs on every instance up to a uniform $30$ ($29$ on the $(8,6)$ row, which has only $29$), and on a cluster $s=2$ and $s=3$ were completed on every instance: $288$ runs each on the seven main-protocol configurations, $312$ with the two size-ceiling rows. Bootstrap $95\%$ confidence intervals on every median are listed in Supplementary Table~S22; on the columns with at least $29$ instances no half-width exceeds $\pm0.033$. Two further rows, with $12$ instances each, probe the size ceiling beyond the $n\le8$ protocol at $s=1$ ($(10,6)$ and $(12,10)$, the deepest amplification in the data set, $r^{*}$ up to $11$), so that $312$ instances enter in total: the single-register oracle shows no width dependence, with the full account in Supplementary Note~13.

\paragraph{Larger samples.} Raising the sampling target from $60$ to $200$ pairwise non-isomorphic instances (same seed; the first $60$ instances of every configuration are reproduced exactly) grows macaque $(6,3)$ to $109$ instances and macaque $(7,3)$, macaque $(7,4)$ and mouse $(8,3)$ to $200$, the other configurations being exhausted at their article counts. With $s=2$ and $s=3$ run on every instance the ranges of main-text Table~2 are unchanged (guaranteed $0.63$--$0.91$ against observed $0.92$--$0.99$ for $s=2$; $0.92$--$0.99$ against $0.93$--$1.00$ for $s=3$). A further size-ceiling configuration, mouse $(11,6)$ with $12$ instances, reaches a median $P(r^{*})=0.95$ with $s=2$, where the $s=2$ bound is vacuous.

\FloatBarrier
\section*{Supplementary Note 10: comparison across models and further directions}\label{S:note:10}

\paragraph{Trade-offs and implications relative to previous work.} Table~\ref{S:tab:models} places the results among the alternatives. The value of a compiled oracle depends entirely on the resources one is prepared to assume: in the query model the oracle is free and our constructions are invisible; in the QRAG model an adjacency bit can be read coherently in $O(1)$ time and a plain Grover search over triples already runs in $\Otil(n^{3/2})$ time \citesupp{S-Allcock2023}, which no compiled oracle can match; in the hard-wired circuit model, the model of every explicit quantum clique circuit published so far \citesupp{S-Chang2018,S-Metwalli2020}, the per-query depth sets the coherence time of a run and the per-query non-Clifford count sets its fault-tolerant cost. The edge-colored schedule reduces the depth of every adjacency-type oracle from $\Theta(n^{2})$ to $\Theta(n)$ at $1.5$--$6$ times the width, exact counting oracles included; only the graph-state oracle reduces the non-Clifford count from $\Theta(m)$ to $\Theta(n)$, and there we trade exactness for a one-sided error of at most $2/27$, which the bounded-error search of \citesupp{S-Hoyer2003} removes asymptotically at no cost in query count. Two implications follow. For fault-tolerant execution, where Clifford gates are cheap and every $T$ gate consumes a distilled magic state \citesupp{S-Bravyi2005,S-Fowler2012,S-Litinski2019}, the non-Clifford part of $\Gam$ is the same circuit for every graph: the graph enters through the Clifford unitary $A$, and a factor-$n$ reduction of the magic-state budget at constant depth is the relevant improvement. For hardware whose native resource is the graph state (measurement-based, photonic and neutral-atom platforms \citesupp{S-Raussendorf2001,S-Hein2006,S-Bartolucci2023,S-Bluvstein2022}), $A$ is a state preparation rather than a circuit, the graph-independent blocks $\IP$ and $S_{0}$ are the only operations that require non-Clifford resources, and the depth accounting of this paper becomes a statement about the number of entangling rounds, which is where the analysis in terms of stabilizer overlaps is likely to be most useful.

\begin{table}[!htbp]
\centering\small
\caption{\textbf{Approaches to clique and triangle detection and the resources they assume.} ``Oracle depth'' is the total depth spent in oracle calls; $\mathcal N=\binom nk$; for triangle finding $k=3$; $b=\lceil\log_{2}k\rceil$.}
\label{S:tab:models}
\resizebox{\textwidth}{!}{\begin{tabular}{@{}llllll@{}}
\toprule
approach & model & per-query cost (depth / width / non-Clifford) & total & guarantee & remark\\
\midrule
matrix multiplication \citesupp{S-Itai1978,S-Alman2021} & classical RAM & -- & $O(n^{2.373})$ time & exact & triangles; $O(n^{\omega k/3})$ for $k$-cliques \citesupp{S-Nesetril1985}\\
Le Gall's quantum walk \citesupp{S-LeGall2014} & query model & free & $\Otil(n^{5/4})$ queries & bounded error & non-query cost open\\
Grover over triples with QRAG \citesupp{S-Grover1996,S-Buhrman2022} & QRAG (random access) & $O(1)$ & $\Otil(n^{3/2})$ time & bounded error & needs random-access hardware\\
Grover with the oracle of \citesupp{S-Metwalli2020} & circuit, graph hard-wired & $\Theta(n^{2})$ / $n+O(\log k)$ / $\Theta(m\log k)$ & $\Theta(n^{2}\sqrt{\mathcal N})$ oracle depth & exact & previous state of the art\\
Grover with the parallel counting oracle (this work) & circuit, graph hard-wired & $\Theta(n)$ / $n(2+b)$ / $\Theta(m\log k)$ & $\Theta(n\sqrt{\mathcal N})$ oracle depth & exact & main-text Proposition~6\\
Grover with $\mathcal{O}_{2}$ (this work) & circuit, graph hard-wired & $\Theta(n)$ / $2(n+q)+4$ / $\Theta(n)$ & $\Theta(n\sqrt{\mathcal N})$ oracle depth, $\Theta(n\sqrt{\mathcal N})$ non-Clifford & error $\le1/3$ & Theorem~11\\
Le Gall + edge oracle (this work) & circuit, graph hard-wired & $\Theta(\chi'(G))$ / $\approx2n$ / $\Theta(m)$ & $\Otil(n^{9/4})$ oracle depth & bounded error & Corollary~\ref{S:cor:triangle}; time open\\
\bottomrule
\end{tabular}}
\end{table}

\paragraph{Future directions.} Several questions are opened rather than closed by this work. (i) \emph{Lowering the worst-case overlap.} The extremal graphs of main-text Theorem~8 are cliques minus one edge and stars, with $|a|=\tfrac12\pm2^{-(K-1)/2}$; a test state supported on weight-two subsets only would give the exact gap $2/\binom K2$ and needs no dummy vertices, but requires a controlled Dicke-state preparation inside the oracle, and it is not known whether a cheap Input Preparator with a smaller worst case exists. (ii) \emph{Exactness versus non-Clifford cost.} main-text Proposition~6 gives an exact oracle at depth and width $\Theta(n)$ for fixed $k$; for $k=\omega(1)$ its width grows as $n\log k$ and it is open whether both can be linear. More interesting is the other axis: the counting argument of main-text Proposition~4 bounds all gates, not the non-Clifford ones, and $\Gam$ shows that the graph can be carried entirely by Clifford gates; whether an \emph{exact} clique oracle can use $o(m)$ non-Clifford gates is open. (iii) \emph{Triangle finding as a time bound.} If the non-query operations of Le Gall's walk can be implemented in polylogarithmic depth per step, Corollary~\ref{S:cor:triangle} becomes an $\Otil(n^{9/4})$ time bound in the hard-wired circuit model; the same question determines whether the Õ$(n^{3/2})$ QRAG bound can be matched without random access. (iv) \emph{Other subgraph patterns.} The reduction of the oracle overlap to Gauss sums of $\mathbb Z_{4}$-valued quadratic forms is not specific to cliques: replacing the complete pattern by another graph $P$ changes the quadratic form and the rank conditions, and the same technique should yield exact error bounds for graph-state tests of induced-subgraph isomorphism. (v) \emph{Cost in fault-tolerant and hardware-native settings.} main-text Proposition~12 counts non-Clifford gates; a full space-time-volume estimate with lattice surgery and magic-state factories \citesupp{S-Litinski2019,S-Beverland2022}, and a measurement-based formulation of $A$ and of the controlled reflection $S_{0}$, would quantify the advantage on the platforms mentioned above; a noise study on small instances ($n=4$--$5$, $18$--$20$ qubits) would determine what error rates the filtered oracle tolerates. (vi) \emph{Algorithmic refinements.} Fixed-point amplitude amplification \citesupp{S-Yoder2014} and amplitude estimation \citesupp{S-Brassard2002} can replace the exponential search for unknown $\mathcal M$; the one-sidedness of the filter (cliques are never missed) may allow the classical verification step to be dropped in applications that only need a witness with bounded error. (vii) \emph{Larger simulations.} The block structure exploited by our simulator scales with $\binom nk 2^{N}$ rather than $4^{N}$ and should allow exact trajectories up to $n\approx12$ on a workstation, which would test the $r^{*}$ scaling directly rather than at $r^{*}\le3.5$.

\FloatBarrier
\section*{Supplementary Note 11: guarantees versus experiments for $s=1,2,3$ estimation registers}\label{S:note:11}

\paragraph{The simulator.} With $s$ estimation registers of $t$ qubits the filtered oracle $\mathcal O_{t}^{(s)}$ performs $s$ independent phase estimations of $Q=\Gam S_{0}$, flips the phase when all $s$ estimates are zero, and undoes the estimations in reverse order (main-text Theorem~10, last sentence). Script \texttt{18} simulates it exactly with $L^{s}=4^{s}$ branches of the block state of Supplementary Note~7; for $s=1$ it reproduces script \texttt{09} bit for bit, and on $\ket{y}\ket{0}\ket{0\cdots0}$ the all-zero component equals $1-2\,\varepsilon_{t}(a_{y})^{s}$ to $10^{-13}$ for $(t,s)\in\{(2,1),(2,2),(2,3),(3,2)\}$, which verifies the ``$s$ copies'' statement of Theorem~10 at circuit level. The cost per iteration grows as $s\,4^{s-1}$ applications of $Q$ per branch, so $s=2$ was run on the first $12$ instances of the four cheaper configurations and on $2$--$4$ instances of the others, and $s=3$ on the first $4$ instances of the cheaper configurations (script \texttt{19}); $r\le2r^{*}+4$ for $s=2$ and $r\le r^{*}+2$ for $s=3$.

\paragraph{Observed against guaranteed.} The constant of main-text Theorem~11(ii) is not made explicit in the analysis of \citesupp{S-Hoyer2003}, and its iteration schedule (amplitude amplification interleaved with error reduction) differs from the Grover count $r^{*}$ simulated here, which is why the $s=1$ columns are covered by (ii) only asymptotically. Table~\ref{S:tab:guar} lists, per configuration, the observed median $P(r^{*})$ for $s=1,2,3$ (for $s=1$ also restricted to the instances on which $s=2$ was run, to make the comparison like for like), the secondary comparisons ($\mathcal O_{3}$, and the best success over $r$ of $\Gam$ and $\mathcal O_{2}$), the lower bound of main-text Theorem~11(i), $P_{\mathrm{ideal}}(r^{*})-4r^{*}\varepsilon^{s/2}$, evaluated with the worst-case constant $\varepsilon=2/27$ and with the instance's own $\varepsilon_{\max}=\max_{y\notin\text{cliques}}\varepsilon_{2}(a_{y})$, and the loss analysis; main-text Table~2 instead uses the better of Theorem~11(i) and the first-order bound of Theorem~11(iii), $P_{\mathrm{ideal}}-4\varepsilon^{s}r^{2}$, on the same instances as each observed column. Two facts stand out. First, the worst-case bound is vacuous for $s=1$ in every row and non-vacuous for $s=3$ at every $r^{*}$ that occurs and for $s=2$ on every main-protocol configuration (it becomes vacuous or weak at the size-ceiling rows, $r^{*}$ up to $11$); the observed values exceed it by $0.36--$1.19$ ($s=2$) and $0.11--$0.32$ ($s=3$). Second, the loss actually incurred at $r^{*}$ for $s=1$ is first order in the error: its median is $0.6--1.6$ times $r^{*}\bar\varepsilon$ (with $\bar\varepsilon$ the mean of $\varepsilon_{2}(a_{y})$ over the non-clique candidates, $\bar\varepsilon=0.022--0.046$ across configurations) but only $0.10--0.21$ times $r^{*}\sqrt{\varepsilon_{\max}}$, the scale at which the hybrid argument charges it. Theorem~11(iii) supplies a probability-level bound that is first order in $\varepsilon$, $P(r)\ge P_{\mathrm{ideal}}(r)-4\varepsilon^{s}r^{2}$, proved by a leakage-and-return accounting in Supplementary Note~14 and verified on every recorded sweep of every instance ($5\,588$ (instance, $s$, $r$) points with the instance-wise $\varepsilon_{\max}$: no violation; minimum slack $0.049$, $0.0009$ and below $10^{-4}$ for $s=1,2,3$; for $s=2,3$ the bound is attained to within $10^{-3}$, so the functional form is right). It makes every $s=3$ guarantee of main-text Table~2, and every $s=2$ guarantee of its main-protocol rows, non-vacuous and covers $s=1$ at $r=1$ (worst-case constant) and $r\le2$ (instance-wise). What it does not capture is the observed linearity in $r$: the quadratic term comes from letting the leaked amplitude return coherently at every later iteration, and the data show it does not.

\begin{conjecture}[First-order loss, linear in the iteration count]\label{S:conj:linear}
For $\mathcal O_{2}^{(1)}$ on every admissible instance and every $r\ge1$, $P(r)\ge P_{\mathrm{ideal}}(r)-4\,r\,\bar\varepsilon$, where $\bar\varepsilon$ is the mean of $\varepsilon_{2}(a_{y})$ over the non-clique candidates.
\end{conjecture} The constant has been tested beyond the recorded sweeps: the largest ratio $\mathrm{loss}/(r\bar\varepsilon)$ is $3.56$ over all recorded $s=1$ sweeps of the $312$ instances, $3.67$ over sweeps extended to $r=4r^{*}+10$ on the same instances (macaque $(6,3)$, $r=13$), and $3.87$ over $10\,000$ random clique-bearing graphs with $n=6$--$9$, half of them with a planted near-clique ($n=8$, $k=3$, $r^{*}=1$, $r=13$; script \texttt{cluster/stress\_s6.py}); all are below $4$, and part (iii) of main-text Theorem~11 holds on every one of these points. Because the loss never exceeds $1$, the ratio is at most $1/(r\bar\varepsilon)$ and can exceed $4$ only for $r<1/(4\bar\varepsilon)$; every sweep above extends past that point, so on all $10\,312$ graphs the constant $4$ holds at every $r$, not only at the iteration counts that were swept.

The constant is verified: over all $312$ instances and every recorded iteration count ($r\le2r^{*}+4$ on the original rows, $r^{*}+1$ on the enlarged uniform samples and $r^{*}+2$ on the $n=10,12$ rows, whose sweeps reach $r=13$), the largest observed value of $[P_{\mathrm{ideal}}(r)-P(r)]/(r\bar\varepsilon)$ is $3.56$ (script \texttt{24}), with medians $0.6$--$1.6$ at $r^{*}$; proving the conjecture, i.e.\ removing the quadratic $r$-dependence of Theorem~11(iii), is the open problem left by this analysis.

\begin{table}[!htbp]
\centering\scriptsize
\caption{Observed and guaranteed success at $r^{*}$ (medians over instances, inter-quartile ranges in brackets). $s=2$ and $s=3$ were run on every instance. ``guar.'': $P_{\mathrm{ideal}}(r^{*})-4r^{*}(2/27)^{s/2}$; ``guar.\ inst.'': the same with the instance's $\varepsilon_{\max}$. Last three columns: median $\bar\varepsilon$ and $\varepsilon_{\max}$ over instances and the ratio of the observed $s=1$ loss to $r^{*}\bar\varepsilon$.}
\label{S:tab:guar}
\resizebox{\textwidth}{!}{\begin{tabular}{@{}cclccccccccccccccc@{}}
\toprule
$n$&$k$&host&$\#$&$s{=}1$&$s{=}2$&$s{=}3$&$\mathcal O_{3}$&$\Gam$ $\max_r$&$\mathcal O_{2}$ $\max_r$&guar.\ $s{=}1$&guar.\ $s{=}2$&guar.\ $s{=}3$&guar.\ inst.\ $s{=}1$&guar.\ inst.\ $s{=}2$&$\bar\varepsilon$&$\varepsilon_{\max}$&loss$/(r^{*}\bar\varepsilon)$\\
\midrule
6&3&macaque&60&0.90 [0.87, 0.93]&0.99 [0.97, 1.00]&1.00 [0.97, 1.00]&0.93 [0.92, 0.94]&0.48&0.91&$<0$&0.57&0.84&$<0$&0.67&0.0223&0.0479&1.47 [1.20, 1.62]\\
6&4&mouse&34&0.83 [0.80, 0.86]&0.92 [0.91, 0.93]&0.93 [0.92, 0.93]&0.87 [0.85, 0.90]&0.52&0.86&$<0$&0.05&0.69&$<0$&0.19&0.0339&0.0726&1.11 [0.82, 1.59]\\
7&3&macaque&60&0.88 [0.86, 0.91]&0.98 [0.95, 0.99]&0.99 [0.96, 0.99]&0.92 [0.90, 0.94]&0.36&0.88&$<0$&0.28&0.74&$<0$&0.49&0.0255&0.0479&1.59 [1.06, 2.21]\\
7&4&macaque&30&0.84 [0.82, 0.86]&0.98 [0.97, 0.98]&0.99 [0.98, 1.00]&0.91 [0.89, 0.93]&0.40&0.85&$<0$&0.10&0.74&$<0$&0.17&0.0349&0.0726&1.51 [1.14, 2.06]\\
7&5&mouse&45&0.84 [0.82, 0.87]&0.99 [0.98, 0.99]&1.00 [1.00, 1.00]&0.92 [0.89, 0.94]&0.40&0.85&$<0$&0.11&0.76&$<0$&0.25&0.0456&0.0726&1.27 [1.11, 1.94]\\
8&3&mouse&30&0.88 [0.87, 0.90]&0.99 [0.98, 0.99]&0.99 [0.98, 0.99]&0.94 [0.92, 0.96]&0.35&0.88&$<0$&$<0$&0.66&$<0$&0.21&0.0307&0.0479&0.73 [0.65, 1.22]\\
8&6&mouse&29&0.88 [0.86, 0.90]&0.97 [0.97, 0.97]&0.98 [0.98, 0.98]&0.87 [0.84, 0.90]&0.32&0.88&$<0$&$<0$&0.66&$<0$&$<0$&0.0454&0.0726&0.57 [0.50, 1.16]\\
10&6&mouse&12&0.64 [0.62, 0.66]&0.96 [0.96, 0.96]&1.00 [1.00, 1.00]&--&0.11&0.65&$<0$&$<0$&0.11&$<0$&$<0$&0.0339&0.0726&1.10 [0.98, 1.12]\\
12&10&macaque&12&0.87 [0.86, 0.89]&0.99 [0.99, 0.99]&1.00 [1.00, 1.00]&--&0.31&0.87&$<0$&$<0$&0.52&$<0$&$<0$&0.0507&0.0690&0.45 [0.43, 0.46]\\
\bottomrule
\end{tabular}}
\end{table}

\paragraph{Error models.} The distinction between error models is load-bearing, so we state it precisely. In the faulty-oracle model of Regev and Schiff~\citesupp{S-Regev2008} each call silently applies the identity with some fixed probability (a classical mixture whose hidden failure record decoheres the algorithm) and their hybrid argument over that record forces $\Omega(\mathcal N)$ queries: no speed-up survives, and no amount of repetition repairs it. The deviation of $\Ot$ is not that channel. It is one fixed, known unitary whose action on a candidate is $\sigma_{y}(1-2\varepsilon_{y}^{\,s})$ on the search sector plus a coherent amplitude of norm at most $2\varepsilon^{s/2}$ on estimation states orthogonal to $\ket0$: flagged, amplitude-bounded, one-sided ($\varepsilon_{y}=0$ on cliques), so no environment records a failure, the Regev--Schiff hybrid has nothing to average over, and the leaked branch remains available to interference, which is exactly what the phase-estimation filter exploits. This is the bounded-error model of H{\o}yer et al.~\citesupp{S-Hoyer2003} that main-text Theorem~11(ii) invokes, and it is why filtering is admissible here while impossible there (Supplementary Note~11; the modern picture of which faulty-oracle models retain a quantum advantage is given by Lolck et al.~\citesupp{S-Lolck2024}). Equation~main-text equation~(7) is verified on the compiled circuits to $8\times10^{-14}$ (main-text Sect.~8.1).

\paragraph{Which guarantee covers which experiment.} Which statement covers which experiment is spelled out once, and the division of labor is deliberate: \emph{the rigorous basis of every claim in this paper is the $s\ge2$ regime}, and the case for the algorithm rests on it. With $s=3$ the better of (i) and (iii) is non-vacuous at every $r^{*}$ that occurs, and with $s=2$ on every configuration of the main protocol, and main-text Table~2 lists it next to the observed median $P(r^{*})$ on the same instances: guaranteed $0.63$--$0.91$ against observed $0.92$--$0.99$ for $s=2$, and $0.92$--$0.99$ against $0.93$--$1.00$ for $s=3$, where the bound is attained to numerical precision (minimum slack below $10^{-4}$ over all recorded sweeps; script \texttt{24}); the price of entering this regime is a factor $s$ in filter cost and $2s-2$ additional qubits, and nothing else. The single-register variant, the cheapest choice and the one used for most of main-text Table~2, sits outside it: (i) is vacuous for every $r\ge1$, (ii) applies only asymptotically (its constant is not explicit in H{\o}yer et al.~\citesupp{S-Hoyer2003} and its schedule differs from the Grover count $r^{*}$ that we simulate), and part (iii), the first probability-level statement for $s=1$, is non-vacuous at $r=1$ with the worst-case constant and up to $r=2$ with the instance's own $\varepsilon_{\max}$; beyond the smallest iteration counts the $s=1$ columns are therefore reported as an \emph{illustration} of the same behavior at half the filter cost, not as evidence for a theorem.

\FloatBarrier
\section*{Supplementary Note 12: the log-depth compilation of $\Gam$ and the fault-tolerant accounting}\label{S:note:12}

Main-text Proposition~13 compiles $\Gam$ to depth $O(\chi'(G)+\log n)$ at $3N+1$ qubits by (a) replacing the dummy CZ gates of $A$ by an in-place CNOT parity tree over the real $\inp$ qubits, $q$ CZ gates on its root and $\binom q2$ CZ gates among the dummies, (b) computing the mod-4 weight in $\IP$ with a tree of in-place two-bit adders whose carries stay in $\lceil N/2\rceil$ ancillas until $\IP^{\dagger}$ cancels them, (c) feeding the $N$ Toffolis of step~3 of $\IP$ from $\rem_{1}$ and $\lfloor N/2\rfloor-1$ CNOT copies of it, and (d) realizing $S_{0}$ (and its controlled version in $\Ot$) as an AND tree on $N-1$ ancillas; the proof is in Supplementary Note~14. Script \texttt{21} builds the circuit, verifies by statevector that $\Gam$ in this compilation is the same unitary as the block-by-block circuit of script \texttt{05} on random superpositions of the search register ($10^{-13}$, no leakage into the ancillas, $n=3,4$), and produces Table~\ref{S:tab:logdepth}. The filtered oracle, which carries the $s\ge2$ guarantees of main-text Theorem~11, is verified directly by scripts \texttt{27} and \texttt{27proj}, and we state the achieved scope exactly. First, a scan of the circuit proves that every gate touches the $\idx$ register only in control positions ($384$ gates), so $\mathcal O_{2}^{(s)}$ is block-diagonal over the $\idx$ basis by construction: it acts on $\ket y\ket0$ as $\ket y\otimes U_{y}\ket0$, where $U_{y}$ is the $19$-qubit reduced circuit obtained by resolving the $\idx$ controls at $y$, and its search-sector action is its diagonal. Second, at $24$ qubits ($n=3$, $s=1$) the full-width log-depth $\mathcal O_{2}^{(1)}$ equals a block-by-block $\mathcal O_{2}^{(1)}$ assembled from the blocks of script \texttt{05}, on the complete zero-ancilla sector and $10^{5}$ uniformly sampled amplitudes of a random product superposition of $\idx$, to $2\times10^{-15}$, with no amplitude outside the returned-ancilla sector (at most $0$) and with diagonal equal to the $\sigma_{y}(1-2\varepsilon_{2}(a_{y}))$ of main-text Theorem~10 to $1\times10^{-14}$. Third, with block-diagonality proven, the diagonal of every candidate block was evaluated exactly on the reduced circuits: for the full two-register stack ($s=2$; the $26$-qubit $\mathcal O_{2}^{(2)}$) over every candidate of all $8$ graphs on three vertices, including the extremal non-clique overlap $a=-5/8$ of Theorem~8, with maximum deviation from $\sigma_{y}(1-2\varepsilon_{2}^{\,2})$ of $5\times10^{-14}$ (the extremal candidate: $2\times10^{-14}$); and for the $27$-qubit $\mathcal O_{2}^{(1)}$ at $n=4$ over every candidate of all $64$ graphs on four vertices ($256$ diagonals, maximum deviation $3\times10^{-14}$). Full-width statevector runs stop at $24$ qubits on our hardware (a $2^{26}$- or $2^{27}$-amplitude state leaves no memory for an equivalence copy); beyond that width the correctness of the log-depth $\mathcal O_{2}^{(s)}$ rests on the proven block structure, the verified blocks and their composition, and we label it so: the resource counts of Tables~\ref{S:tab:logdepth} and~\ref{S:tab:tdepth} at larger $n$ are counts on the constructed circuit, not simulated ones. On $K_{n}$ the depth of $\Gam$ falls from $89N-1$ to $\approx13.7n+403$ and that of $\mathcal O_{2}$ from $622N-13$ to $\approx91.6n+2\,847$; on sparse graphs the depth follows the number of color classes of $G$ at $\approx9$ layers per class plus $\approx500$ layers for the logarithmic-depth blocks, whose constant is set by the four Input Preparators (controlled-Hadamards, adder trees, fan-outs) and the AND tree. The $T$-count rises from $72N-10$ to $78N-14$ because the AND tree costs $2(N-1)$ Toffolis where the dirty-ancilla multi-controlled gate cost fewer $T$ gates, and every non-Clifford rotation remains a $T$ gate. The comparison of the main text assumes that magic-state distillation dominates the space-time volume; the $\Theta(n^{2})$ Clifford gates of $A$ still cost $O(d)$ code cycles each in lattice surgery \citesupp{S-Litinski2019} but consume no magic states, and on graph-state platforms they are absorbed into the resource state. Table~\ref{S:tab:tdepth} gives the Toffoli, controlled-Hadamard and $T$ counts, the $T$-count with measurement-based un-computation of every Toffoli pair, and the $T$-depth as scheduled for all oracles: the $T$-depth of $\Gam$ in this compilation is $145$--$169$ for $n=6$--$20$ and $145$--$193$ for $n=6$--$40$ (logarithmic, because the CZ layers carry no $T$ gates), against $79$--$207$ for the parallel counter (one Toffoli layer per color class) and $412$--$5\,242$ for the sequential counter.

\paragraph{Comparison with the binary-encoded QAA oracle of Perriello \citesupp{S-Perriello2025}.} The closest external construction is the pure-QAA $k$-clique oracle of Perriello (CF~'25), which searches the same Dicke superposition over $\binom nk$ candidate sets but tests the clique predicate by binary arithmetic: a register of $n$ counters of $\lceil\log_{2}n\rceil$ qubits accumulates the selected row sums of the adjacency matrix through one controlled adder per edge, interlaced so that up to $n$ adders run in parallel, followed by comparators against $k-1$. From the leading terms of his Table~1, one $U_{f}$ costs $2n^{2}\lceil\log_{2}n\rceil$ Toffolis at depth $2n\lceil\log_{2}n\rceil$ and width $n(\lceil\log_{2}n\rceil+2)$, and the oracle is $U_{f}^{\dagger}R_{1^{\perp}}U_{f}$; with his own Clifford$+T$ conversions (Toffoli: four $T$ gates at $T$-depth one with one ancilla) and no rotations anywhere in the oracle; rotation synthesis enters only his diffusion, which both algorithms share as the same Dicke reflection. One oracle call at $n=20$ costs $8\,236$ Toffolis $=3.3\times10^{4}$ $T$ gates, $T$-depth $\approx440$ and $\approx160$ qubits, independent of $k$. One call of $\mathcal O_{2}^{(1)}$ costs $545N-162$ $T$ gates at $3N+3$ qubits, i.e.\ $1.13\times10^{4}$ $T$ and $66$ qubits at $(n,k)=(20,10)$ ($N=21$) and $1.29\times10^{4}$ $T$ and $75$ qubits at $(20,3)$ ($N=24$), with measured $T$-depth $1\,240$ at $n=20$ (Table~\ref{S:tab:tdepth}). The per-query comparison is therefore two-sided: his oracle is $2.8$ times shallower in $T$-depth and exact, ours $2.6$--$2.9$ times cheaper in $T$ gates at $2.1$--$2.4$ times fewer qubits, and $T$-depth$\times$width agree to within $16$--$32\%$. Because his non-Clifford count is $\Theta(m\log n)$ per query and ours $\Theta(n)$, the $T$-count ratio grows linearly in $n$ (already $2.9$ at $n=20$ on dense graphs), and extrapolating his $T$-depth $4n\lceil\log_{2}n\rceil+2n$ against the fit $13.7n+966$ to our measured values puts the $T$-depth crossover near $n\approx70$. Three caveats: his Table~1 states dominant terms only; our numbers are transpiled circuits; and plain depth is not comparable across the NCT and $\{R_{z},\sqrt X,X,\mathrm{CX}\}$ bases without re-transpilation. Script \texttt{26} computes both sides and the crossover.

\begin{table}[!htbp]
\centering\small
\caption{The log-depth compilation of $\Gam$ and $\mathcal O_{2}$ (main-text Proposition~13; script \texttt{21}; settings of Table~\ref{S:tab:nonclifford}). Left: complete graphs, $k=3$; every non-Clifford rotation is a $T$ gate ($T$-count $78N-14$ for $\Gam$, $545N-162$ for $\mathcal O_{2}$). Right: $\Gam$ on random graphs $G(n,p)$ with the rounds given by a greedy coloring of $G$ (``coloring'') or by the $K_{n}$ 1-factorization (``1-fact.''); the block-by-block compilation of Table~\ref{S:tab:nonclifford} costs $89N-1$ on every graph.}
\label{S:tab:logdepth}
\begin{minipage}[t]{0.55\textwidth}\centering
\resizebox{\textwidth}{!}{\begin{tabular}{@{}cclccccc@{}}
\toprule
$n$&$N$&oracle&qubits&depth&CX&gates&$T$\\
\midrule
6&10&$\Gam$&31&487&774&2\,097&766\\
6&10&$\mathcal O_{2}$&33&3\,421&5\,255&14\,064&5\,288\\
8&12&$\Gam$&37&505&952&2\,595&922\\
8&12&$\mathcal O_{2}$&39&3\,529&6\,455&17\,393&6\,378\\
10&14&$\Gam$&43&523&1\,138&3\,081&1\,078\\
10&14&$\mathcal O_{2}$&45&3\,637&7\,703&20\,650&7\,468\\
12&16&$\Gam$&49&561&1\,332&3\,595&1\,234\\
12&16&$\mathcal O_{2}$&51&3\,877&8\,999&24\,075&8\,558\\
16&20&$\Gam$&61&654&1\,744&4\,627&1\,546\\
16&20&$\mathcal O_{2}$&63&4\,561&11\,735&30\,949&10\,738\\
20&24&$\Gam$&73&688&2\,188&5\,691&1\,858\\
20&24&$\mathcal O_{2}$&75&4\,765&14\,663&38\,015&12\,918\\
\bottomrule
\end{tabular}}
\end{minipage}\hfill
\begin{minipage}[t]{0.43\textwidth}\centering
\resizebox{\textwidth}{!}{\begin{tabular}{@{}cccccccc@{}}
\toprule
$n$&$p$&$m$&$\Delta$&rounds&coloring&1-fact.&$T$\\
\midrule
12&0.15&8&3&3&493&503&1\,234\\
12&0.40&26&8&8&517&533&1\,234\\
12&1.00&66&11&11&561&561&1\,234\\
16&0.15&15&4&5&572&582&1\,546\\
16&0.40&47&8&10&592&600&1\,546\\
16&1.00&120&15&15&654&654&1\,546\\
20&0.15&39&6&7&578&600&1\,858\\
20&0.40&84&13&14&614&626&1\,858\\
20&1.00&190&19&19&688&688&1\,858\\
\bottomrule
\end{tabular}}
\end{minipage}
\end{table}

\begin{table}[!htbp]
\centering\scriptsize
\caption{Fault-tolerant accounting of the $k$-clique oracles on $K_{n}$, $k=3$ (script \texttt{22}; settings of Table~\ref{S:tab:nonclifford}). Toffoli = two-controlled X gates before transpilation; mcx = multi-controlled gates with at least three controls (synthesized by Qiskit; the block-by-block $S_{0}$ and the counter of the sequential oracle as compiled); CH = controlled-Hadamards; $T$ = odd multiples of $\pi/4$ after transpilation, which equals $7\,\mathrm{Toffoli}+2\,\mathrm{CH}$ when no mcx remains; $\pi/8$ = non-$T$ rotations; $T_{\mathrm{AND}}$ = $T$-count with a temporary logical AND \citesupp{S-Jones2013,S-Gidney2018} for every Toffoli pair ($2\,\mathrm{Toffoli}+2\,\mathrm{CH}$), given only when no mcx remains; $T$-depth = longest path of the transpiled circuit counting $T$ gates only, with the gate order kept (an upper bound on the $T$-depth). The $n$-dependent layers of $\Gam$ in the compilation of Proposition~13 are Clifford, so its $T$-depth is logarithmic, whereas that of the parallel counter grows with $\chi'(G)$ and that of the sequential counter with $m$.}
\label{S:tab:tdepth}
\resizebox{\textwidth}{!}{\begin{tabular}{@{}clccccccccc@{}}
\toprule
$n$&oracle&qubits&depth&Toffoli&mcx&CH&$T$&$\pi/8$&$T_{\mathrm{AND}}$&$T$-depth\\
\midrule
6&sequential counter, as compiled&9&1\,043&31&30&0&422&225&--&243\\
6&sequential counter, Toffoli-only&9&1\,150&121&0&0&847&0&242&412\\
6&parallel counter (Prop.~6)&35&216&94&0&0&658&0&188&79\\
6&$\Gam$, block by block&22&889&80&1&40&710&0&--&322\\
6&$\Gam$, Prop.~13&31&487&98&0&40&766&0&276&145\\
6&$\mathcal O_{2}$, block by block&24&6\,207&480&12&240&4\,766&0&--&2\,234\\
6&$\mathcal O_{2}$, Prop.~13&33&3\,421&696&0&240&5\,288&0&1\,872&1\,048\\
8&sequential counter, as compiled&11&1\,932&57&56&0&777&420&--&449\\
8&sequential counter, Toffoli-only&11&2\,136&225&0&0&1\,575&0&450&766\\
8&parallel counter (Prop.~6)&47&258&158&0&0&1\,106&0&316&96\\
8&$\Gam$, block by block&26&1\,067&96&1&48&854&0&--&386\\
8&$\Gam$, Prop.~13&37&505&118&0&48&922&0&332&145\\
8&$\mathcal O_{2}$, block by block&28&7\,451&576&12&288&5\,726&0&--&2\,678\\
8&$\mathcal O_{2}$, Prop.~13&39&3\,529&840&0&288&6\,378&0&2\,256&1\,048\\
10&sequential counter, as compiled&13&3\,097&91&90&0&1\,240&675&--&719\\
10&sequential counter, Toffoli-only&13&3\,434&361&0&0&2\,527&0&722&1\,232\\
10&parallel counter (Prop.~6)&59&319&238&0&0&1\,666&0&476&119\\
10&$\Gam$, block by block&30&1\,245&112&1&56&998&0&--&450\\
10&$\Gam$, Prop.~13&43&523&138&0&56&1\,078&0&388&145\\
10&$\mathcal O_{2}$, block by block&32&8\,695&672&12&336&6\,686&0&--&3\,122\\
10&$\mathcal O_{2}$, Prop.~13&45&3\,637&984&0&336&7\,468&0&2\,640&1\,048\\
12&sequential counter, as compiled&15&4\,538&133&132&0&1\,811&990&--&1\,053\\
12&sequential counter, Toffoli-only&15&5\,044&529&0&0&3\,703&0&1\,058&1\,810\\
12&parallel counter (Prop.~6)&71&359&334&0&0&2\,338&0&668&135\\
12&$\Gam$, block by block&34&1\,423&128&1&64&1\,142&0&--&514\\
12&$\Gam$, Prop.~13&49&561&158&0&64&1\,234&0&444&148\\
12&$\mathcal O_{2}$, block by block&36&9\,939&768&12&384&7\,646&0&--&3\,566\\
12&$\mathcal O_{2}$, Prop.~13&51&3\,877&1\,128&0&384&8\,558&0&3\,024&1\,072\\
16&sequential counter, as compiled&19&8\,248&241&240&0&3\,277&1\,800&--&1\,913\\
16&sequential counter, Toffoli-only&19&9\,200&961&0&0&6\,727&0&1\,922&3\,302\\
16&parallel counter (Prop.~6)&95&441&574&0&0&4\,018&0&1\,148&168\\
16&$\Gam$, block by block&42&1\,779&160&1&80&1\,430&0&--&642\\
16&$\Gam$, Prop.~13&61&654&198&0&80&1\,546&0&556&169\\
16&$\mathcal O_{2}$, block by block&44&12\,427&960&12&480&9\,566&0&--&4\,454\\
16&$\mathcal O_{2}$, Prop.~13&63&4\,561&1\,416&0&480&10\,738&0&3\,792&1\,240\\
20&sequential counter, as compiled&23&13\,062&381&380&0&5\,175&2\,850&--&3\,029\\
20&sequential counter, Toffoli-only&23&14\,604&1\,521&0&0&10\,647&0&3\,042&5\,242\\
20&parallel counter (Prop.~6)&119&542&878&0&0&6\,146&0&1\,756&207\\
20&$\Gam$, block by block&50&2\,135&192&1&96&1\,718&0&--&770\\
20&$\Gam$, Prop.~13&73&688&238&0&96&1\,858&0&668&169\\
20&$\mathcal O_{2}$, block by block&52&14\,915&1\,152&12&576&11\,486&0&--&5\,342\\
20&$\mathcal O_{2}$, Prop.~13&75&4\,765&1\,704&0&576&12\,918&0&4\,560&1\,240\\
\bottomrule
\end{tabular}}
\end{table}

\begin{table}[!htbp]
\centering\small
\caption{$K=7$, all $32\,767$ admissible non-cliques (script \texttt{23}): number of graphs, maximum of $|a|$, the parity-refined bound $2^{-r/2}+2^{-\lfloor r'/2\rfloor}$ and the values of $c(F)$ in each rank class $(r,r')$.}
\label{S:tab:rankclasses}
\begin{tabular}{@{}ccrccl@{}}
\toprule
$r$&$r'$&graphs&$\max|a|$&bound&$c(F)$\\
\midrule
2&5&80&0.2500&0.7500&-1.414, 0\\
2&6&256&0.6250&0.6250&-1.000, 0, +1.000\\
2&7&315&0.6250&0.6250&-1.414, 0, +1.414\\
4&3&16&0.2500&0.7500&0\\
4&4&320&0.5000&0.5000&-1.000, 0, +1.000\\
4&5&2400&0.5000&0.5000&-1.414, 0, +1.414\\
4&6&7680&0.3750&0.3750&-1.000, 0, +1.000\\
4&7&7812&0.3750&0.3750&-1.414, 0, +1.414\\
6&2&1&0.6250&0.6250&+1.000\\
6&3&15&0.1250&0.6250&0\\
6&4&300&0.3750&0.3750&-1.000, 0, +1.000\\
6&5&1860&0.1250&0.3750&0, +1.414\\
6&6&5952&0.2500&0.2500&-1.000, 0, +1.000\\
6&7&5760&0.2500&0.2500&-1.414, 0, +1.414\\
\bottomrule
\end{tabular}
\end{table}
\begin{table}[!htbp]
\centering\small
\caption{The rank-$2$ family, exact: maximum of $|a|$ over $F=K_{\alpha,\beta,\gamma}$ plus $\delta\ge1$ isolated vertices (all compositions of $K$), the maximizer, the conjectured value $\tfrac12+2^{-(K-1)/2}$, and $a(K_{K}-e)$.}
\label{S:tab:rank2}
\begin{tabular}{@{}cclccc@{}}
\toprule
$K$&compositions&$\max|a|$ (at $\alpha,\beta,\gamma,\delta$)&$\tfrac12+2^{-(K-1)/2}$&$a(K_{K}-e)$&equal\\
\midrule
7&16&$5/8$ at (1, 1, 0, 5)&$5/8$&$-5/8$&yes\\
11&56&$17/32$ at (3, 2, 0, 6)&$17/32$&$-15/32$&yes\\
15&132&$65/128$ at (1, 1, 0, 13)&$65/128$&$-65/128$&yes\\
19&255&$257/512$ at (3, 2, 0, 14)&$257/512$&$-255/512$&yes\\
23&435&$1025/2048$ at (1, 1, 0, 21)&$1025/2048$&$-1025/2048$&yes\\
27&683&$4097/8192$ at (3, 2, 0, 22)&$4097/8192$&$-4095/8192$&yes\\
31&1010&$16385/32768$ at (1, 1, 0, 29)&$16385/32768$&$-16385/32768$&yes\\
\bottomrule
\end{tabular}
\end{table}

\begin{table}[!htbp]
\centering\small
\caption{\textbf{Sample sizes and bootstrap confidence intervals for Table~2}. For every configuration: the number of pairwise non-isomorphic instances, the mean $r^{*}$, and the median $P(r^{*})$ of the filtered oracle with one, two and three estimation registers, each followed by the inter-quartile range where shown in Table~2 and by the bootstrap $95\%$ confidence interval of the median ($2\,000$ resamples). Run counts for $s=2,3$ are in parentheses; the two size-ceiling rows were run at $s=1$ only.}
\label{S:tab:m6stats}
\resizebox{\textwidth}{!}{\begin{tabular}{@{}cclcclll@{}}
\toprule
$n$&$k$&host&$\#$&$\langle r^{*}\rangle$&$\mathcal O_{2}^{(1)}$: med [IQR] [CI]&$\mathcal O_{2}^{(2)}$: med ($\#$) [CI]&$\mathcal O_{2}^{(3)}$: med ($\#$) [CI]\\
\midrule
6&3&macaque&60&1.6&0.90 [0.87, 0.93] [0.89, 0.93]&0.99 (60) [0.97, 0.99]&1.00 (60) [0.97, 1.00]\\
6&4&mouse&34&2.4&0.83 [0.80, 0.86] [0.81, 0.85]&0.92 (34) [0.92, 0.92]&0.93 (34) [0.93, 0.93]\\
7&3&macaque&60&2.3&0.88 [0.86, 0.91] [0.88, 0.90]&0.98 (60) [0.97, 0.98]&0.99 (60) [0.98, 0.99]\\
7&4&macaque&30&2.6&0.84 [0.82, 0.86] [0.83, 0.85]&0.98 (30) [0.97, 0.98]&0.99 (30) [0.98, 0.99]\\
7&5&mouse&45&2.6&0.84 [0.82, 0.87] [0.83, 0.86]&0.99 (45) [0.98, 0.99]&1.00 (45) [1.00, 1.00]\\
8&3&mouse&30&3.7&0.88 [0.87, 0.90] [0.87, 0.89]&0.99 (30) [0.98, 0.99]&0.99 (30) [0.98, 0.99]\\
8&6&mouse&29&3.6&0.88 [0.86, 0.90] [0.87, 0.89]&0.97 (29) [0.97, 0.97]&0.98 (29) [0.98, 0.98]\\
10&6&mouse&12&9.8&0.64 [0.62, 0.66] [0.62, 0.67]&0.96 (12) [0.96, 0.96]&1.00 (12) [0.99, 1.00]\\
12&10&macaque&12&5.5&0.87 [0.86, 0.89] [0.86, 0.89]&0.99 (12) [0.99, 0.99]&1.00 (12) [1.00, 1.00]\\
\bottomrule
\end{tabular}}
\end{table}

\paragraph{Verification of the compiled circuits.} The proof is in Supplementary Note~14; the compiled circuit was verified to be the same unitary as the block-by-block one to $10^{-13}$ on random superpositions of the search register (script \texttt{21}), and the filtered oracle is verified directly rather than by composition alone: a scan of the circuit proves it block-diagonal over the $\idx$ basis, the full-width $24$-qubit $\mathcal O_{2}^{(1)}$ equals its block-by-block form to $2\times10^{-15}$, and, block-diagonality being proven, the diagonal $\sigma_{y}(1-2\varepsilon_{2}^{\,s}(a_{y}))$ of main-text Theorem~10 is certified below $10^{-13}$ for every candidate block of the two-register stack ($s=2$, all graphs on three vertices, including the extremal $a=-5/8$) and of the $27$-qubit oracle at $n=4$ (all $64$ graphs); beyond these widths correctness rests on the verified blocks and their composition (script \texttt{27}; Supplementary Note~12). Fig.~\ref{S:fig:primitivesSI} and main-text Table~1 give the compiled numbers on the complete graph ($k=3$) for the sequential counting oracle of Metwalli et al.~\citesupp{S-Metwalli2020} built from its published description, the parallel counting oracle of main-text Proposition~6, and $\Gam$ and $\mathcal O_{2}$ in the compilation of main-text Proposition~13, all under the same transpiler settings (main-text Sect.~8.1; full data, including the $\Theta(n)$-depth compilation of $\Gam$ at $2N+2$ qubits, in Supplementary Tables~S6, S7 and~S18). On $K_{n}$ the depth of one $\Gam$ query is $487$--$688$ for $n=6$--$20$ ($933$ at $n=40$; $\approx13.7n+403$ over $n=6$--$40$, against $89N-1=89n+355$ block by block), and that of one $\mathcal O_{2}$ query $3\,421$--$4\,765$ ($6\,373$ at $n=40$; $\approx91.6n+2\,847$ over $n=6$--$40$); on random graphs the depth of $\Gam$ follows the number of color classes of $G$ ($578$ layers at $\Delta=6$ against $688$ on $K_{20}$; Supplementary Fig.~S15c and Supplementary Note~12).

\paragraph{The two-sided comparison.} The comparison is two-sided and we state it as such. In depth on $K_{n}$ the parallel counter ($\approx21.9n+92$ over $n=6$--$40$) and $\Gam$ are within a factor $2.3$ of each other up to $n=36$, and $\Gam$ is the shallower of the two from $n=40$ ($933$ against $965$), and $\mathcal O_{2}$ is shallower than the sequential counter from $n\approx11$ and about nine times deeper than the parallel one (four times asymptotically); in non-Clifford cost the order reverses, $\Gam$ below both counters from $n=8$ and $\mathcal O_{2}$ falling below the sequential counter from $n=24$ and below the parallel one just above $n=40$ (both measured on compiled circuits up to $n=40$); these crossovers are illustrative (a different Toffoli synthesis shifts them by a few vertices, not the $\Theta(n)$-against-$\Theta(n^{2})$ ordering), and beyond them the ratio grows linearly in $n$. The closest external comparison, the QRAM-free QAA oracle of Perriello~\citesupp{S-Perriello2025}, is exact and $2.8$ times shallower in $T$-depth per query at moderate $n$, while $\mathcal O_{2}$ is $2.9$ times cheaper in $T$ gates at $2.4$ times fewer qubits and carries the proven one-sided error; the full head-to-head, including the $T$-depth crossover near $n\approx70$, is in Supplementary Note~12. Neither family dominates the other: the counting oracles are the constructions of choice when width is the scarce resource (the sequential counter) or exactness is required at the smallest depth (the parallel counter), the graph-state oracle when non-Clifford gates are.

\FloatBarrier
\section*{Supplementary Note 13: complementary compiled details (the edge detector, the counting oracle and the size ceiling)}\label{S:note:13}

This note gives the compiled resource counts of the Part~I constructions and the single-register size-ceiling benchmarks.

\paragraph{The edge detector, compiled.} Transpiled to $\{R_{z},\sqrt{X},X,\mathrm{CX}\}$, the depth of the partitioned circuit on $K_{n}$ is exactly $32n-49$ for $n\ge8$, against $\approx3.9n^{2}$ for the naive circuit; on random graphs a greedy coloring of $G$ itself gives a further factor of about two over the $K_{n}$ layering, as main-text Lemma~3 predicts (Supplementary Note~7, Fig.~S8). Two qualifications apply: the OR block costs a $\lfloor n/2\rfloor$-controlled X whose synthesis does not depend on $m$, so the layering pays off only from $n\approx7$, and the gain is a depth-for-width trade, $\lfloor n/2\rfloor$ ancillas against a single output qubit. By main-text Proposition~4 the construction is optimal to within a logarithmic factor at its width. Inside Le Gall's triangle-finding algorithm \citesupp{S-LeGall2014} it gives $\Otil(n^{5/4})$ oracle calls of depth $O(n)$ each; we state this as a bound on oracle depth and not as a time complexity, because the non-query cost of the nested quantum walk has not been analyzed in the circuit model (Supplementary Note~8).

\begin{figure}[!htbp]
\centering
\includegraphics[width=0.9\textwidth]{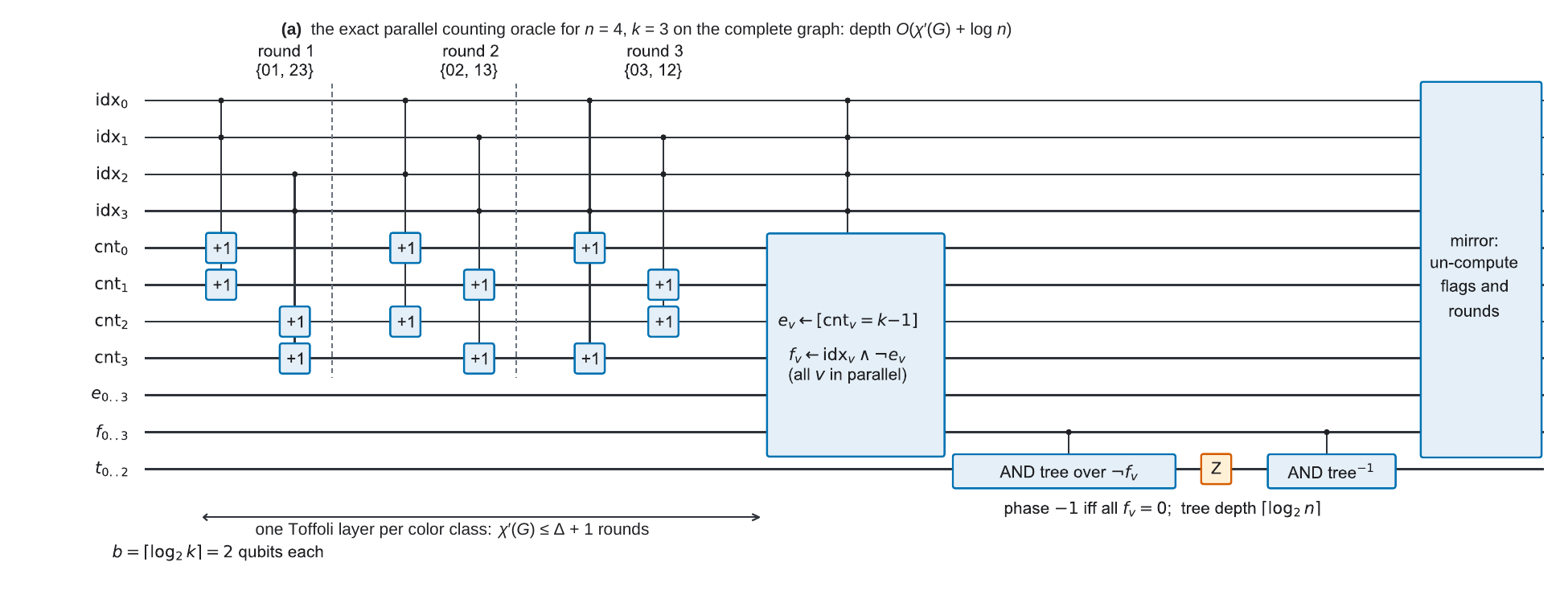}\\[2pt]
\includegraphics[width=0.96\textwidth]{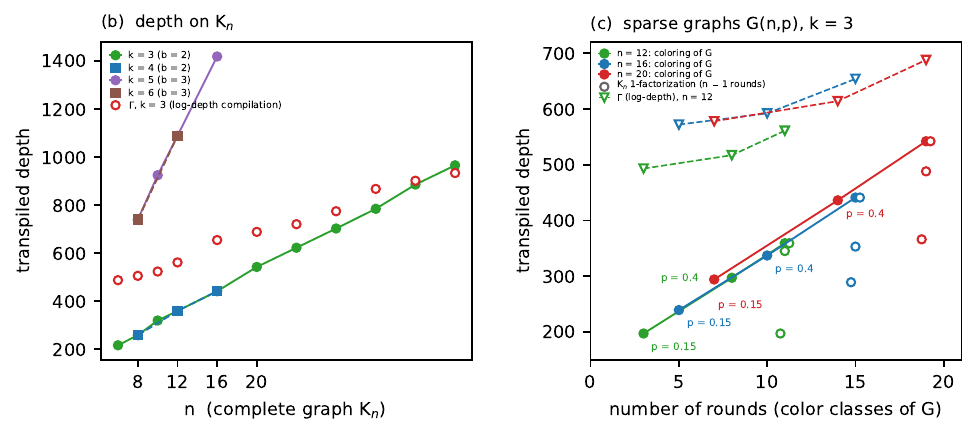}
\caption{\textbf{The exact parallel counting oracle (main-text Proposition~6).} \textbf{a}, Circuit for $n=4$, $k=3$ on the host $K_{4}$: in each of the $\chi'(K_{4})=3$ rounds every edge $uv$ of the matching adds $\idx_{u}$ to the counter of $v$ and $\idx_{v}$ to the counter of $u$ (boxes $+1$, controls on $\idx$); all increments of a round act on disjoint qubits and form one layer. The comparison bits $e_{v}=[\mathtt{cnt}_{v}=k-1]$, the flags $f_{v}=\idx_{v}\wedge\neg e_{v}$, the AND tree on the ancillas $t$ and the $Z$ on its root apply the phase $-1$ exactly when $\supp(y)$ is a clique; the second half of the circuit un-computes everything. \textbf{b}, Transpiled depth on $K_{n}$ for $k=3$--$6$ ($n\le20$; $k=3$ and $\Gam$ measured up to $n=40$) (basis $\{R_{z},\sqrt X,X,\mathrm{CX}\}$, optimization level 1, all-to-all coupling), compared with the graph-state oracle $\Gam$ of Part~II in the log-depth compilation of main-text Proposition~13 ($k=3$). \textbf{c}, Depth on random graphs $G(n,p)$, $k=3$, against the number of rounds: with a proper edge coloring of $G$ (filled, greedy coloring with at most $2\Delta-1$ classes) the depth follows $\chi'(G)$; with the $K_{n}$ 1-factorization (open, always $n-1$ rounds) it does not; triangles: $\Gam$ in the compilation of main-text Proposition~13 on the same graphs, which follows $\chi'(G)$ with a larger constant.}
\label{S:fig:counting}
\end{figure}

\paragraph{The counting oracle, compiled.} The proof is in Supplementary Notes~4 and~14; both variants were verified by statevector simulation on graphs with and without $k$-cliques. Fig.~\ref{S:fig:counting}b,c gives the compiled cost with the settings used throughout the paper. On $K_{n}$ the depth is $216$--$542$ for $n=6$--$20$ and $k=3$ ($\approx23n+76$; identical for $k=4$) and $739$--$1\,418$ for $k=5,6$ ($b=3$), against $89n+355$ for the graph-state oracle of Part~II compiled block by block and $\approx13.7n+403$ in its log-depth compilation (main-text Proposition~13): the counting oracle is the shallowest construction in this paper for $n\le20$, by a factor of $1.1$--$2.3$ relative to the latter. On random graphs the depth follows the number of color classes, $\approx21$ layers per class plus $\approx150$ for flags and tree at $n=20$, so that a graph with $\Delta=6$ costs $294$ layers against $542$ on $K_{20}$ (Supplementary Table~S7). Main-text Proposition~6 settles the depth question for counting oracles: for fixed $k$ nothing is gained in depth or width by giving up exactness, and the regime left open by main-text Proposition~4 is only $k=\omega(1)$, where the counters cost $\Theta(n\log k)$ width and $\Theta(n\,\mathrm{poly}\log k)$ depth.

\paragraph{The size ceiling at $s=1$.} Two rows probe the size ceiling beyond the $n\le8$ protocol: twelve instances each at $(10,6)$ (mouse; $r^{*}$ up to $11$, the deepest amplification in the data set) and $(12,10)$ (macaque; the mouse graph contains no $10$-clique in $4\times10^{5}$ samples). The single-register oracle shows no width dependence: median $P(r^{*})=0.64$ ($95\%$ CI $0.62$--$0.67$) at $\langle r^{*}\rangle\approx10$ and $0.87$ ($0.86$--$0.89$) at $\langle r^{*}\rangle\approx5.5$, the loss tracking the iteration count within the $4\,r\bar\varepsilon$ of Conjecture~S6, not the width, while the plain oracle collapses (median $\Gam$ of $0.09$ and $0.29$), so the filter's advantage widens with size exactly as the near-clique leakage of main-text Theorem~8 predicts. Exact simulation is what bounds the table: an $(11,6)$ instance costs about fifteen minutes of block-simulator time and $s\ge2$ at $n\ge9$ exceeds our single node, so statements beyond these widths rest on the constructions and the guarantees, not on statevectors. Main-text Corollary~9 was checked exactly on $4\,500$ clique-free subgraphs drawn by the same procedure ($300$ per configuration): the smallest phase $1-2a^{2}$ applied by $\Gam$ to any candidate was $0.21875=7/32$ and the largest $|a|$ was $0.625=5/8$, i.e.\ the bounds of main-text Corollary~9 and main-text Theorem~8 are attained but never violated, and no candidate is ever fully marked. The instances are small ($n\le8$, at most $28$ qubits) and the simulations are noiseless; at these sizes the exact counting oracles are cheaper than $\mathcal O_{2}$ in every metric, and the asymptotic statements rest on Theorems~8, 10 and~11, not on main-text Table~2 (Supplementary Note~9).

\FloatBarrier
\section*{Supplementary Note 14: proofs of Theorem~11 and of Propositions~4, 6 and~13}\label{S:note:14}

This note proves Theorem~11(i) and (iii) and Propositions~4, 6 and~13 of the main text; notation follows the main text.

\paragraph{Proof of Theorem~11(i) of the main text.}
Let $O^{*}$ be the ideal phase oracle and $D$ the diffuser. The ideal evolution keeps the ancillas in $\ket0$, and on states $\sum_{y}c_{y}\ket{y,0}$ the operator $\Ot^{(s)}-O^{*}$ acts on orthogonal sectors with norms $\|-2\varepsilon_{y}\ket{y,0}+\ket{g_{y}}\|=2\sqrt{\varepsilon_{y}}$ (and $0$ on cliques), so the hybrid argument gives $\|(D\Ot^{(s)})^{r}\ket{\sigma,0}-(DO^{*})^{r}\ket{\sigma,0}\|\le2r\sqrt{\varepsilon_{t}^{\,s}}$, and a projective probability changes by at most twice the state distance. Part (ii) is the main theorem of \citesupp{S-Hoyer2003} applied to the base algorithm that runs $\mathrm{QPE}_{2}$ and writes $[\est=0]$ into an output qubit, which has one-sided error at most $2/27<1/3$.

\paragraph{Proof of Theorem~11(iii) of the main text.}
Write $O$ for $\Ot^{(s)}$, $P_{0}$ for the projector onto the search sector $\mathrm{span}\{\ket y\otimes\ket0_{\mathrm{anc}}\}$, and $P_{\perp}=I-P_{0}$. Three structural facts. The diffuser is $D_{0}\otimes I_{\mathrm{anc}}$, so it commutes with $P_{0}$ and never moves amplitude between the sectors. $O$ is block-diagonal over the $\idx$ basis, $O=\sum_{y}\ket y\!\bra y\otimes O_{y}$, because every gate of $\Gam$ and of the estimation acts on the ancillas or is controlled by $\idx$. And within the block of $y$, main-text Theorem~10 gives $\bra0O_{y}\ket0=\sigma_{y}(1-2\varepsilon_{y}^{\,s})$ with $\sigma_{y}=-1$ for cliques ($\varepsilon_{y}=0$) and $+1$ otherwise, so unitarity of $O_{y}$ gives $\|P_{\perp}O_{y}\ket0\|=\|\bra0O_{y}P_{\perp}\|=2\sqrt{\varepsilon_{y}^{\,s}(1-\varepsilon_{y}^{\,s})}\le2\varepsilon^{s/2}$, where $\varepsilon$ bounds every non-clique $\varepsilon_{y}$ (worst case $2/27$, or the instance's own maximum). Split $\psi_{r}=(DO)^{r}\psi_{0}$ into $\varphi_{r}=P_{0}\psi_{r}$ and $\chi_{r}=P_{\perp}\psi_{r}$; then $\varphi_{r+1}=D\,O_{\mathrm{id}}(I-2E)\varphi_{r}+D\rho_{r}$ with $E=\sum_{y}\varepsilon_{y}^{\,s}\ket y\!\bra y\otimes\ket0\!\bra0$, $O_{\mathrm{id}}=\sum_{y}\sigma_{y}\ket y\!\bra y\otimes I$ and $\rho_{r}=P_{0}OP_{\perp}\chi_{r}$, and $\chi_{r+1}=D(P_{\perp}O\varphi_{r}+P_{\perp}O\chi_{r})$. Hence $\|\chi_{r}\|\le2\varepsilon^{s/2}r$ and $\|\rho_{r}\|\le2\varepsilon^{s/2}\|\chi_{r}\|\le4\varepsilon^{s}r$. Comparing with the ideal trajectory $\psi^{\mathrm{id}}_{r}=(DO_{\mathrm{id}})^{r}\psi_{0}$, which stays in the search sector, $\Delta_{r}=\varphi_{r}-\psi^{\mathrm{id}}_{r}$ obeys $\|\Delta_{r+1}\|\le\|\Delta_{r}\|+2\|E\varphi_{r}\|+\|\rho_{r}\|\le\|\Delta_{r}\|+2\varepsilon^{s}+4\varepsilon^{s}r$, so $\|\Delta_{r}\|\le2\varepsilon^{s}r^{2}$. Since $P(r)\ge\|\Pi_{\mathrm{good}}\varphi_{r}\|^{2}$ and both states have norm at most one, $P(r)\ge P_{\mathrm{ideal}}(r)-2\|\Delta_{r}\|=P_{\mathrm{ideal}}(r)-4\varepsilon^{s}r^{2}$. Only the per-candidate bounds $\varepsilon_{y}\le\varepsilon$ enter, so $\varepsilon$ may be the maximum over the instance's non-clique candidates. The bound was checked against every recorded sweep of every instance: $5\,588$ (instance, $s$, $r$) points with the instance-wise constant, including the additional two- and three-register runs, with no violation and minimum slack $0.049$, $0.0009$ and below $10^{-4}$ for $s=1,2,3$ (script \texttt{24}). \qed

\paragraph{Proof of Proposition~6 of the main text.}
Fix a proper edge coloring of $G$ with classes $C_{1},\dots,C_{\chi'}$. Within a class every vertex meets at most one edge, so the $2|C_{c}|$ controlled increments of class $c$ ($\mathtt{cnt}_{v}\mathrel{+}=\idx_{u}$ and $\mathtt{cnt}_{u}\mathrel{+}=\idx_{v}$ for $uv\in C_{c}$) act on pairwise disjoint qubit sets and form one layer; on $K_{n}$ the round-robin 1-factorization gives $n-1$ layers. Increments commute, so after all classes $\mathtt{cnt}_{v}\equiv|N(v)\cap H|\pmod{2^{b}}$ with $H=\supp(y)$. For $v\in H$ this count is at most $k-1<2^{b}$, so no wrap-around occurs where it matters, and $v\in H$ is adjacent to all of $H\setminus\{v\}$ iff $\mathtt{cnt}_{v}=k-1$, i.e.\ iff $e_{v}=1$; for $v\notin H$ the flag $f_{v}=\idx_{v}\wedge\neg e_{v}$ is $0$ regardless of the counter. Hence all flags vanish iff $H$ is a clique; the AND tree computes $\bigwedge_{v}\neg f_{v}$ into its root in $\lceil\log_{2}n\rceil$ Toffoli layers on $n-1$ clean ancillas, the $Z$ on the root applies the phase $-1$ exactly then, and since every step is a permutation of the computational basis the mirror sequence restores tree, flags and counters exactly (a counter that wrapped for some $v\notin H$ is un-wrapped by the same increments in reverse). A controlled $b$-bit increment is $b$ gates with $1,\dots,b$ controls besides the control $\idx_{u}$, of depth $O(\mathrm{poly}\,b)$; for $b\le3$ every such gate has at most two controls once the idle flag $f_{v}$ is used as a clean ancilla for the top bit, so a round is a constant number of Toffoli layers and no multi-controlled gate other than the tree occurs. This gives the depth $O(\chi'(G)\,\mathrm{poly}\log k+\log n)$ and the count of $n(4+b)-1$ qubits ($b$ counter bits, $e_{v}$, $f_{v}$ per vertex and $n-1$ tree ancillas); replacing the tree by an $n$-controlled $Z$ on dirty ancillas \citesupp{S-Iten2016} saves the $n-1$ ancillas at depth $O(n)$ for the final test (Supplementary Note~4, which also lists the compiled cost for $k=3$--$6$ and on sparse graphs). Both variants were verified by statevector simulation for $n=4$, $k\in\{3,4\}$ on $K_{4}$, $K_{4}$ minus an edge, a 4-cycle with a chord and two random graphs: phase exactly $\pm1$ on every weight-$k$ input, with the correct sign and all ancillas restored, to $3\times10^{-13}$ (script \texttt{12}). \qed

\paragraph{Proof of Proposition~13 of the main text.}
Every replacement is a diagonal or classical reversible sub-circuit that acts on the computational basis as the block it replaces. (a) On $\ket{x}$ the dummy gates of $A$ apply $\prod_{d}\prod_{u\in R}(-1)^{x_{d}x_{u}}\prod_{d<d'}(-1)^{x_{d}x_{d'}}=\prod_{d}(-1)^{x_{d}\,\hw(x_{R})}\prod_{d<d'}(-1)^{x_{d}x_{d'}}$; the CNOT tree writes $\hw(x_{R})\bmod2$ into its root without changing the other real qubits' contribution to later gates once it is undone, the $q$ CZ gates on the root apply the first product and the $\binom q2$ CZ gates the second. Its depth is $2\lceil\log_{2}n\rceil+q+O(1)$ and it is Clifford. (b) A two-bit mod-4 counter of a pair $(x_{i},x_{j})$ is $(x_{i}\oplus x_{j},\,x_{i}x_{j})$, computed by a Toffoli into a clean ancilla and a CNOT; two counters $(s_{a},c_{a})$, $(s_{b},c_{b})$ add in place by $c_{b}\mathrel{\oplus}=s_{a}s_{b}$, $c_{b}\mathrel{\oplus}=c_{a}$, $s_{b}\mathrel{\oplus}=s_{a}$; a binary tree of these steps has $\lceil\log_{2}N\rceil$ levels of constant depth, its root $(S,C)$ is $\hw(x)\bmod4$, which is added into $\rem$ by a Toffoli and two CNOTs, and undoing the CNOTs alone restores $\inp=x$ while the carries, a fixed function of $x$, stay in the $\lceil N/2\rceil$ ancillas. Since $A$ acts on $\inp$ only, $\IP^{\dagger}$ finds the same $x$ and the same carries, re-forms the sums with the CNOTs, subtracts from $\rem$ and undoes the whole tree, which returns the ancillas to $\ket0$; the Toffoli count per pass is $N-1$, as for the chain. (c) The $N$ Toffolis of step~3 share the control $\rem_{1}$; copying $\rem_{1}$ onto $\lfloor N/2\rfloor-1$ ancillas with a CNOT tree lets each of the $\lfloor N/2\rfloor$ controls serve at most three Toffolis, and the copies are undone by the same tree. (d) $S_{0}$ is the phase $-1$ on $\inp=0^{N}$: after $X^{\otimes N}$ an AND tree of Toffolis on $N-1$ clean ancillas computes $\bigwedge_{i}\bar x_{i}$ in $\lceil\log_{2}N\rceil$ layers, a $Z$ on the root (a CZ from $\est$ for the controlled version) applies the phase, and the tree and the $X$ gates are undone. The ancillas of (b), (c) and (d) are used at different times and share one register of $N-1$ qubits. The depth of one $\Gam$ query is therefore $2\chi'(G)$ CZ layers plus $O(\log n)$, the width $3N+1$, and the non-Clifford gates are the same Toffolis and controlled-Hadamards as before plus $2(N-1)$ Toffolis for the AND tree in place of the $O(N)$ Toffolis of the multi-controlled gate. The compiled circuit agrees with the block-by-block one on random superpositions of the search register to $10^{-13}$ (script \texttt{21}). \qed

\paragraph{Proof of Proposition~4 of the main text.}
Over a gate set with $s$ gates of arity at most $c$, a circuit of $g$ gates on $w$ qubits is specified by at most $g\log_{2}(s\,w^{c})$ bits. The $2^{\binom n2}$ graphs give $2^{\binom n2}$ distinct oracle unitaries, so $g\log_{2}(sw^{c})\ge\binom n2$ for some $G$, i.e.\ $g=\Omega(n^{2}/\log w)$; since $g\le d\,w$ the claim follows. A binary-to-one-hot decoder on $2\lceil\log_{2}n\rceil$ address qubits costs $O(\log n)$ depth and $O(n)$ ancillas, so our oracle also accepts binary-encoded queries at no asymptotic cost in depth.

\paragraph{Remarks on main-text Proposition~4.} main-text Proposition~4 is a Shannon-type counting argument \citesupp{S-Shannon1949,S-Basu2023}: the number of circuits of a given size over a finite gate set is finite, while the number of distinct oracle unitaries is $2^{\binom n2}$. Its proof is in Supplementary Note~14. It extends verbatim to $k$-clique oracles: adding one universal vertex to every graph on $n-1$ vertices keeps all $2^{\binom{n-1}2}$ graphs distinguishable by their $k$-subsets, so any family of $k$-clique oracles of width $w$ also has depth $\Omega(n^{2}/(w\log n))$ for some graph.

\FloatBarrier
\section*{Supplementary Note 15: computational protocols}\label{S:note:15}

This note gives the complete protocols summarized in main-text Sect.~8.1.

\paragraph{Exact statevector simulation.} $\idx$ appears in $\Gam$ (and in $\Ot$) only as a control, so $\Gam=\sum_{y}\ket y\bra y\otimes G_{y}$ is block-diagonal in $\idx$, and the diffuser $I-2\ket\sigma\bra\sigma$ acts on $\idx$ alone. The state therefore always has the form $\sum_{y\in\supp\sigma}\ket y\otimes v_{y}$ with $v_{y}\in\mathbb C^{2^{t}\cdot4\cdot2^{N}}$, and one iteration costs one application of $G_{y}$ per row and one rank-one update; within $G_{y}$ the counter, the conditional complement and $S_{0}$ are basis permutations or diagonal, and only the controlled Hadamards require arithmetic. The simulator agrees with a gate-for-gate Qiskit \citesupp{S-Qiskit2024} statevector run of the full circuits to $|\Delta P|<8.1\times10^{-14}$ on all output distributions tested ($n=4,5$, $r=1,2$, both initializations discussed below; on a cluster also up to $(n,k)=(9,7)$ with $28$ qubits, where the amplitudes agree to $1.9\times10^{-16}$), and is about $2000$ times faster. All probabilities reported are exact; $\mathcal{O}_{t}$ is simulated with the $\est$ register carried explicitly.

\paragraph{Host graphs, sampling and success criteria.} Statistics of the two host graphs were computed from the raw edge lists after removing duplicate lines and self-loops (Supplementary Table~S10); the densities quoted by the repository ($0.486$ and $0.998$) count duplicates. Both graphs are used with their full vertex sets ($91$ and $1\,076$ vertices). For each $(n,k)$ and host, $n$-subsets of the vertex set were drawn uniformly at random with a fixed seed (script \texttt{15}), in batches of $20\,000$; each induced subgraph was relabeled and its number $\mathcal M$ of $k$-cliques computed exactly. Subgraphs with $\mathcal M=0$ were set aside for the false-positive check ($300$ per configuration); subgraphs with $\mathcal M/\mathcal N\ge1/2$ were counted as degenerate and excluded, because for them $\theta\ge\pi/4$ and the ideal iteration count is zero; the remaining subgraphs were de-duplicated up to graph isomorphism, and sampling stopped at $60$ pairwise non-isomorphic instances or at $4\times10^{5}$ samples; for $(7,4)$ and $(8,3)$, the two configurations with the most expensive filtered-oracle sweeps, the first $30$ instances were simulated. The numbers of samples, clique-bearing, degenerate and distinct subgraphs per configuration are stored with the instances. For every instance, $\theta=\arcsin\sqrt{\mathcal M/\mathcal N}$, $r^{*}=\lfloor\pi/(4\theta)\rfloor$ (the integer nearest to $\pi/(4\theta)-\tfrac12$, i.e.\ the optimum of the ideal oracle; $r^{*}\ge1$ once degenerate instances are excluded), and $P(r)$ was computed for $r\le3r^{*}+6$ with $\Gam$, for $r\le2r^{*}+4$ with $\mathcal O_{2}$ and for $r\le r^{*}+2$ with $\mathcal O_{3}$, whose maxima lie at $r^{*}$ or its neighbor; $\mathcal O_{3}$, the secondary comparison, was run on all instances of the cheaper configurations and on the first $20$, $12$ and $6$ instances of $(7,3)$, $(7,4)$ and $(8,3)$ (script \texttt{16}). Success is the exact probability that a measurement of $\idx_{0..n-1}$ returns a $k$-clique; a \emph{false positive} would be a candidate that is fully marked without being a $k$-clique, and the check of main-text Corollary~9 computes $\min_{y}\lambda_{y}$ and $\max_{y}|a_{y}|$ over all candidates of every stored clique-free subgraph with the initialization of the algorithm. The initial shot-sampled benchmark (with a ``top-ten'' success criterion, the search register initialized as $\ket{D_{N,k+q}}$ on all $N$ qubits, and the mouse-retina subgraphs drawn from vertex ids $0$--$90$ only) is analyzed in Supplementary Notes~5 and~6; the exact simulations of the main text replace it.

\paragraph{Transpilation and resource counts.} Resource counts use Qiskit's transpiler \citesupp{S-Qiskit2024} with basis $\{R_{z},\sqrt X,X,\mathrm{CX}\}$, optimization level 1 and a fixed seed, under all-to-all coupling or a linear chain. Multi-controlled gates are synthesized by Qiskit's default high-level-synthesis plugin, which borrows qubits of the circuit that are idle at the time of the gate as clean or dirty ancillas \citesupp{S-Iten2016,S-Khattar2025}; in $\Gam$ and in the edge detector the idle $\idx$ (and $\rem$) qubits serve as dirty ancillas for $S_{0}$ and for the OR block, which is what makes their compiled depth linear (a standalone $13$-controlled X, with no idle qubits to borrow, transpiles to depth $1\,789$ against $106$ inside a circuit with $13$ idle qubits). The V-chain numbers of Fig.~\ref{S:fig:mcx} use $c-2$ clean ancillas. All Clifford$+T$ numbers of the article use one normalization: every multi-controlled gate is re-synthesized into Toffolis (three Toffolis and a clean ancilla for the three-controlled gates of the sequential counter), and non-Clifford gates are then counted after transpilation as $R_{z}$ rotations whose angle is not a multiple of $\pi/2$, each an odd multiple of $\pi/4$, i.e.\ a $T$ gate ($7$ per Toffoli, $2$ per controlled-Hadamard). For $\Gam$, $\mathcal O_{2}$ and the parallel counting oracle the circuits are already of this form (script \texttt{13}); for the sequential counter of \citesupp{S-Metwalli2020} the re-synthesis gives $T$-counts of exactly $28n(n-1)+7$ ($1\,575$, $3\,703$ and $6\,727$ at $n=8,12,16$), which is what main-text Fig.~2b and the crossovers of main-text Sect.~7 use. The circuit Qiskit emits without the re-synthesis mixes in $\pi/8$ rotations ($420$ of $1\,197$ non-Clifford rotations at $n=8$, $1\,800$ of $5\,077$ at $n=16$), each of which costs many $T$ gates; those mixed-basis counts appear only in Supplementary Note~12. The oracle of \citesupp{S-Metwalli2020} was re-implemented from its published description; $\mathcal O_{2}$ was built gate by gate from Fig.~\ref{S:fig:gammaSI}b--c and its diagonal elements agree with main-text equation~(7) to $6.5\times10^{-14}$ on an $n=4$ instance (script \texttt{12}); the log-depth compilation of main-text Proposition~13 (script \texttt{21}) was compared with it by statevector on random superpositions of $\idx$ with all ancillas returning to $\ket0$.

\FloatBarrier
\section*{Supplementary Note 16: circuit diagrams of the constructions}\label{S:note:16}

This note collects the gate-level circuit diagrams of main-text Sects.~3 and~4 and the derivation of the Input Preparator state.

\begin{figure}[!htbp]
\centering
\begin{minipage}[t]{0.48\textwidth}\centering\includegraphics[width=\linewidth]{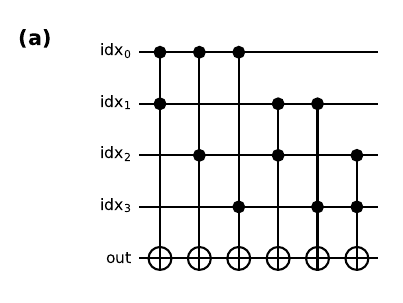}\end{minipage}\hfill
\begin{minipage}[t]{0.50\textwidth}\centering\includegraphics[width=\linewidth]{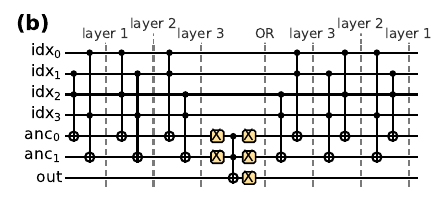}\end{minipage}\\[6pt]
\begin{minipage}[t]{0.52\textwidth}\centering\includegraphics[width=\linewidth]{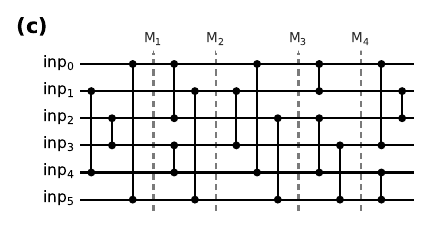}\end{minipage}\hfill
\begin{minipage}[t]{0.46\textwidth}\centering\includegraphics[width=\linewidth]{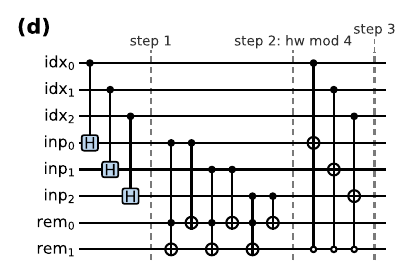}\end{minipage}
\caption{\textbf{Circuit primitives.} \textbf{a}, Naive edge-detection oracle for $K_{4}$: one Toffoli per edge on the common target $\mathtt{out}$ (depth $m=6$); after conjugating $\mathtt{out}$ by Hadamards this is a hypergraph-state unitary. \textbf{b}, The same oracle scheduled by the $\chi'(K_{4})=3$ perfect matchings $\{01,23\},\{02,13\},\{03,12\}$ (main-text Definition~2; dashed barriers): within a layer the Toffolis act on disjoint qubits and write into distinct ancillas; a multi-controlled X with inverted controls and an X on $\mathtt{out}$ (together an ``OR'') set $\mathtt{out}$; the un-computation block, the gate-by-gate mirror image of the first block, restores the ancillas. \textbf{c}, The graph-state gate Alpha, $A_{K_{6}}=\prod_{uv\in E}\mathrm{CZ}_{uv}$ (main-text Definition~2): 15 CZ gates in $\chi'(K_{6})=5$ layers given by the round-robin 1-factorization (dashed barriers $M_{1}$--$M_{4}$ separate consecutive layers); a sparse example with $\Delta+1$ color classes is in Fig.~\ref{S:fig:sparse}. \textbf{d}, Input Preparator for $n=3$: controlled Hadamards (step 1), a two-bit modulo-4 counter of the weight of $\inp$ on $\rem$ (step 2), and Toffolis controlled on $\idx_{j}$ and on $\rem_{1}=0$ (open circle, step 3) that complement $\inp$ within $H$; the circuit is graph-independent. Throughout, filled dots are controls, open circles inverted controls, $\oplus$ targets and gray dashed lines barriers between layers.}
\label{S:fig:primitivesSI}
\end{figure}

\begin{figure}[!htbp]
\centering
\includegraphics[width=0.82\textwidth]{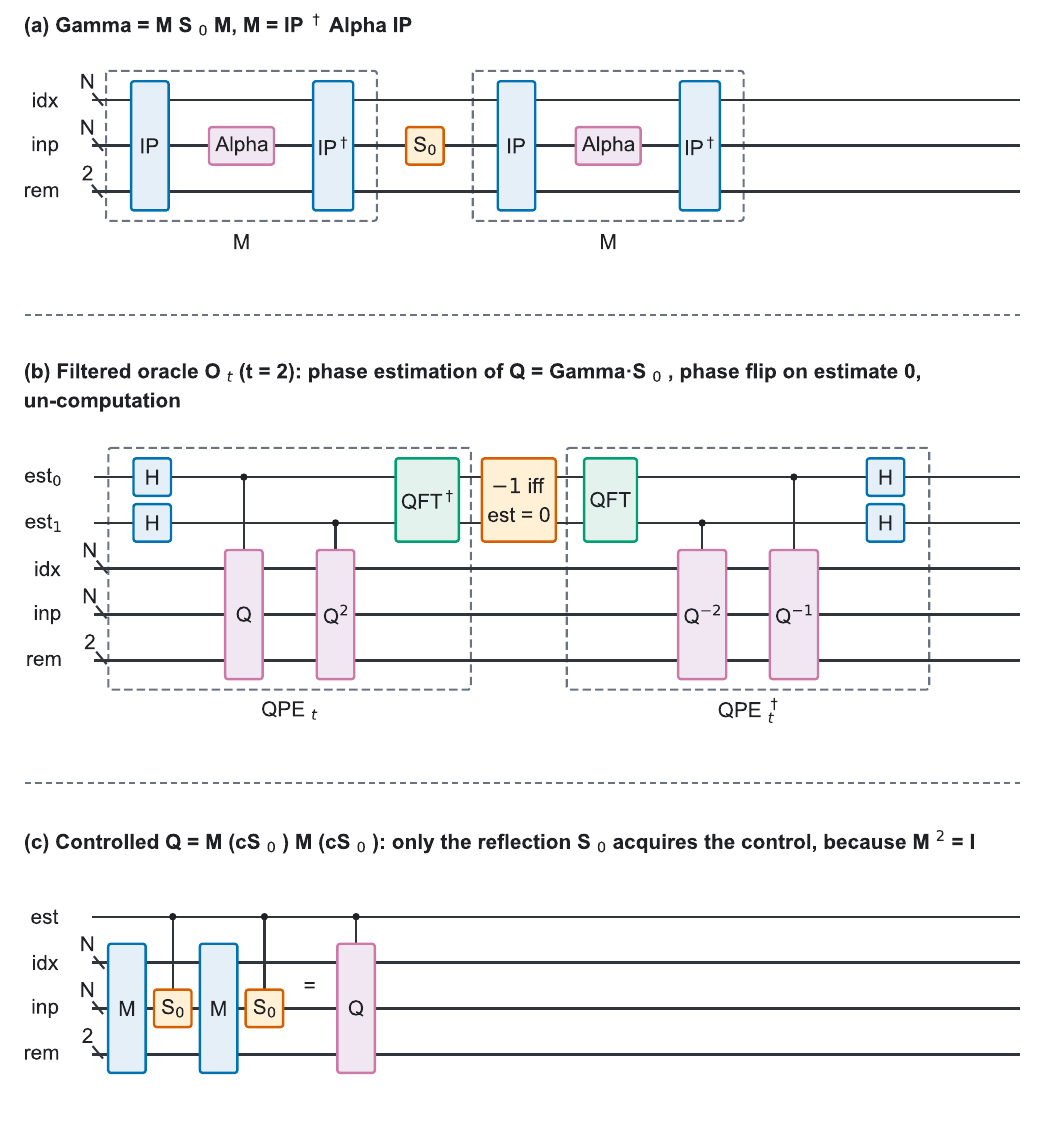}
\caption{\textbf{The clique oracle and the phase-estimation filter.} \textbf{a}, $\Gam=MS_{0}M$ with $M=\IP^{\dagger}A\,\IP$ and $S_{0}$ the inverted-control multi-controlled Z on $\inp$; because $A^{2}=I$, $M^{2}=I$ and $\Gam$ is a reflection on the invariant plane (main-text Proposition~7). \textbf{b}, The filtered oracle $\Ot$ for $t=2$: $t$-bit phase estimation of $Q=\Gam S_{0}$, a phase flip conditioned on the estimate being $0$, and un-computation. The eigenphase of $Q$ is exactly $0$ for a clique and at least $102.6^{\circ}$ in modulus otherwise (main-text Theorem~8), so $\Ot$ marks cliques with phase exactly $-1$ and any other $H$ with probability $\varepsilon_{t}\le2/27$ (main-text Theorem~10). \textbf{c}, Only the reflection $S_{0}$ needs the control qubit: $M(\mathrm{c}S_{0})M(\mathrm{c}S_{0})$ equals $Q$ when the control is $1$ and $M^{2}=I$ when it is $0$.}
\label{S:fig:gammaSI}
\end{figure}

\paragraph{Edge detection.} The naive circuit places one Toffoli gate per edge, controlled on $\idx_{a},\idx_{b}$ and targeting $\mathtt{out}$ (Fig.~\ref{S:fig:primitivesSI}a). After conjugating the target by Hadamards it is the third-order hypergraph-state unitary $\prod_{ab\in E}\mathrm{CCZ}_{a,b,\mathtt{out}}$ \citesupp{S-Rossi2013,S-Qu2013,S-Guhne2014}, and its depth is $m$ because every gate shares the target. Our circuit (Fig.~\ref{S:fig:primitivesSI}b) is scheduled by the matchings of main-text Lemma~3: the Toffolis of one matching write into $\lfloor n/2\rfloor$ \emph{distinct} ancillas $\mathtt{anc}$, one multi-controlled X with inverted controls followed by an X on $\mathtt{out}$ (together an OR) collapses the ancilla register onto $\mathtt{out}$, and the gate-by-gate mirror image of the first block un-computes the ancillas.

\paragraph{The Input Preparator state.} For $H=\supp(y)$ with $|H|=K\equiv3\pmod4$, step 1 of $\IP$ creates the uniform superposition over the $2^{K}$ subsets $x\subseteq H$ on $\inp$; step 2 writes $\hw(x)\bmod4$ on $\rem$ ($\rem_{1}$ the most significant bit); step 3 replaces $x$ by $H\setminus x$ when $\rem_{1}=0$, i.e.\ when $\hw(x)\equiv0,1\pmod4$, which gives $\hw(H\setminus x)\equiv3,2$. Every admissible subset is thus reached twice, from itself and from its complement, with two different $\rem$ values, which yields main-text equation~(1) with total weight $2\cdot2^{-K}=2^{-(K-1)}$ per subset and main-text equation~(2) because $A$ does not act on $\rem$.

\FloatBarrier
\bibliographystylesupp{quantum}
\bibliographysupp{refs_supp}

\end{document}